\documentclass[12pt]{article}
\usepackage{enumerate}
\usepackage{graphicx}
\usepackage{mathtools}
\begin{document}
\def\be{\begin{equation}}
\def\bea{\begin{eqnarray}}
\def\ee{\end{equation}}
\def\eea{\end{eqnarray}}
\def\d{\partial}
\def\eps{\varepsilon}
\def\la{\lambda}
\def\b{\bigskip}
\def\nn{\nonumber \\}
\def\p{\partial}
\def\t{\tilde}
\def\h{{1\over 2}}
\def\be{\begin{equation}}
\def\bea{\begin{eqnarray}}
\def\ee{\end{equation}}
\def\eea{\end{eqnarray}}
\def\Ab{{\bar{A}}}
\def\Cb{{\bar{C}}}
\def\rb{{\bar{r}}}
\def\b{\bigskip}
\def\u{\uparrow}
\newcommand{\comment}[2]{#2}

\begin{center}

\vspace{1cm}
{\Large\bf
Charged wormholes in higher dimensions
}
\vspace{1.4cm}

{Rhucha Deshpande\footnote{rdeshpande@albany.edu} and Oleg Lunin\footnote{olunin@albany.edu} }\\

\vskip 0.5cm

{\em  Department of Physics,\\
University at Albany (SUNY),\\
Albany, NY 12222, USA
 }

\vskip 0.08cm

\vspace{.2cm}

\end{center}

{\baselineskip 11pt
\begin{abstract}
\noindent
%AT
We construct regular solutions of Einstein's equations connecting two copies of the $AdS_d$ space. Our geometries approach $AdS_{d-p} \times S^p$ in the interior, and topologically they have wormhole-like structures with non-contractible spheres $S^p$. Being inspired by charged branes, these solutions are supported by higher form gauge fields and a $d$-dimensional cosmological constant. Our ansatz also covers regular wormholes connecting two copies of $AdS_{q} \times S^{d-q}$ with various values 
of $q$, and we uncover an interesting phase structure of initial conditions leading to different interpolations.
\end{abstract}
}

%\maketitle

\newpage

\setcounter{tocdepth}{2}
\tableofcontents

\newpage

\section{Introduction}

Maximally symmetric spaces and their products have played important roles in improving 
our understanding of string theory, physics of black holes, and strongly coupled systems via AdS/CFT duality. In string theory, one can obtain products of spheres and AdS spaces by taking near horizon limits of black branes \cite{Mald}. By applying such limits to multiple stacks, one can find geometries that interpolate between $AdS_p\times S^q$ solutions with different radii \cite{CoulLars,LLM}, and even the dimensions $(p,q)$ of the resulting ingredients may be different, as in the case of bubbling geometries in M theory \cite{LLM}. Most results for such interpolations have been obtained in supersymmetric settings. In this article we will explore non-supersymmetric geometries interpolating between $AdS_p\times S^q$ with different values of $(p,q)$, and we will uncover interesting wormhole--like structures of the resulting space--times.

Our study is motivated by two types of questions.  First, we are interested in general mechanisms of resolving singularities by dissolution of branes into fluxes. The most famous example of this scenario in the regular near--horizon limit of D3 branes that gave rise to the AdS/CFT correspondence \cite{Mald,AdScft}. For geometries with 
$AdS_p\times S^q$ asymptotics, there is another ``bubbling'' mechanism that ensures smooth ending of space due to contraction of spheres \cite{LLM}. In both scenarios branes are replaced by fluxes, and supergravity solutions are regular and horizon--free. Extending the bubbling construction to four dimensions, one finds a different mechanism for regularization: instead of terminating due to a collapse of a sphere, fluxes create wormhole--type structures that interpolate between two spacial infinities \cite{BblAdS2}. This is not very surprising since the $AdS_2$ space itself has a disconnected boundary, and this fact has interesting consequences for the field theory dual \cite{AdS2dual}. It is natural to ask whether one can extend this regularization mechanism involving several boundaries to higher dimensions, and in this article we will do that, although in a limited setting. The configurations of D3 branes discussed in \cite{BblAdS2} were supersymmetric, so they had an entire moduli space, which resulted in a rich structure of supergravity solutions. Here we will focus on regular solutions without supersymmetry, so they will be specified by a small number of constants. 

The second motivation for our study comes from the desire to understand dynamics of branes in the presence of a cosmological constant. In contrast to geometries describing asymptotically flat black branes, which are known explicitly \cite{HorStrom}, solutions with cosmological constant have been found only numerically \cite{HorMann,DeLu}\footnote{Article 
\cite{DeLu} also presented analytical results for some special cases, but they are not nearly as explicit as the Horowitz--Strominger solutions \cite{HorStrom} with flat asymptotics.}.
As expected, most of these neutral solutions had a singularity hidden behind a horizon. In 
the asymptotically flat case, regularization occurred only for the extremal branes, so it is natural to look for charged extensions of solutions found in \cite{HorMann,DeLu}. Although some charged black strings with cosmological constant have been constructed in \cite{RaduCharge}, these solutions still have horizons, and it is not clear whether the ansatz presented in these articles admits any regular horizon--free geometries. In contrast to this earlier work, we use inspiration from the asymptotically--flat extremal solution to require the charged geometries to approach $AdS_a\times S^b$ in the interior while having $AdS_p\times S^q$ asymptotics. For 
$(a,b)=(p,q)$, we recover some known solutions, but more generally, we will construct nontrivial flows between different $AdS\times S$ spaces in the presence of fluxes and cosmological constant. For 
$\Lambda=0$, we recover the well--known near horizon limits of various extremal branes. 

Although our main motivation comes from studying black branes and regularization mechanisms, our results also contribute to the existing literature on wormhole geometries. Study of wormholes has a long history \cite{WormOld}, and recently these objects attracted additional attention in the context of the ER=EPR conjecture \cite{EREPR}. While many results have been obtained in two and four dimensions \cite{4dWH,4dWHQ}, the higher dimensional case has not been explored as much, with few exceptions \cite{WHdim}. Our article fills this gap by constructing several classes of higher dimensional wormhole geometries. 

This paper has the following organization. In section \ref{SecSetup} we introduce the ansatz used throughout the article, write equations of motion, and present several exact solutions. The remaining part of the paper is dedicated to construction of geometries interpolating between such solutions. Section \ref{SecFlowSumry} presents the summary of such interpolations. In section \ref{SecMagn} we use a combination of analytical and numerical techniques to construct all regular interpolating solutions with magnetic charges and a cosmological constant. All these geometries have wormhole--like structures. Section \ref{SecElectr} is dedicated to study of regular solutions supported by a cosmological constant and an electric charge. In this case, geometries end via smooth collapses of spheres, in contrast to the wormhole--like mechanisms encountered in sections \ref{SecMagn} and \ref{SecDyon}. Finally, in section \ref{SecDyon} we construct solutions which carry both magnetic and electric charges, and we uncover a rich phase structure for parameters governing such geometries. Some technical details are presented in the Appendix.

\section{Setup and summary}
\label{SecSetup}

\renewcommand{\theequation}{2.\arabic{equation}}
\setcounter{equation}{0}

Over the last three decades, Anti-De-Sitter space has played a very important role in string theory as a near--horizon limit of branes \cite{Mald} and as a part of the exact Freund--Rubin--type geometries 
$AdS_m\times X$ supported by fluxes \cite{FrRub}. Alternatively, $AdS_d$ arises as a maximally--symmetric solution of $d$--dimensional Einstein's equations with a negative cosmological constant. In this article we will explore an interplay between these scenarios. Specifically, we will study $d$--dimensional systems with a cosmological constant and various fluxes, and we will construct regular geometries interpolating between $AdS_m\times S^n$ in the interior and $AdS_{m+n}$ or $AdS_{m-1}\times S^{n+1}$ at infinity. Our solutions exhibit an interesting topological interpolation between two copies of infinity through a regular $AdS_m\times S^n$ near horizon region. This structure can be interpreted as a higher dimensional version of wormholes. In this section we will present the general setup and the summary of our results. The detailed analysis of various flows is performed in the remaining part of this article. 

\subsection{Ansatz and Einstein's equations}

We begin with recalling the geometry of the $AdS_d$ space with radius $L$. In global coordinates, the metric can be written as
\bea\label{AdSmetr1}
ds^2=L^2\left[-(1+r^2)dt^2+\frac{dr^2}{r^2+1}+r^2 d\Omega^2_{d-2}\right],
\eea
and the range of $r$ depends on $d$. For $d>2$, collapse of the sphere ensures termination of the $r$ direction at zero, so $r\in [0,\infty)$ covers the entire geometry. In contrast to this, for $d=2$, $r$ ranges from negative to positive infinity, so $AdS_2$ has a disconnected boundary. If $AdS_2$ is obtained as a near--horizon geometry of the extremal Reissner--Nordstrom black hole,
\bea\label{AdSmetr2RN}
ds^2=-f dt^2+\frac{dr^2}{f}+r^2 d\Omega_2^2,\quad 
f=1-\frac{2GM}{r}+\frac{G q^2}{r^2},\quad M=\frac{q}{\sqrt{G}}\,,
\eea
then only the Poincare patch of the $r>0$ region is recovered\footnote{The same Poincare patch arises in the near--horizon limit of intersecting D3 branes \cite{D3intrst}.}. If one can modify the geometry (\ref{AdSmetr1}) at small values of $r$ to prevent a collapse of the sphere, then it might be possible to introduce a second asymptotic region at negative $r$ even for $d>2$, making the situation similar to the two--dimensional case. 
The resulting geometry can be interpreted as a wormhole interpolating between two copies of $AdS_d$ asymptotics at $r=\infty$ and $r=-\infty$. In this article we will explore a more general version of such a wormhole by imposing an ansatz 
\bea\label{AnstzMetr1}
ds^2=A\, d\Sigma_m^2+\frac{dr^2}{B}+C d\Omega^2_{n},
\eea
where $d\Sigma_m^2$ is the metric of an $m$--dimensional AdS space with unit radius. For $m=1$ and a specific choice of functions $(A,B,C)$, our ansatz recovers the metric (\ref{AdSmetr1}). The wormhole geometries of the form (\ref{AnstzMetr1}) with $AdS_{m+n+1}$ asymptotics approach
\bea\label{BounInftyPM}
A\approx L^2(r^2+1),\quad B\approx \frac{1}{L^2}(r^2+1), \quad C\approx  L^2 r^2\quad 
\mbox{at}\quad r=\pm\infty,
\eea
while having non--vanishing and non--singular functions $(A,B,C)$ in the interior. 

The metrics (\ref{AnstzMetr1}) with the boundary conditions (\ref{BounInftyPM}) have been explored in \cite{DeLu}, and here we briefly summarize the results of that investigation. Assuming that there were no matter fields apart from the cosmological constant, it was shown that the ansatz (\ref{AnstzMetr1}) with the boundary conditions (\ref{BounInftyPM}) at $r=+\infty$ necessarily led to singularity in the interior. In contrast to this, the ansatz inspired by black branes\footnote{This metric can be viewed as taking a decompactification limit for
$AdS_m$ in (\ref{AnstzMetr1}) and breaking the resulting $R^{1,m-1}$ into a product $R_t\times R^{m-1}$.},
\bea\label{DeLuAnstz}
ds^2=-A\, dt^2+D[dx_1^2+\dots+dx_{m-1}^2]+\frac{dr^2}{B}+C d\Omega^2_{n}
\eea
admitted the unique regular solution (with a sphere collapsing at some point, i.e., with $C(r_0)=0$) and a family of black brane geometries (with 
$A$ vanishing at the horizon). Prior to the analytical construction of the regular solution (\ref{DeLuAnstz}) and black branes in the presence of the cosmological constant \cite{DeLu}, both types of geometries had been found numerically in the $m=2$ case \cite{HorMann}. Since either $A$ or $C$ vanish for non--singular geometries (\ref{DeLuAnstz}) with or without a horizon, coordinate $r$ cannot be extended to negative infinity. This is not very surprising: since the warp factor $C$ monotonically decreases as one goes to the interior, eventually one reaches either $C=0$ (regular geometry) or some finite value of $C$ and $A=0$ (regular horizon of a black brane). 

To avoid the problem described in the last paragraph, one should create an obstruction for $C$ to decrease. This can be accomplished by putting a magnetic flux through the sphere: as $C$ is getting smaller, the density of such flux grows, preventing the sphere from collapsing.  A well-known example of this phenomenon is the Freund--Rubin solution \cite{FrRub}, for which the magnetic flux keeps the radius of $S^n$ constant. In this article we will consider the metric (\ref{AnstzMetr1}) supported by a cosmological constant and fluxes respecting the symmetries of (\ref{AnstzMetr1})\footnote{One can also add scalar fields 
$\phi_i(r)$ and couple them to $m$-- and $n$--form fluxes, but we will not explore this direction. Some examples of wormhole geometries supported by scalars can be found in \cite{WormScalar}. Four dimensional case of solutions (\ref{TheAnsatz1}) without scalar fields has been discussed in \cite{Example4d}.}
\bea\label{TheAnsatz1}
ds^2&=&A\, d\Sigma_m^2+\frac{dr^2}{A}+C d\Omega^2_{n},\\
F_n&=&q_1 d\Omega_{n},\quad F_m=q_2 d\Sigma_m.\nonumber
\eea 
Going from (\ref{AnstzMetr1}) to (\ref{TheAnsatz1}) we fixed the freedom in reparameterising $r$ by setting $B=A$.  Fluxes satisfy the Maxwell's equations for arbitrary constant values of $(q_1,q_2)$, and the Einstein's equations are 
\bea\label{FullEinstein}
&&d_-g^2+\frac{n_- q_1^2}{2(d-2)C^n}-\frac{m_- q_2^2}{2(d-2)A^m}-\frac{n\dot A\dot C}{4C}+\frac{nA{\dot C}^2}{4C^2}-\frac{m}{2}\ddot A-\frac{nA\ddot C}{2C}=0,
\nn
&&d_-g^2+\frac{n_-}{C}-\frac{mq_1^2}{2(d-2)C^n}-\frac{m_-q_2^2}{2(d-2)A^m}-\frac{m_+{\dot A}{\dot C}}{4C}+\frac{(n-2)A{\dot C}^2}{4C^2}-
\frac{A\ddot C}{2C}=0,\nn
&&d_-g^2-\frac{m_-}{A}+\frac{n_-q_1^2}{2(d-2)C^n}+\frac{nq_2^2}{2(d-2)A^m}-\frac{m_-}{4}
\frac{{\dot A}^2}{A}-\frac{n{\dot A}{\dot C}}{4C}-\frac{1}{2}{\ddot A}=0.
\eea
Here we wrote the negative cosmological constant $\Lambda$ in terms of a convenient parameter $g$ and introduced shortcuts $(m_-,n_-,d_-)$:
\bea\label{LamDefG}
\Lambda=-\frac{(d-1)(d-2)}{2}g^2,\quad m_-=m-1,\quad n_-=n-1,\quad d_-=d-1.
\eea
In the remaining part of this article we will analyze the solutions of the system (\ref{FullEinstein}) and explore  topological properties of the corresponding geometries (\ref{TheAnsatz1}). 
 
\subsection{Exact solutions and asymptotic behavior}
\label{SecExactSoln}
\label{SecExact}

In this subsection we discuss some exact solutions of the system (\ref{FullEinstein}), as well as several types of asymptotic behavior of functions $(A,C)$ at large values of $r$. The resulting geometries (\ref{TheAnsatz1}) can be viewed as fixed points of the system  (\ref{FullEinstein}), and in the next subsection we will discuss flows between them.

\bigskip
\noindent
{\bf A. $AdS_d$ geometry}

As shown in \cite{DeLu}, the only regular neutral solution of the system (\ref{FullEinstein}) is the $AdS_d$ space. In the gauge (\ref{TheAnsatz1}), the geometry is given by
\bea\label{AdSexact}
ds^2=\frac{1}{g^2}\left[(g^4 r^2+1)d\Sigma_m^2+\frac{g^4 dr^2}{g^4 r^2+1}+g^4 r^2d\Omega_n^2\right],\quad F_n=0,\quad F_m=0.
\eea
In particular, functions $A$ and $C$ grow quadratically in $r$, so the expressions
\bea\label{AdSexactAsymp}
A\simeq g^2 r^2+\frac{1}{g^2},\quad C\simeq g^2 r^2
\eea
give an asymptotic solution of the system (\ref{FullEinstein}) at large $r$ even in the presence of 
charges\footnote{The contributions to (\ref{FullEinstein}) containing $(q_1,q_2)$ are suppressed in the limit (\ref{AdSexactAsymp}) since the densities of the electric and magnetic fluxes go to zero.} $(q_1,q_2)$. We will call the asymptotic behavior (\ref{AdSexactAsymp}) the ``$AdS_d$ fixed point''. 

\bigskip
\noindent
{\bf B. $AdS_{m+1}\times S^n$ geometry (magnetic solution)}

For $q_2=0$, the system (\ref{FullEinstein}) admits an exact solution with constant $C$:
\bea\label{AdS32exct1}
A=A_\star r^2+\frac{1}{A_\star}, \quad C=C_\star,\quad A_\star=\frac{d-1}{m}g^2+\frac{n-1}{2m(d-2)}\frac{q_1^2}{C_\star^n}\,.
\eea
Here $C_\star$ is a root of an algebraic equation
\bea\label{AdS32exct2}
(d-1)g^2-\frac{mq_1^2}{2(d-2)C_\star^{n}}+\frac{n-1}{C_\star}=0.
\eea
Substituting (\ref{AdS32exct1}),  (\ref{AdS32exct2}) into the metric (\ref{TheAnsatz1}), one finds the $AdS_{m+1}\times S^n$ geometry with 
\bea
\frac{1}{R_{AdS}^2}=A_\star,\quad \frac{1}{R_{S}^2}=\frac{1}{C_\star}.
\eea
To write the expressions for the radii in a symmetric form, it is convenient to rescale the charge by 
$C_\star^{n/2}$:
\bea\label{AdS32exct3}
ds^2&=&R_{AdS}^2\left[\left(\frac{r^2}{R_{AdS}^4}+1\right) d\Sigma_m^2+\frac{dr^2}{r^2+R_{AdS}^4}\right]+R_S^2\, d\Omega^2_{n},\\
F_n&=&Q_1 R_S^n\, d\Omega_{n},\quad F_m=0.\nonumber
\eea 
In this notation, the radii of the AdS and the sphere are given by
\bea\label{AdS32exct3a}
\frac{1}{R_{AdS}^2}=\frac{(n-1)}{2m(d-2)}Q_1^2+\frac{d-1}{m}g^2,\quad 
\frac{1}{R_{S}^2}=\frac{mQ_1^2}{2(d-2)(n-1)}-\frac{d-1}{n-1}g^2\,.
\eea
Solutions (\ref{AdS32exct3})--(\ref{AdS32exct3a}) were first discussed in \cite{LiuSabra}. 

For $g=0$, one recovers the standard Freund--Rubin solutions, and the ratio
\bea
\frac{R_{S^n}}{R_{AdS_{m+1}}}=\frac{n-1}{m}
\eea
reproduces the well--known results for $AdS_p\times S^q$ with 
\bea
(p,q)=(4,7),\  (7,4),\ (5,5),\ (3,3),\ (2,2).\nonumber
\eea
For $Q_1=0$, the expressions (\ref{AdS32exct3a}) give imaginary values of $R_S$, but upon analytic continuation of the sphere coordinates, one finds the product $AdS_{m+1}\times H_n$ with the correct signature and the correct values of the cosmological constant for AdS and hyperbolic spaces. We are interested in keeping a sphere $S^n$ rather than the hyperbolic space, and for a fixed value of $g$, this imposes a lower bound on $Q_1^2$. Note that the original charge $q_1$ can still vary from zero to infinity. 
We will refer to the solution (\ref{AdS32exct1})--(\ref{AdS32exct2}) as the ``$AdS_{m+1}\times S^n$ fixed point''. 

Since function $A$ in (\ref{AdS32exct1}) grows quadratically at large values of $r$, one concludes that the expressions
\bea\label{AdS32apprx}
A\simeq A_\star r^2, \quad C\simeq C_\star,\quad A_\star=\frac{d-1}{m}g^2+\frac{n-1}{2m(d-2)}\frac{q_1^2}{C_\star^n}\,
\eea
would give an {\it asymptotic} solution of the system (\ref{FullEinstein}) even for nonzero $q_2$. However, in the presence of the electric charge, the solution would deviate from (\ref{AdS32exct1}) in the interior. We will call the asymptotic behavior (\ref{AdS32apprx}) 
the ``$AdS_{m+1}\times S^n$ fixed point''. 

\bigskip
\noindent
{\bf C. $AdS_{m}\times S^{n+1}$ geometry (electric solution)}

If the magnetic charge is switched off, then the system (\ref{FullEinstein}) admits an exact $AdS_{m}\times S^{n+1}$ solution. In the $B=A$ gauge, the warp factors are given by
\bea\label{AdS2xS3}
A=A_*,\quad C=\frac{1}{A_* u^2}\sin^2[u r],\quad 
u=\left[\frac{m_-q_2^2}{2(d-2)nA^{m+1}_*}-\frac{d_-g^2}{nA_*}\right]^{\frac{1}{2}}\,.
\eea
Parameter $A_*$ is a solution of an algebraic equation
\bea\label{AdS2xS3star}
d_- g^2-\frac{m_-}{A_*}+\frac{nq_2^2}{2(d-2)A_*^m}=0.
\eea
The radii of the sphere and AdS are given by
\bea\label{RadAdS2S3}
R_{AdS}^2=A_*,\quad R_S^2=\frac{1}{A_* u^2}\,.
\eea
In an alternative gauge, this solution can be written in a form similar to (\ref{AdS32exct1}):
\bea\label{AdS2xS3alt}
A=A_*,\quad B=C=\frac{1}{C_*}-C_* r^2,\quad C_*=\frac{m_-q_2^2}{2n(d-2)A_*^m}-\frac{d_-g^2}{n}\,.
\eea
Here $A_*$ is still the smallest real solution of equation (\ref{AdS2xS3star}). In both gauges, the sphere $S^n$ collapses at two points, $r=(0,\frac{\pi}{u})$ for (\ref{AdS2xS3}) and $r=\pm \frac{1}{C_*}$ for (\ref{AdS2xS3alt}), forming a non--contractible $S^{n+1}$. 

In contrast to (\ref{AdS32exct2}), which leads to positive $C_\star$ for all values of $q_1$, equation (\ref{AdS2xS3star}) admits physically interesting solutions only if $q_2$ is below some maximal value, 
$q_2<q_{max}$, and reality of $u$ in (\ref{AdS2xS3}) imposed an additional constraint $q_2<q_{cr}$. Before showing this for a general case, it is useful to perform explicit calculations for the $m=2$ example. 

If $m=2$, then the solutions of the quadratic equation (\ref{AdS2xS3star}) are
\bea\label{Astar22}
m=2:&&A_*=\frac{1}{2d_-g^2}\left[1- \sqrt{1-\frac{2nd_-g^2 q_2^2}{d-2}}\right]\,,\\
&&A_{\bullet}=\frac{1}{2d_-g^2}\left[1+ \sqrt{1-\frac{2nd_-g^2 q_2^2}{d-2}}\right]\,.\nonumber
\eea 
This gives the maximal allowed value of the charge $q_2$:
\bea
m=2:\quad q_{max}=\frac{1}{g}\sqrt{\frac{n-1}{2n^2}}\,.
\eea
Furthermore, only $A_*$ from (\ref{Astar22}) may give rise to the $AdS_m\times S^{n+1}$ solution (\ref{AdS2xS3}), while $A_{\bullet}$ formally leads to an imaginary $u$ and 
$AdS_m\times H_{n+1}$, which will be discussed as the example D below. Even for $A_*$, parameter $u$ becomes imaginary for a sufficiently large charge, so the $AdS_2\times S^3$ solution (\ref{AdS2xS3}) exists only if
\bea
m=2:\quad q_2< q_{cr},\quad q_{cr}=\left[\frac{2}{(d-2)d_-g^2}\right]^{\frac{1}{2}}\,.
\eea
The value of $q_{cr}$ is determined by simultaneously solving  (\ref{AdS2xS3}), (\ref{AdS2xS3star}), and the condition $u=0$. 
This $m=2$ example illustrates the generic properties of the equation (\ref{AdS2xS3star}) and its solutions, and now we will prove them for a general $m$. 

To find $q_{max}$ for a general $m$, we first observe that the left hand side of equation (\ref{AdS2xS3star})\footnote{Here we use $a$ instead of $A_\star$ to stress that we are looking at the left hand side of (\ref{AdS2xS3star}) as a function of its argument, so it doesn't necessarily vanish for an arbitrary $a$.}, 
\bea
f(a)=d_- g^2-\frac{m_-}{a}+\frac{nq_2^2}{2(d-2)a^m}
\eea 
becomes positive when $a$ approaches zero or infinity, and it has only one real positive extremum located at 
\bea
a_{extr}=\left[\frac{mnq_2^2}{2(d-2)m_-}\right]^{\frac{1}{m-1}}.
\eea
This means that equation (\ref{AdS2xS3star}) has two positive solutions if $f(a_{extr})<0$, and it has none if 
$f(a_{extr})>0$. The transition point, where $f(a_{extr})=0$, determines the value of $q_{max}$:
\bea\label{AdS2xS3qMax}
A_*^{m-1}=\frac{nmq_{max}^2}{2(d-2)m_-},\quad d_-g^2=\frac{m^2_-}{mA_*}\quad
\Rightarrow\quad 
q_{max}^2=\frac{2(d-2)m_-}{nmg^{2(m-1)}}\left[\frac{m_-^2}{d_-m}\right]^{m-1}.
\eea
If $q_2>q_{max}$, then equation (\ref{AdS2xS3star}) has no real positive solutions. For example, at very large $q_2$, the approximate (complex) solutions are
\bea
A_*\simeq \left[-\frac{nq_2^2}{2(d-2)d_-g^2}\right]^{\frac{1}{m}}\,,\nonumber
\eea
and they are unphysical for both even and odd $m$. At $q_2=q_{\max}$, the solutions of (\ref{AdS2xS3star}) are degenerate:
\bea
A_*=A_\bullet=\left[\frac{m_-q_{cr}^2}{2(d-2)d_-g^2}\right]^{\frac{1}{m}},
\eea
but they correspond to imaginary $u$:
\bea
u^2=
\left[\frac{g^2}{nA_*}-\frac{d_-g^2}{nA_*}\right]=-\frac{(d-2)g^2}{nA_*}<0,
\eea
so they describe $AdS_m\times H_{n+1}$ rather than  $AdS_m\times S^{n+1}$.

While positivity of $A_\star$ gives an absolute upper bound on $q_2$, reality of $u$ in (\ref{AdS2xS3}) gives even stricter bound on the same charge\footnote{As we will see in examples D and E below, equations  (\ref{AdS2xS3}) describe interesting geometries even if this  stricter bound is violated. In contrast to this, solutions (\ref{AdS2xS3}) with $q_2>q_{max}$ are unphysical.}: $q_2<q_{cr}$, where
\bea\label{qCrit}
q_{cr}=\sqrt{2m_-}\left[\frac{(m_-)^2}{(d-2)d_-g^2}\right]^{\frac{m-1}{2}}.
\eea
Indeed, since the critical value corresponds to $u=0$, equation (\ref{AdS2xS3}) gives a simple relation for the corresponding $A_*$:
\bea
A^{(cr)}_*=\left[\frac{m_-q_{cr}^2}{2(d-2)d_-g^2}\right]^{\frac{1}{m}}\quad 
\Rightarrow\quad
q_{cr}^2=\frac{2(d-2)d_-g^2 A^{m}_*}{m_-}.
\eea
Substituting this into (\ref{AdS2xS3star}), one arrives at (\ref{qCrit}). 

\bigskip
\noindent
To summarize, we have found that equation (\ref{AdS2xS3star}) has at most two real positive roots, $A_*$ and $A_\bullet$. The $AdS_{m}\times S^{n+1}$ solutions exist if the electric charge $q_2$ is bounded from above,  $|q_2|<q_{cr}$, and the initial conditions are given 
by $A_*$. When the charge exceeds $q_{cr}$, or when the initial condition $A_\bullet$ is chosen, one can still find a counterpart of (\ref{AdS2xS3}), but the warp factor $C$ grows without bound. This situation will be analyzed in the next two examples. Once $q_2^2$ exceeds 
$q_{max}^2$ given by (\ref{AdS2xS3qMax}), there are no solutions with constant warp factor $A$. In contrast to examples A and B above, solution (\ref{AdS2xS3}) has finite warp factors 
$(A,C)$, and it cannot be extended to $r=\infty$. Therefore, (\ref{AdS2xS3}) exists as an exact solution for $q_1=0$, but one can't flow to it asymptotically in the presence of the magnetic charge.

\bigskip
\noindent
{\bf D. $AdS_{m}\times R^{n+1}$ (electric solution).}

For the special value of the charge, $q_2=q_{cr}$, the coefficient $u$ in (\ref{AdS2xS3}) approaches zero, leading to the 
$AdS_{m}\times R^{n+1}$ solution: 
\bea\label{AdS2xR4}
A=A_*,\quad C=\frac{r^2}{A_*},\quad q_2=q_{cr}=\left[\frac{2(d-2)d_-g^2 A^{m}_*}{m_-}\right]^{\frac{1}{2}},\quad A_*=\frac{(m_-)^2}{(d-2)d_-g^2}\,.
\eea
As in the sphere example C, the stress tensor of the electric field cancels the effect of the negative cosmological constant along some directions, and in the present case this cancellation leads to the  $R^{n+1}$ factor. 

Solution (\ref{AdS2xR4}) gives rise to asymptotic geometries 
\bea\label{AdS2xR4as}
A\simeq A_*,\quad C\simeq\frac{r^2}{A_*}
\eea
for $q_2=q_{cr}$ and arbitrary values of $q_1$. Indeed, due to the growth of function $C$, the terms proportional to $q_1^2$ in (\ref{FullEinstein}) are suppressed at infinity, so they give only subleading contributions to (\ref{AdS2xR4as}). 
We will refer the asymptotic solutions (\ref{AdS2xR4as}) as 
the ``$AdS_{m}\times R^{n+1}$ fixed points''. 

\bigskip
\noindent
{\bf E. $AdS_{m}\times H_{n+1}$ (electric solutions).}

The product of AdS and a maximally symmetric hyperbolic space with positive signature can be obtained by a simple analytic continuation of (\ref{AdS2xS3}):
 \bea\label{AdS2xH3}
A=A_{\bullet},\quad C=\frac{1}{A_{\bullet} v^2}\sinh^2[v r],\quad 
v^2=\frac{d_-g^2}{nA_{\bullet}}-\frac{m_-q_2^2}{2(d-2)nA^{m+1}_{\bullet}}.
\eea
Here $A_{\bullet}$ is a solution of equation (\ref{AdS2xS3star}),
\bea\label{AdS2xH3star}
d_- g^2-\frac{m_-}{A_{{\bullet}}}+\frac{nq_2^2}{2(d-2)A_{{\bullet}}^m}=0,
\eea
which gives a positive $v^2$. 
For future use, we also rewrite $v^2$ from (\ref{AdS2xH3}) purely in terms of $A_{\bullet}$, eliminating $q_2$ by solving equation (\ref{AdS2xH3star}):
 \bea\label{AdS2xH3vv}
v^2=\frac{d_-(d-2)g^2}{n^2A_{\bullet}}-\left[\frac{m_-}{nA_{\bullet}}\right]^2\,.
\eea
The resulting geometry (\ref{TheAnsatz1}) describes the product of $AdS_{m}$ and the maximally symmetric space of negative curvature, $H_{n+1}$. This solution exists even if $q_2=0$, when 
it simplifies to
\bea\label{AdS2xH3q0}
A=\frac{m_-}{d_- g^2},\quad C=\frac{n}{d_-g^2}\sinh^2\left[\frac{d_-g^2 r}{\sqrt{nm_-}}\right]\,.
\eea
In the alternative $B=C$ gauge, the  $AdS_{m}\times H_{n+1}$ solution still has the form (\ref{AdS2xS3alt}), with a replacement $A_*\rightarrow A_\bullet$, but now parameter $C_*$ is negative. 

The analysis presented in the example C implies that the solution (\ref{AdS2xH3}) exists only if $q_2$ is smaller than $q_{max}$ given by (\ref{AdS2xS3qMax}). However, in contrast to the 
$AdS_{m}\times S^{n+1}$ case, which does not have solutions for $q_2\ge q_{cr}$, one finds two 
$AdS_{m}\times H_{n+1}$ geometries with different radii for every $q_2$ in this range. Indeed, when the charge $q_2$ is larger than the critical value given by equation (\ref{AdS2xR4}), parameter $u$ in (\ref{AdS2xS3}) becomes imaginary, leading to unbounded $C$ even for $A_*$:
 \bea\label{AdS2xH3*}
A=A_*,\quad C=\frac{1}{A_* v^2}\sinh^2[v r],\quad 
v^2=\frac{d_-g^2}{nA_*}-\frac{m_-q_2^2}{2(d-2)nA^{m+1}_*}.
\eea
Although equation (\ref{AdS2xH3}) defines an exact solution only in the absence of the magnetic charge, its asymptotic form,
\bea\label{AdS2genD}
A\simeq A_{\bullet},\quad C\simeq c_0\exp[2v r],
\eea
is applicable even if $q_1\ne 0$. Due to the growth of function $C$, the terms proportional to $q_1^2$ in (\ref{FullEinstein}) are highly suppressed at infinity, so they give only subleading contributions to (\ref{AdS2genD}). For $q_2=0$, the asymptotic form (\ref{AdS2genD}) simplifies to
\bea\label{AdS2genDg0}
A\simeq\frac{m-1}{(d-1)g^2},\quad 
C\simeq c_0\exp\left[\frac{2(d-1)g^2}{\sqrt{n(m-1)}}r\right].
\eea
We will refer the asymptotic solutions (\ref{AdS2genD}) and (\ref{AdS2genDg0}) as 
the ``$AdS_{m}\times H_{n+1}$ fixed points''. 

\bigskip
\noindent
{\bf F. $AdS_{m}\times S^{n}\times R$ geometry (dyonic solution).}

The only remaining option for constant sphere or AdS radii is $AdS_m\times S^n\times R$. 
The warp factors describing this geometry are
\bea\label{SpecSolnNoRun}
A=aq_2^{2/m_-},\quad C=cq_1^{2/n_-},\quad
\frac{m_-}{a^m}q_2^{-2/m_-}-\frac{n_-}{c^n}q_1^{-2/n_-}=2(d-2)d_- g^2.
\eea
Parameters $a$ and $c$ are defined as
\bea\label{SpecSolnNRac}
a=\left(2m_-\right)^{-1/m_-},\quad c=\left(2n_-\right)^{-1/n_-}.
\eea
For example, the $AdS_2\times S^2\times R$ space has
\bea\label{SpecSolnNoRun22}
A=\frac{q_2^2}{2},\quad C=\frac{q_1^2}{2},\quad \frac{1}{q_2^2}-\frac{1}{q_1^2}=6g^2.
\eea
Since values of warp factors $(A,C)$ in (\ref{SpecSolnNoRun}) remain finite, the electric and magnetic contributions are never suppressed in (\ref{FullEinstein}), so in contrast to options 
A, B, D, E above, it is impossible to flow to the $AdS_{m}\times S^{n}\times R$ geometry by starting from charges which don't satisfy the second relation in (\ref{SpecSolnNoRun}). Therefore, 
$AdS_{m}\times S^{n}\times R$ is an isolated exact geometry which cannot be viewed as an asymptotic boundary condition for a larger class of geometries.

\subsection{Summary of regular geometries and phase diagrams}
\label{SecFlowSumry}

The remaining part of this article is dedicated to the study of interpolations between various asymptotics listed above. It is useful to separate the discussion into three cases: the magnetic, the electric, and the dyonic solutions. The detailed analysis of these situations will be presented in sections \ref{SecMagn}, \ref{SecElectr}, \ref{SecDyon}, and here we summarize the results. 

\subsubsection{Magnetic solutions}
\label{SecSubSumMgn}

\begin{figure}
\begin{center}
 \includegraphics[width=1 \textwidth]{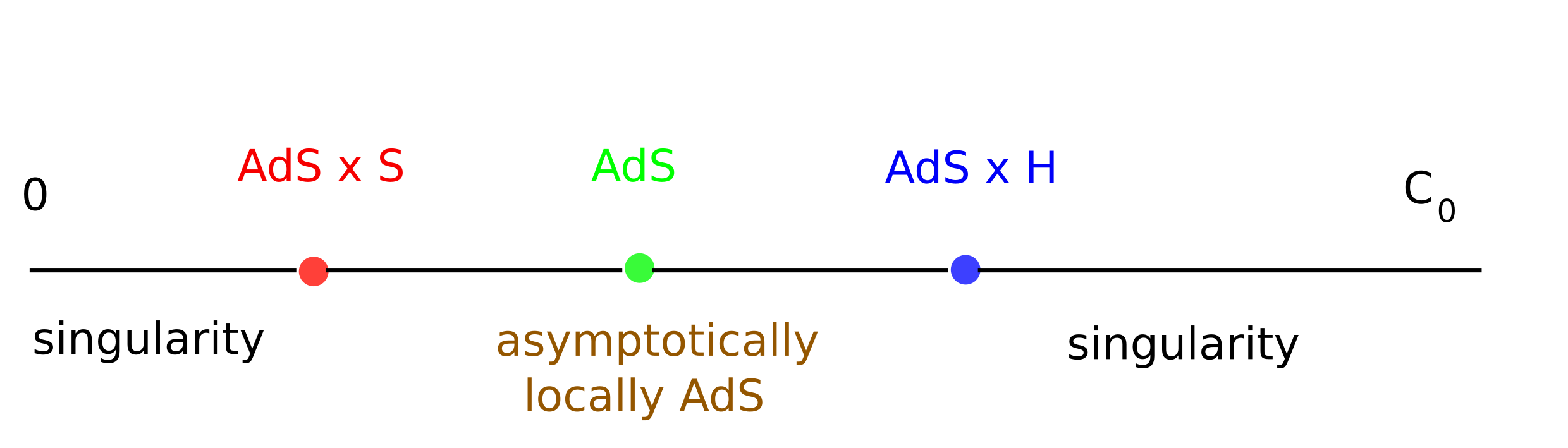}
 \end{center}
\caption{Correspondence between the values of $C_0$ and asymptotic boundary conditions for the magnetic solutions.}
\label{FigSumMagn}
\end{figure}

We begin with summarizing solutions with $q_2=0$, which are analyzed in section \ref{SecMagn}. In this purely magnetic case, the flux of the gauge field prevents the sphere $S^n$ from collapsing, so regular solutions of the system (\ref{FullEinstein}) must have a wormhole--like structure of interpolation between two asymptotic regions. As shown in section \ref{SecMagnBound}, regularity implies that there must exist some value of the coordinate $r$, where derivatives of both $A$ and $C$ vanish. Choosing the origin of $r$ at that point, we arrive at the boundary conditions:
\bea\label{ACbcWH}
{\dot A}(0)={\dot C}(0)=0,\quad A(0)=A_0,\quad C(0)=C_0\,.
\eea
Since the system (\ref{FullEinstein}) is invariant under reflection of $r$, these conditions imply that the resulting solutions are even,
\bea\label{ACeven}
A(-r)=A(r),\quad C(-r)=C(r),
\eea
so they interpolate between two identical asymptotic regions at $r=\pm\infty$. These asymptotics can be given by options A, B, E from the previous subsection. Eliminating second derivatives from equations from the system (\ref{FullEinstein}), one arrives at a first order equation, which, for  $q_2=0$, has the form (\ref{FirstOrdGen}). Application of this equation to $r=0$, gives a constraint on the values of $A_0$ and $C_0$, leading to the expression (\ref{FirstOrdGenCnstr}) for $A_0$:
\bea\label{A0smry}
{A_0}=mm_-\left[\frac{nn_-}{C_0}-\frac{q_1^2}{2C_0^n}+(d-2)d_-g^2\right]^{-1}.
\eea
Therefore, the magnetic solutions of the system (\ref{FullEinstein}) with boundary conditions (\ref{ACbcWH}) are controlled by one parameter $C_0$. All regular magnetic solutions fall in this class, but for some values of $C_0$, geometries may develop singularities away from $r=0$. The detailed analysis carried out in section \ref{SecMagn} gives the following options:
\begin{enumerate}[(a)]
\item $C_0\le C_{min}$: unphysical branch with a negative warp factor $A$.

The critical point $C_{min}$ is determined by solving the algebraic equation of $n$--th degree:
\bea\label{CminDefIntro2}
\frac{nn_-}{C_{min}}-\frac{q_1^2}{2(C_{min})^n}+(d-2)d_-g^2=0.
\eea

\item $C_{min}<C_0<C_\star$: the geometry ends with a naked curvature singularity at $r=r_1>0$. The critical value $C_\star$ is given by equation (\ref{AdS32exct2}). 

\item $C_0=C_\star$: the $AdS_{m+1}\times S^n$ solution (\ref{AdS32exct1}), (\ref{AdS32exct3}), (\ref{AdS32exct3a}).

\item $C_\star<C_0<C_\#$: the flow between  $AdS_{m+1}\times S^n$ and an asymptotically locally 
$AdS_d$ space (ALA). The full geometry describes a regular interpolation between two copies of ALA at $r=\pm \infty$ through a wormhole with a non--contractible $S^n$. The definition of the critical point $C_\#$ and properties of the asymptotic geometry are discussed in section \ref{SecMagn}.

\item  $C_0=C_\#$: the flow between  $AdS_{m+1}\times S^n$ and the standard $AdS_d$ at infinity. The resulting space describes a regular interpolation between two copies of $AdS_d$ at $r=\pm \infty$ through a wormhole with a non--contractible $S^n$.

\item $C_\#<C_0<C_\bullet$: the flow between  $AdS_{m+1}\times S^n$ and  asymptotically locally 
$AdS_d$ space (ALA). The full geometry describes a regular interpolation between two copies of ALA at $r=\pm \infty$ through a wormhole with a non--contractible $S^n$.

\item $C_0=C_\bullet$: the flow between  $AdS_{m+1}\times S^n$ and $AdS_m\times H_{n+1}$ at infinity.\\
The resulting space describes a regular interpolation between two copies of $AdS_m\times H_{n+1}$ at 
$r=\pm \infty$ through a wormhole with a non--contractible $S^n$. The critical value $C_\bullet$ is discussed in section \ref{SecMagn}.
\item $C_0>C_\bullet$: a flow to a naked curvature singularity at $r=r_1>0$.

\end{enumerate}
A pictorial representation of this summary is presented in figure \ref{FigSumMagn}, and the values of 
$(C_\#,C_\bullet)$ in various dimensions are found in section \ref{SecMagn}.

\subsubsection{Electric solutions}
\label{SecSubSumElctr}

In contrast to geometries with nonzero magnetic charge, the purely electric solutions of the the system (\ref{FullEinstein}) may contain a point with collapsing $S^n$ (i.e., with $C=0$).  As demonstrated in section \ref{SecElectrBC}, regular electric solutions must have a point where
\bea
A(0)=A_0,\quad C(0)=0.
\eea
Solutions of the system (\ref{FullEinstein}) are fully specified by $A_0$. For example, the first few terms in the small $r$ expansion are given by (\ref{ElctrSeries}). 

In contrast to the magnetic case, where the picture presented in Figure \ref{FigSumMagn} was the same for all values of charge, the structure of the electric solutions depends on $q_2$. A similar charge dependence will be present in the dyonic case as well. To proceed we recall the critical values $(q_{cr},q_{max})$ from (\ref{qCrit}) and (\ref{AdS2xS3qMax}), and we define $A_*$ and $A_\bullet$ as the real positive solutions of the equation (\ref{AdS2xS3star})\footnote{By differentiating this equation, one can show that there are at most two such solutions. We assume that $A_*\le A_\bullet$.}:
\bea\label{AdS2xS3starCopy}
d_- g^2-\frac{m_-}{A}+\frac{nq_2^2}{2(d-2)A^m}=0.
\eea
As shown in section \ref{SecElectrAnlz}, one encounters the following options for various values of $q_2$:
\begin{enumerate}[A.]
\item $q_2<q_{cr}$
\begin{enumerate}[(a)]
\item $A_0<A_*$: the geometry ends in a curvature singularity, where $S^n$ collapses and the AdS warp factor diverges.

\item $A_0=A_*$: the $AdS_m\times S^{n+1}$ solution (\ref{AdS2xS3}). 

\item $A_*<A_0<A_\bullet$: the geometry ends in a curvature singularity, where $S^n$ collapses and the AdS warp factor diverges. The situation is qualitatively similar to branch (a).

\item $A_0=A_\bullet$: the $AdS_m\times H_{n+1}$ solution (\ref{AdS2xH3}).

\item $A_\bullet<A_0<A_\#$: asymptotically locally $AdS_d$ space.

In contrast to branches (d) and (f) from section \ref{SecSubSumMgn}, in the electric case there is only one spacial infinity. Space ends at $r=0$ where the collapsing sphere and the radial direction combine to give a regular patch of $R^{m+1}$. 

\item $A_0=A_\#$: regular space with the standard $AdS_d$ asymptotics.

This solution can be interpreted as the unique geometry produced by an electric charge $q_2$ in the asymptotically $AdS_d$ space. The geometry interpolates between a regular patch of flat space at $r=0$ and the standard $AdS_d$ at infinity. In contrast to the magnetic option (e), there a smooth limit of vanishing charge, where the standard $AdS_d$ space in global coordinates is recovered.

\item $A_0>A_\#$: asymptotically locally $AdS_d$ space, similar to branch (d).

\end{enumerate}
\item $q_2=q_{cr}$
\begin{enumerate}[(a)]
\item $A_0<A_*$: a flow to $AdS_m\times R^{n+1}$ asymptotics. 

Regardless of the initial value $A_0$, the warp factor $A(r)$ flows to $A_*$ at infinity, and 
$C(r)$ becomes quadratic at large $r$. While approaching the asymptotic solution (\ref{AdS2xR4as}), which serves as a stable attractor point, the warp factors $A$ and $C$ go through oscillations around their final values. 

\item $A_0=A_*$: the exact $AdS_m\times R^{n+1}$ solution (\ref{AdS2xR4}).

\item  $A_*< A_0<A_\bullet$: solution with $AdS_m\times R^{n+1}$ asymptotics, similar to (a).
\end{enumerate}
The regions (d)--(g) are identical to their counterparts from option A. 
\item $q_{cr}<q_2< q_{max}$

\begin{enumerate}[(a)]
\item $A_0<A_*$: a flow to $AdS_m\times H_{n+1}$ asymptotics.

Regardless of the initial value $A_0$, the warp factor $A(r)$ flows to $A_*$ at infinity, and 
$C(r)$ becomes exponential at large $r$. While approaching the asymptotic solution (\ref{AdS2genD}), which serves as a stable attractor point, the warp factors $A$ and $C$ go through oscillations around their final values\footnote{To be precise, oscillations take place in the regime 
$q_{cr} < q_2 < q_2^{(trns)}$, where $q_2^{(trns)}$ is given by (\ref{q2Trans}). For larger values of $q_2$, the approach to asymptotic values is monotonic. See section \ref{SecElectr} for a detailed discussion of this point.}. 

\item $A_0=A_*$: the exact $AdS_m\times H_{n+1}$ solution (\ref{AdS2xH3}).

\item $A_*< A_0<A_\bullet$: solution with $AdS_m\times H_{n+1}$ asymptotics, similar to (a)

\end{enumerate}
The regions (d)--(g) are identical to their counterparts from option A. 

\item $q_2= q_{max}$

\begin{enumerate}[(a)]
\item $A_0<A_*=A_\bullet$: a flow to $AdS_m\times H_{n+1}$ asymptotics, similar to C(a).
\end{enumerate}
The regions (d)--(g) are identical to their counterparts from option A. 

\item $q_2>q_{max}$ 

Once the electric charge exceeds the maximal value (\ref{AdS2xS3qMax}), any initial condition $A(0)=A_0$ generates a flow to a locally $AdS_d$ space. Specifically, any $A_0<A_\#$ results in the 
option A(e), while $A_0=A_\#$ and $A_0>A_\#$ lead to branches (f) and (g). 
\end{enumerate}
\begin{figure}
\begin{center}
 \includegraphics[width=0.6 \textwidth]{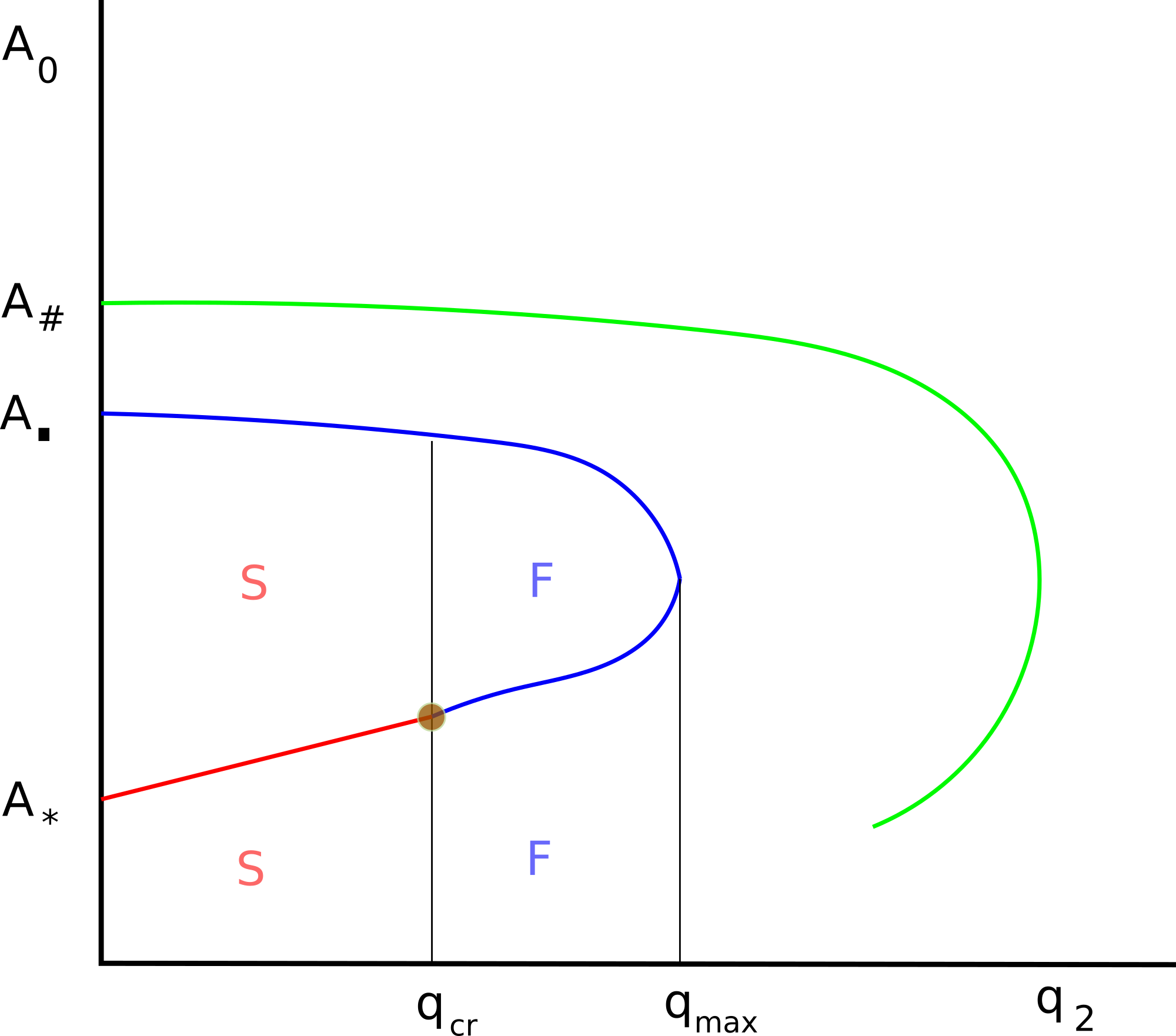}
 \end{center}
\caption{Correspondence between the values of $A_0$ and asymptotic boundary conditions for the electric solutions. The color scheme and notation are described in the text.}
\label{FigSumElectr}
\end{figure}
The graphical summary of these results in presented in Figure \ref{FigSumElectr}. Let us describe the notation used in this graph:
\begin{itemize}
\item The red line describes the exact  $AdS_m\times S^{n+1}$ solution (\ref{AdS2xS3}).
\item The blue lines describe the exact  $AdS_m\times H_{n+1}$ solutions (\ref{AdS2xH3}).
\item The green line describes the flow to  $AdS_{d}$.
\item The gold point at $q_2=q_{cr}$ describes the exact $AdS_m\times R^{n+1}$ solution (\ref{AdS2xR4}).
\item The vertical black line at $q_2=q_{cr}$ describes flows to $AdS_m\times R^{n+1}$.
\item Boundary conditions in the regions marked by $S$ correspond to flows leading to naked singularities.
\item Boundary conditions in the regions marked by $F$ describe flows to $AdS_m\times H_{n+1}$.
\item Boundary conditions in the unmarked regions describe flows to asymptotically locally $AdS_d$.
\end{itemize}
The details of this construction, the numerical values of  $A_\#$ and profiles of functions $(A,C)$ are presented in section \ref{SecElectr}.

\subsubsection{Dyonic solutions}
\label{SecSubSumDyon}

Once both electric and magnetic charges are turned on, the solutions of the system (\ref{FullEinstein}) exhibit an interesting and a fairly complicated phase structure, but all regular geometries have wormhole-type properties, just like their magnetic counterparts. Due to this qualitative similarity, it is convenient to fix the values of $(g,q_1)$ and analyze various flows as functions of the electric charge $q_2$. We refer to section \ref{SecDyon} for the detailed discussion of the phase diagram, and here we summarize the main results.

For $q_2=0$, one recovers the three special points $(C_*,C_\#,C_\bullet)$ encountered in section \ref{SecSubSumMgn} and the structure shown in Figure \ref{FigSumMagn}. When $q_2$ is turned on, requirement of regularity still leads to the boundary conditions (\ref{ACbcWH}) and interpolation between two identical asymptotic regions implied by (\ref{ACeven}), but now the expression for $A_0$ becomes more complicated than (\ref{A0smry}). Specifically, this parameter satisfies a nonlinear equation (\ref{BounCnstrDyon}),
\bea\label{BounCnstrDyonSmry}
\frac{mm_-}{A_0}-\frac{nn_-}{C_0}+\frac{q_1^2}{2C_0^n}-\frac{q_2^2}{2A_0^m}-(d-2)d_-g^2=0,
\eea
which may have more than one root. As demonstrated in section \ref{SecDyon}, the last equation has two real positive solutions for $A_0$ if
\bea\label{Dec18ineq}
2(d-2)d_- g^2+\frac{2nn_-}{C_0}-\frac{q_1^2}{C_0^n}< 
2(m_-)^2\left[\frac{2m_-}{q_2^2}\right]^{\frac{1}{m-1}},
\eea
it has one solution when the last relation becomes an equality, and it has no solutions when the sign in reversed\footnote{For obvious reasons, we are focusing only on real values of $C_0$, so the left hand side of 
(\ref{Dec18ineq}) is always real, and comparison to the right hand side makes sense.}. The inequality (\ref{Dec18ineq}) divides the $(q_2,C_0)$ plane into physical and unphysical regions separated by a branch cut. We will label the physical solutions of (\ref{BounCnstrDyonSmry}) as $(A_+,A_-)$, and the explicit expressions for the $m=2$ case are given by 
(\ref{Apm22}). 

In the dyonic case, the only option for the exact solution from section \ref{SecExactSoln} is 
$AdS_{m}\times S^{n}\times R$ given by (\ref{SpecSolnNoRun})--(\ref{SpecSolnNRac}). It occurs when the charges satisfy the relation\footnote{Parameters $(a,c)$ are defined in (\ref{SpecSolnNRac}).}
\bea\label{SecChargeSmrySec}
\frac{m_-}{a^m}q_2^{-2/m_-}-\frac{n_-}{c^n}q_1^{-2/n_-}=2(d-2)d_- g^2,
\eea
and a direct calculation shows that the corresponding values of $(g,q_1,q_2,C_0)$ fall on the branch cut separating (\ref{Dec18ineq}) from the unphysical region. At fixed values of $(g,q_1)$, the $AdS_{m}\times S^{n}\times R$ is represented by a point in the $(q_2,C_0)$ plane. All other points in this plane correspond to nontrivial flows or to singular solutions.

\begin{figure}
\begin{center}
 \includegraphics[width=0.8 \textwidth]{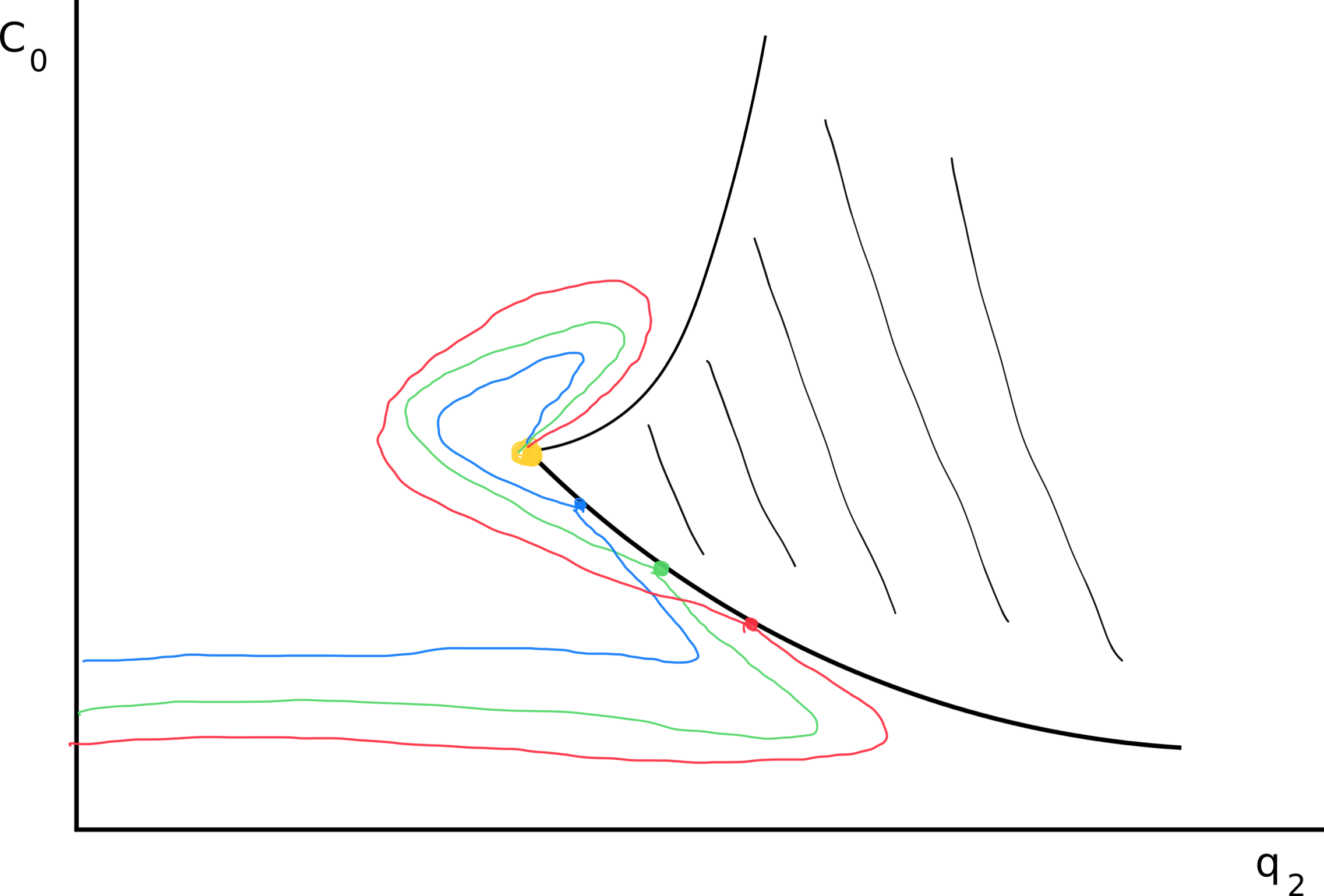}
 \end{center}
\caption{Summary of the phases for the dyonic solutions at fixed $(g,q_1)$. The color scheme and notation are 
described in the text.}
\label{FigSumDyon}
\end{figure}
\bigskip
\noindent
The analysis of the phase structure carried out in section \ref{SecDyon} can be summarized in a picture presented in Figure \ref{FigSumDyon}. It has the following features:
\begin{itemize}
\item For fixed values of $(g,q_1)$, the boundary conditions for the system (\ref{FullEinstein}) are specified by two numbers, $(q_2,C_0)$ and a discrete choice between $A_+$ and $A_-$, the two real positive solutions of (\ref{BounCnstrDyonSmry}).
\item Solutions with positive real $(A,C)$ exist only if the condition (\ref{Dec18ineq}) is satisfied. This region is bounded by the branch cut, which is shown as the black line in Figure \ref{FigSumDyon}.
\item For nonzero $q_1$ and $q_2$, both $A$ and $C$ are nontrivial functions of $r$ unless the charges satisfy the constraint (\ref{SecChargeSmrySec}). In the latter case, the system admits the $AdS_{m}\times S^{n}\times R$ (\ref{SpecSolnNoRun})--(\ref{SpecSolnNRac}). This special point is shown as a gold dot in Figure \ref{FigSumDyon}.
\item When the electric charge approaches zero, the structure of flows reproduces the picture shown in Figure \ref{FigSumMagn}. In particular, the three special points $(C_*,C_\#,C_\bullet)$ on the vertical axis of Figure \ref{FigSumDyon} describe the $AdS_{m+1}\times S_{n}$ solution and flows to $AdS_d$ and 
$AdS_m\times H_{n+1}$. Following the color conventions from Figure \ref{FigSumMagn}, these points are shown in red, green, and blue.
\item For positive values of the electric charge $q_2$, the exact $AdS_{m+1}\times S_{n}$ solution disappears, but for certain boundary conditions $C_0$, it is possible to flow to $AdS_{m+1}\times S_{n}$ at $r=\pm \infty$. The corresponding values of $C_0$ at various $q_2$ are depicted as the red curve in Figure \ref{FigSumDyon}.
\item The other curves in Figure \ref{FigSumDyon} identify the boundary conditions leading to wormholes with $AdS_d$ (green) and $AdS_m\times H_{n+1}$ (blue) asymptotics. Recall that the boundary conditions (\ref{ACbcWH}) derived from regularity implies that solutions must have the same asymptotic behavior at $r=\infty$ and $r=-\infty$.
\item The red, green, and blue lines start at the $q_2=0$ axis, and they end at the gold point. In the process of going there, every curve touches the branch cut once, where $A_+$ solution of the constraint (\ref{BounCnstrDyonSmry}) is exchanged with $A_-$. The $A_+$ branch connects the $q_2=0$ axis and the branch cut, while the $A_-$ branch connects the branch cut and the gold point. 
\item As can be seen from the schematic Figure \ref{FigSumDyon}, the number of branches and their order depend on $q_2$: for example, for sufficiently large $q_2$, there are no regular flows at all, and any initial condition $C_0$ leads to a naked singularity. Following the analysis performed in section \ref{SecDyon}, we present separate discussions of the solutions with $A_+$ and $A_-$.
\item Solutions with $A_+$. As the value of $C_0$ increases for a fixed $q_2$, one encounters the following phases:
\begin{enumerate}[a.]
\item Below the bottom edge of the red curve there are no regular solutions; all flows lead to naked singularity. This situation is similar to options (a)-(b) from section \ref{SecSubSumMgn}.
\item Between the red and the blue curves, one finds flows to asymptotically locally $AdS_d$, which are similar to options (d) and (f) in section \ref{SecSubSumMgn}. At the location of the green curve one encounters the flow to the standard 
$AdS_d$. 
\item The blue curve separates the flow to asymptotically locally $AdS_d$ space (ALA) and a flow to a naked singularity, a counterpart of item (h) in section \ref{SecSubSumMgn}. 
\item In contrast to the situation described in section \ref{SecSubSumMgn}, for sufficiently large $q_2$ it is possible to cross the blue line as $C_0$ increases. For example, one can encounter a sequence
\bea
\mbox{sing.}\ 
\xrightarrow[\mbox{red}]\qquad\mbox{ALA} \xrightarrow[\mbox{blue}]\qquad \mbox{sing.}\ \xrightarrow[\mbox{blue}]\qquad \mbox{ALA}
\ \rightarrow\ \mbox{branch cut} \nonumber
\eea
The singularity and the asymptotically locally $AdS_d$ (ALA) are always separated either by the blue line or by the red one.
\end{enumerate}
The green curve is crossed in the middle of the ALA regions, when the $A/C$ ratio approaches one at infinity.
\item Solutions with $A_-$. As the value of $C_0$ increases for a fixed $q_2$, one encounters the phases in the same order as in the $A_+$ case, but now the red, green, and blue curves explore the space above the branch cut as well. The longest sequence of $A_-$ phases is encountered when $q_2$ is slightly above the $q_*$. As $C_0$ increases from zero, one finds
\bea
&&\hskip -1cm\mbox{sing.}\ 
\xrightarrow[\mbox{red}]\qquad\mbox{ALA} \xrightarrow[\mbox{blue}]\qquad \mbox{sing.}\ 
\xrightarrow[\mbox{brnch cut}]\ \quad \mbox{unphys. reg.}\ \xrightarrow[\mbox{brnch cut}]\
\mbox{sing.}\ \xrightarrow[\mbox{red}]\ \nn
&&\xrightarrow[\mbox{red}]\qquad \mbox{ALA} \xrightarrow[\mbox{blue}]\ \mbox{sing.}
\xrightarrow[\mbox{blue}]\qquad \mbox{ALA}\xrightarrow[\mbox{red}]\ \mbox{sing.} \nonumber
\eea
As in the $A_+$ case, the green curve is crossed in the middle of the ALA regions, when the $A/C$ ratio approaches one at infinity.
\end{itemize}
The details of this construction, the numerical data for the locations of the curves, and typical profiles of functions $(A,C)$ are presented in section \ref{SecDyon}.

\section{Magnetic solutions}
\label{SecMagn}
\renewcommand{\theequation}{3.\arabic{equation}}
\setcounter{equation}{0}

In this section we will perform a detailed analysis of equations (\ref{FullEinstein}) with vanishing $q_2$. and nonzero $q_1$. As discussed in section \ref{SecExactSoln}, in this range of parameters the system admits an exact $AdS_{m+1}\times S^n$ solution given by (\ref{AdS32exct1})--(\ref{AdS32exct2}), as well as two asymptotic fixed points:  $AdS_d$ given by (\ref{AdSexactAsymp}) and $AdS_{m}\times H_{n+1}$ given by (\ref{AdS2genD}).
We will now construct the flows between these fixed points and demonstrate the existence of two types of wormholes.

\subsection{The boundary problem for regular solutions}
\label{SecMagnBound}

Since equations (\ref{FullEinstein}) with vanishing $q_2$ will be extensively used in this section, we begin with quoting them for future reference:
\bea\label{FullEinsteinM}
&&d_-g^2+\frac{n_- q_1^2}{2(d-2)C^n}-\frac{n\dot A\dot C}{4C}+\frac{nA{\dot C}^2}{4C^2}-\frac{m}{2}\ddot A-\frac{nA\ddot C}{2C}=0,
\nn
&&d_-g^2+\frac{n_-}{C}-\frac{mq_1^2}{2(d-2)C^n}-\frac{m_+{\dot A}{\dot C}}{4C}+\frac{(n-2)A{\dot C}^2}{4C^2}-
\frac{A\ddot C}{2C}=0,\\
&&d_-g^2 A-m_-+\frac{(n-1)q_1^2 A}{2(d-2)C^n}-\frac{m_-}{4}{\dot A}^2-\frac{nA{\dot A}{\dot C}}{4C}-\frac{1}{2}A{\ddot A}=0.\nonumber
\eea
Throughout this section we will assume that $q_1\ne 0$. Although the system (\ref{FullEinstein}) contains two parameters, $g$ and $q_1$, one of them can be absorbed in rescaling of the warp factors $(A,C)$ and the radial coordinate, i.e., in a slight modification of the ansatz (\ref{TheAnsatz1}). Such a rescaling identifies the single physical order parameter and reduces unnecessary clutter in analyzing the system (\ref{FullEinsteinM}). Specifically, if we write
\bea\label{ACrescaleBar}
A=q_1^{\mu}{\bar A}[r q_1^{-\mu}],\quad C=q_1^{\mu}{\bar C}[r q_1^{-\mu}],\quad \mu=\frac{2}{n-1},
\eea
then equations for $({\bar A},{\bar C})$ would contain $q$ and $g$ only in one combination 
\bea\label{gBarDef}
{\bar g}=gq_1^{\frac{1}{n-1}}.
\eea 
Equations for the rescaled variables are
\bea\label{RescaledSystemEqn}
&&{\bar g}^2 d_-+\frac{n_- }{2(d-2){\bar C}^n}-\frac{n\dot {\bar A}\dot{\Cb}}{4\Cb}+\frac{n\Ab{\dot \Cb}^2}{4\Cb^2}-\frac{m}{2}\ddot \Ab-\frac{n\Ab\ddot \Cb}{2\Cb}=0,
\nn
&&{\bar g}^2 d_-+\frac{n_-}{\Cb}-\frac{m}{2(d-2)\Cb^n}-\frac{m_+{\dot \Ab}{\dot \Cb}}{4\Cb}-\frac{(n-2)\Ab{\dot \Cb}^2}{4\Cb^2}-
\frac{\Ab\ddot \Cb}{2\Cb}=0,\\
&&{\bar g}^2 d_-\Ab-m_-+\frac{(n-1)\Ab}{2(d-2)\Cb^n}-\frac{m_-}{4}{\dot \Ab}^2-\frac{n\Ab{\dot \Ab}{\dot \Cb}}{4\Cb}-\frac{1}{2}\Ab{\ddot \Ab}=0.\nonumber
\eea
The dot above functions $(\Ab,\Cb)$ denotes the derivatives with respect to the rescaled radial coordinate, 
${\bar r}=r q_1^{-\mu}$, while the dot above functions $(A,C)$ denotes the derivative with respect to $r$. 
The analysis presented in this section will refer to both the original system (\ref{FullEinsteinM}) and its rescaled version (\ref{RescaledSystemEqn}).

\bigskip

We begin with observing that for a nonzero value of $q_1$, the warp factor $C$ in equations (\ref{FullEinsteinM}) cannot become arbitrarily small. This feature has a simple physical origin: since $F_n$ has a constant flux over the sphere, the flux density is proportional to $C^{-\frac{n}{2}}$, and it cannot become infinite. Therefore the flux acts as an effective potential preventing the sphere from shrinking. For the analysis that follows it is also useful to present a formal argument for this phenomenon. Combining equations (\ref{FullEinsteinM}) to eliminate second derivatives, one finds the relation
\bea\label{FirstOrdGen}
\frac{nn_- A{\dot C}^2}{4C^2}+\frac{mn {\dot A}{\dot C}}{2C}+\frac{mm_-{\dot A}^2}{4A}+\frac{mm_-}{A}-\frac{nn_-}{C}+\frac{q_1^2}{2C^n}-(d-2)d_-g^2=0.
\eea
Assuming that $C$ becomes arbitrarily small, the leading contributions to the last equation give\footnote{Note that vanishing $A$ would lead to a curvature singularity in the metric (\ref{TheAnsatz1}), so 
for all physically interesting solutions, $1/A$ must remain finite and positive.} 
\bea
\frac{nn_- A{\dot C}^2}{4C^2}-\frac{nn_-}{C}+\frac{q_1^2}{2C^n}\approx 0.
\eea
This equation cannot be satisfied for arbitrarily small $C$ unless $q_1=0$. In the latter case, the equation can be integrated to give
\bea
C=a (r-r_0)^2+O((r-r_0)^3),
\eea
and the $AdS_d$ geometry (\ref{AdSexact}) represents a special case of such solution. 

In this section we are interested in nonzero $q_1$, so the warp factor $C$ is bounded from below:
\bea
C \ge C_0>0.
\eea
If the minimal value $C_0$ is reached at some finite point, then without loss of generality we can assume that this happens at $r=0$.\footnote{Recall that the system (\ref{FullEinsteinM}) has a symmetry under shifting $r$.} Then ${\dot C}(0)=0$, and equations (\ref{FullEinsteinM})  fully determine the Taylor expansions of $(A,C)$ near zero in terms of two parameters, $C(0)$ and ${\dot A}(0)$. In particular, $A(0)$ can be determined from (\ref{FirstOrdGen}). There are only six logical options for solutions of (\ref{FullEinstein}) with nonzero $q_1$:
\begin{enumerate}[1)]
\item Both $A$ and $C$ decrease monotonically as $r$ goes from infinity to negative infinity. There are no solutions of this type.

Indeed, in such a situation, functions $(A,C)$ must approach positive asymptotic values $(A_0,C_0)$ at $r=-\infty$, so at large negative values of $r$ they can be written as 
\bea
A=A_0+\eps A_1(r),\quad C=C_0+\eps C_1(r),
\eea
with a small parameter $\eps$. Substituting these expansions into (\ref{FullEinsteinM}), and focusing on the leading order in $\eps$, one finds a system of three {\it algebraic} equations for $(A_0,C_0)$, which has no solutions.

\item Function $C$ monotonically decreases as $r$ goes from infinity to the interior, and function $A$ reaches a local minimum at $r=r_0$. This behavior is inconsistent with equations (\ref{FullEinsteinM}).

Specifically, numerical study of the system (\ref{FullEinsteinM}) indicates that if $A(r)$ reaches a local minimum at some point $r_0$, then ${\dot C}$ must vanish either at that point or elsewhere. Combining this with the statement 1) above, we conclude that the system (\ref{FullEinsteinM}) does not admit solutions with monotonic $C$, and there must be a point where ${\dot C}=0$.

\item Function $C$ reaches a local minimum at $r=r_0$ with ${\dot C}(r_0)=0$, ${\dot A}(r_0)\ne 0$, $C(r_0)<C_\star$. This option leads to a naked curvature singularity.

Specifically, numerical integration of the system (\ref{FullEinsteinM}) indicates that for ${\dot A}(r_0)< 0$, function $A(r)$ diverges at some $r=r_1>r_0$, and $C(r)$ goes to zero at that point. The leading behavior of the warp factors near the singularity is given by
\bea\label{Aug9}
C\simeq c (r_1-r)^\mu,\quad A\simeq\frac{2mq_1^2}{(d-2)n_- \mu^2 c^n}(r_1-r)^{2-n\mu},\quad
\mu=\frac{2m}{n(m-1)+1}\,.
\eea
If ${\dot A}(r_0)> 0$, then such singularity occurs at $r=r_1<r_0$.

\item Function $C$ reaches a local minimum at $r=r_0$ with ${\dot C}(r_0)=0$, ${\dot A}(r_0)\ne 0$, $C(r_0)>C_\star$, where $C_\star$ is a solution of (\ref{AdS32exct2}). This option leads to a naked curvature singularity.

Specifically, numerical integration of the system (\ref{FullEinsteinM}) indicates that for ${\dot A}(r_0)< 0$, function $C(r)$ diverges at some $r=r_1>0$, and $A(r)$ goes to zero at that point. If ${\dot A}(r_0)> 0$, then such singularity occurs at $r=r_1<r_0$. The leading behavior of the warp factors near the singularity is given by
\bea\label{Aug9L}
C\simeq c (r_1-r)^{-\mu},\quad A\simeq a(r_1-r)^{\frac{2+n\mu}{m+1}},\quad
\mu=2\frac{(m+1)\sqrt{mn(d-2)}-2mn}{n[(m+1)^2+nm_-]}\,.
\eea
In contrast to (\ref{Aug9}), the asymptotic expansion around $r=r_1$ does not fix the coefficient $a$ in $A$. The three integration constants $(a,c,r_1)$ in (\ref{Aug9L}) can be related to the boundary conditions at $r=r_0$ via numerical integration of the system (\ref{FullEinsteinM}).

\item Function $C$ reaches a local minimum at $r=r_0$ with ${\dot C}(r_0)=0$, ${\dot A}(r_0)\ne 0$, $C(r_0)=C_\star$. This option leads to the unique solution (\ref{AdS32exct1})--(\ref{AdS32exct2}), and the only remaining freedom is the shift of $r$. In particular, ${\dot C}$ vanishes everywhere, and there exists one point where ${\dot A}$ vanishes as well. Therefore one encounters a special case of the option 6) below. 

\item Functions $C$ and $A$ reach local minima at the same point $r=r_0$: ${\dot A}(r_0)={\dot C}(r_0)=0$. This option leads to several interesting solutions, and it will be discussed in detail in the remaining part of this section. 

\end{enumerate}
\noindent
To summarize, we have demonstrated that integration of the system (\ref{FullEinsteinM}) leads to singular solutions unless functions $C$ and $A$ reach local minima at the same point $r=r_0$: ${\dot A}(r_0)={\dot C}(r_0)=0$. Since the system (\ref{FullEinsteinM}) is invariant under a shift of $r$, we will choose $r_0=0$ without loss of generality. Furthermore, starting with the boundary conditions
\bea\label{BCAug9}
C(0)=C_0,\quad A(0)=A_0,\quad {\dot C}(0)={\dot A}(0)=0,
\eea
and observing that the system (\ref{FullEinsteinM}) is invariant under the sign reversal of $r$, we conclude that the unique solution satisfying the conditions (\ref{BCAug9}) is even:
\bea
A(-r)=A(r),\quad C(-r)=C(r).
\eea
Therefore it is sufficient to construct functions $(A,C)$ in the $r\ge0$ region and ensure regularity of the geometry (\ref{TheAnsatz1}) there. Then the proper behavior at all values of $r$ will follow. Finally, we note that the parameter $A_0$ in (\ref{BCAug9}) is not independent, but rather it is determined by application of equation (\ref{FirstOrdGen}) to $r=0$:
\bea\label{FirstOrdGenCnstr}
{A_0}=mm_-\left[\frac{nn_-}{C_0}-\frac{q_1^2}{2C_0^n}+(d-2)d_-g^2\right]^{-1}.
\eea
The remaining part of this section is dedicated to the study of the system (\ref{FullEinsteinM}) with boundary conditions (\ref{BCAug9}) and (\ref{FirstOrdGenCnstr}).

\subsection{Wormhole geometries from regular flows}
\label{SecSubMagnFlow}

Equations (\ref{FullEinsteinM}) can be integrated with the boundary conditions (\ref{BCAug9}) and (\ref{FirstOrdGenCnstr}) for any value of $C_0$, but generically the resulting solutions lead to singular metrics (\ref{TheAnsatz1}). In section \ref{SecExact} we discussed an exception, the 
$AdS_{m+1}\times S^n$ geometry (\ref{AdS32exct1}), and in this subsection we will construct additional regular solutions with fewer isometries. In particular, we will be interested in flows that connect the 
$AdS_{m+1}\times S^n$ approximation near the origin with the $AdS_d$ asymptotics (\ref{AdSexactAsymp}) at infinity.

The solutions of the system (\ref{FullEinsteinM}), (\ref{BCAug9}), (\ref{FirstOrdGenCnstr})  are fully specified by one parameter $C_0$, so we will analyze various options for its value.

\begin{enumerate}[(a)]
\item $C_0< C_{min}$: unphysical branch with negative $A_0$.

Equation (\ref{FirstOrdGenCnstr}) implies that unless $C_0$ is sufficiently large, the initial value $A_0$ of the AdS warp factor becomes negative. The critical point $C_{min}$ is determined by solving the algebraic equation of $n$--th degree:
\bea\label{CminDef}
\frac{nn_-}{C_{min}}-\frac{q_1^2}{2(C_{min})^n}+(d-2)d_-g^2=0.
\eea
For example, if $n=2$, then 
\bea
n=2:\quad C_{min}=\frac{1}{(d-2)d_-g^2}\left[\sqrt{1+\frac{(d-2)d_-(q_1g)^2}{2}}-1\right].
\eea
The limiting case, $C_0= C_{min}$, is also unphysical since $A(0)=0$.

\item $C_{min}<C_0<C_\star$: naked curvature singularity at $r=r_1>0$. Recall that $C_\star$ is given by equation (\ref{AdS32exct2}). 

Although in this case equations (\ref{FullEinsteinM}) can be integrated in the vicinity of zero to give a locally regular solution, function $C(r)$ is monotonically decreasing, and at some point $r=r_1>0$ it hits zero, while $A(r)$ goes to infinity. The leading behavior near this singularity is given by (\ref{Aug9}). The distance between this singular point and the origin,
\bea
\int_0^{r_1}\frac{dr}{\sqrt{A}}\,,\nonumber
\eea
is finite, so this branch leads to a naked curvature singularity. 

\begin{figure}
 \includegraphics[width=1 \textwidth]{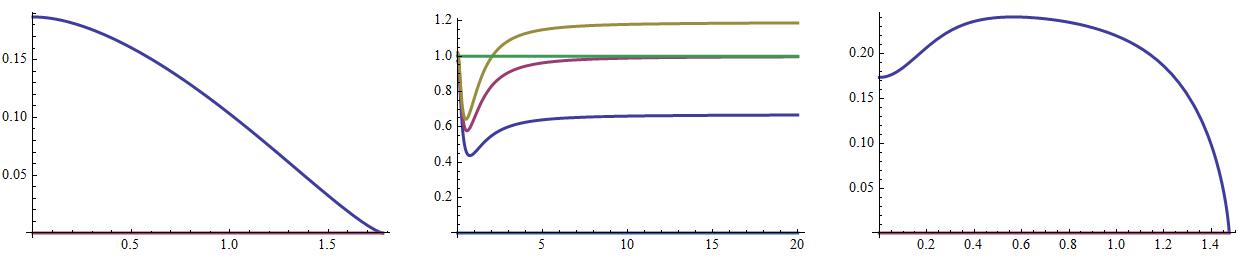}
 \begin{center}
(a)\qquad \qquad \qquad \qquad \qquad \qquad \qquad(b)\qquad \qquad \qquad \qquad \qquad
 \qquad  (c)
\end{center}
\caption{Profiles summarising various options for magnetic solutions:\newline
(a) Function $C(r)$ for $C_{min}<C_0<C_\star$: geometry terminates at naked singularity;
\newline
(b) Ratios $C(r)/A(r)$ for three values of $C_0$ in the range $C_{\star}<C_0<C_\bullet$: asymptotically locally AdS spaces;\newline
(c) Function $A(r)$ for $C_0>C_\bullet$: geometry terminates at naked singularity.
}
\label{FigQ1Sing}
\end{figure}

\item $C_0=C_\star$: the $AdS_{m+1}\times S^n$ solution (\ref{AdS32exct1}). 

The resulting geometry is (\ref{AdS32exct3}), and the radii of the AdS space and the sphere are given by (\ref{AdS32exct3a}). 
\label{optDpg}
\item $C_\star<C_0<C_\#$: the flow between  $AdS_{m+1}\times S^n$ and an asymptotically locally 
$AdS_d$. The resulting space describes a regular interpolation between two copies of $AdS_d$ at $r=\pm \infty$ through a wormhole with a non--contractible $S^n$.

Numerical integration of the system (\ref{FullEinsteinM}) with this range of initial conditions leads to growing functions $A$ and $C$ at infinity, and the leading contributions are given by
\bea\label{ACstarLead}
A\simeq (g{\hat r})^2,\quad C\simeq b(g{\hat r})^2,\quad {\hat r}=r-r_0.
\eea
The values of two constants, $b$ and $r_0$ are fully determined by $C_0$, and parameter $b$ varies between zero and one as $C_0$ varies between $C_\star$ and $C_\#$. The critical point $C_\#$ is defined by the relation 
\bea\label{DefCnmbr}
b(C_\#)=1.
\eea
It is instructive to look at subleading terms in (\ref{ACstarLead}) and to analyze the properties of the corresponding geometries (\ref{TheAnsatz1}). For all values of $(m,n)$, equations (\ref{FullEinsteinM})  fully specify the expansions of $(A,C)$ near infinity in terms of three constants $(b,r_0,a)$. These parameters are independent if one focuses only on solutions at large $r$, but integration of the system  (\ref{FullEinsteinM}) from zero with initial conditions (\ref{BCAug9}), (\ref{FirstOrdGenCnstr}) determines everything in terms of $C_0$. The detailed expansions of $A$ and $C$ depend on the dimension of space, and here we consider two situations to illustrate some qualitative features. 

For example, in the $m=n=2$ case, the asymptotic expansion near infinity reads 
\bea\label{ACasymp22}
(2,2):&&A=(g{\hat r})^2+\frac{1+2b}{3bg^2}+\frac{b^2-1+3g^2q_1^2}{18(b{\hat r})^2 g^6}\ln\frac{a}{\hat r}+o({\hat r}^{-3}),\\
&&C=b(g{\hat r})^2+\frac{b-1}{6g^2}-\frac{b^2-1+3g^2q_1^2}{36b{\hat r}^2 g^6}\ln\frac{a}{\hat r}+
\frac{1-b^2}{36b{\hat r}^2 g^6}+o({\hat r}^{-3}),\nonumber
\eea
and all subsequent terms are fully determined in terms of $(b,r_0,a)$. The corresponding answer for the 
 $(m,n)=(3,2)$ and $(2,3)$ cases are
\bea\label{ACasymp32}
(2,3):&&A=(g{\hat r})^2+\frac{1+b}{2bg^2}-\frac{(b-1)(b+2)}{18(b{\hat r})^2 g^6}+\frac{a^4}{{\hat r}^3}+o({\hat r}^{-4}),\nn
&&C=b(g{\hat r})^2+\frac{b-1}{6g^2}-
\frac{a^4 b}{3{\hat r}^3}+o({\hat r}^{-4});\\
(3,2):&&A=(g{\hat r})^2+\frac{1+5b}{6bg^2}-\frac{8(b-1)(2b+1)+27g^2q_1^2}{216(b{\hat r})^2 g^6}+\frac{a^4}{{\hat r}^3}+o({\hat r}^{-4}),\nn
&&C=b(g{\hat r})^2+\frac{b-1}{6g^2}+\frac{4(b-1)(2b+1)+27g^2q_1^2}{216(b{\hat r})^2 g^6}-
\frac{a^4 b}{{\hat r}^3}+o({\hat r}^{-4}).\nonumber
\eea
This illustrates the general pattern: in addition to the parameters $(b,r_0)$, which we have already encountered in (\ref{ACstarLead}), a new integration constant $a$ appears in front of ${\hat r}^{d-3}$. For even $d$, the expansions contain only inverse powers of ${\hat r}$, while for odd dimensions one encounters logs as well. The leading terms containing $q_1$ are proportional to $[q_1^2/{\hat r}^{2(n-1)}]$ in even dimensions, and to $[(q_1^2 \ln{\hat r})/{\hat r}^{2(n-1)}]$ for odd $d$. 

To understand the properties of the resulting geometries (\ref{TheAnsatz1}), it is instructive to evaluate the curvature tensors. While the qualitative picture is the same in all dimensions, the details depend on the values of $(m,n)$, and here we focus on the $(2,2)$ case. The Einstein's equations ensure that 
\bea
{R_\mu}^\nu&=&-4g^2\delta_\mu^\nu+o({\hat r}^{-3}).
\eea
To compare the Riemann tensor with its counterpart for the maximally symmetric $AdS_5$ space, it is convenient to introduce a tensor built from the metric:
\bea\label{DefSRiem}
S_{\mu\nu\la\sigma}\equiv g_{\mu\la}g_{\nu\sigma}-g_{\nu\la}g_{\mu\sigma}\,.
\eea
For the AdS space, the Riemann tensor is proportional to $S$, while expressions (\ref{ACasymp22}) lead to corrections, and the leading contributions have a very simple form:
\bea\label{RiemAsnpAdS}
R_{\mu\nu\la\sigma}&=&-g^2S_{\mu\nu\la\sigma}+\frac{b-1}{3b{\hat r}^2 g^2}\left[S^{(sph)}_{\mu\nu\la\sigma}+S^{(AdS)}_{\mu\nu\la\sigma}+\frac{1}{2}S^{(mixed)}_{\mu\nu\la\sigma}\right]+\dots
\eea
Here $S^{(sph)}$ is the part of the tensor $S$ with all indices in the sphere directions, $S^{(AdS)}$ has all indices in $AdS_m$, and $S^{(mixed)}$ contains mixed components with two indices on AdS and two on the sphere\footnote{Definition (\ref{DefSRiem}) and the ansatz (\ref{TheAnsatz1}) ensure that only such mixings between the AdS and the sphere are possible.}. The expression (\ref{RiemAsnpAdS}) ensures that the solutions (\ref{ACasymp22}) lead to  ``asymptotically locally AdS spaces'', as defined in \cite{AsympLocAdS}, for all values of $b$. Integration of (\ref{ACasymp22}) to the interior leads to regular wormhole geometries, as long as $(a,b)$ are specific functions of $C_0$ and $q_1$ is nonzero. For a vanishing charge, the asymptotic expansions (\ref{RiemAsnpAdS}) necessarily lead to singularities in the interior of the geometry (\ref{AnstzMetr1}) unless $b=1$, when one find the standard $AdS_5$ metric sliced by $AdS_2$ and $S^2$. Even though we have focused on the $(2,2)$ case, analysis of other $(m,n)$ values leads to similar conclusions. 

Asymptotic expansions (\ref{RiemAsnpAdS}) indicate a special role played by $b=1$ since for this value one gets the standard $AdS_d$ space both in the leading and first subleading order. In (\ref{DefCnmbr}) we denoted the value of $C_0$ leading to $b=1$ by $C_\#$. Since $b$ is a monotonic function of $C_0$, the 
$C_\star<C_0<C_\#$ range corresponds to $0<b<1$.
\label{optEpg}
\item  $C_0=C_\#$: the flow between  $AdS_{m+1}\times S^n$ and the standard $AdS_d$ at infinity. The resulting space describes a regular interpolation between two copies of $AdS_d$ at $r=\pm \infty$ through a wormhole with a non--contractible $S^n$.

The discussion of option (d) applies to this case upon setting $b=1$. Due to physical importance of this flow, it is interesting to determine the value of $C_\#$ as a function of $g$ and $q_1$. Unfortunately we were able to do this only numerically, and the results for the (2,2) system are presented in figure \ref{FigQ1dots}. However, for small and large charges, the system (\ref{FullEinsteinM}) simplifies, and in the next subsection we will derive analytical expressions for $C_\#$ in these approximations. This discussion will also serve as an analytical proof of existence of the $AdS_d$ wormhole. 

\begin{figure}
 \includegraphics[width=1 \textwidth]{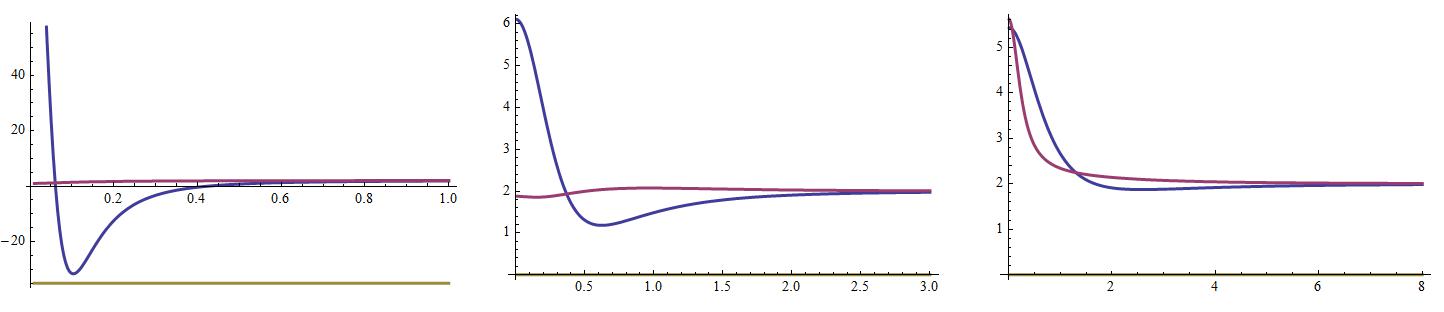}
 \begin{center}
(a)\qquad \qquad \qquad \qquad \qquad \qquad \qquad(b)\qquad \qquad \qquad \qquad \qquad
 \qquad  (c)
\end{center}
\caption{Plots of ${\ddot A}$ (blue) and ${\ddot C}$ (purple) profiles describing $m=n=2$ magnetic solutions with $AdS_5$ asymptotics for $q_1=0.1$ (a), $q_1=1$ (b), $q=10$. In all three cases, $g=1$.}
\label{FigQ1AdS5}
\end{figure}

\item $C_\#<C_0<C_\bullet$: the flow between  $AdS_{m+1}\times S^n$ and  an asymptotically locally 
$AdS_d$.

As in options (d) and (e), the resulting space describes a regular interpolation between two copies of $AdS_d$ at $r=\pm \infty$ through a wormhole with a non--contractible $S^n$, but now $b>1$. When $C_0$ reaches a critical value $C_\bullet$, parameter $b$ goes to infinity, and the leading $AdS_d$ asymptotic behavior (e.g., the structure of the expansions (\ref{ACasymp22})--(\ref{ACasymp32})), breaks down.

\item $C_0=C_\bullet$: the flow between  $AdS_{m+1}\times S^n$ and $AdS_m\times H_{n+1}$ at infinity\footnote{We recall that $H_{n+1}$ denotes the maximally symmetric $(n+1)$--dimensional hyperbolic space.}. The resulting space describes a regular interpolation between two copies of $AdS_m\times H_{n+1}$ at 
$r=\pm \infty$ through a wormhole with a non--contractible $S^n$.

\begin{figure}
 \includegraphics[width=1 \textwidth]{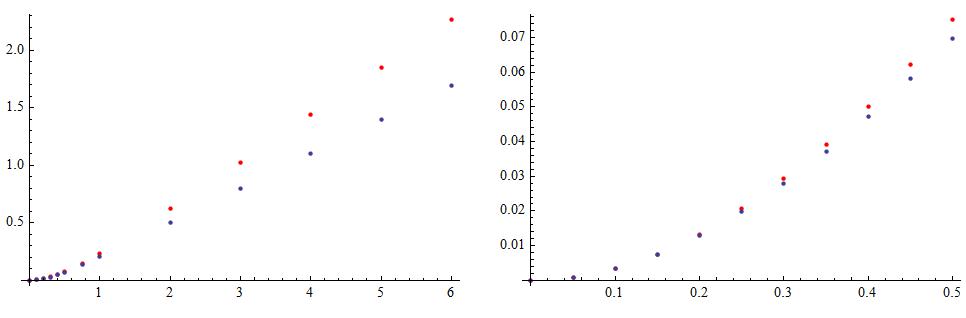}
 \begin{center}
(a)\qquad \qquad \qquad \qquad \qquad \qquad \qquad  \qquad \qquad  (b)
\end{center}
\caption{Plots of $C_\#$ (blue) and $C_\bullet$ (red) as functions of $q_1$ for $m=n=2$, $g=1$. Graph (b) magnifies the small charge region.}
\label{FigQ1dots}
\end{figure}

When $C_0$ reaches $C_\bullet$, both $A$ and $C$ continue to run above their $AdS_{m+1}\times S^n$ values at small $r$, but eventually $A$ saturates to a constant, while $C$ develops an exponential growth. An example of such flow for $m=n=2$ is presented in figure \ref{FigQ1AdSH}. At large values of $r$, the leading terms in $(A,C)$ are given by (\ref{AdS2genD}):
\bea\label{AdS2genDmain}
A\simeq\frac{m-1}{d-1}\frac{1}{g^2},\quad C\simeq c_0\exp\left[\frac{2(d-1)}{\sqrt{n(m-1)}}g^2 r\right],
\eea
where $c_0$ is a free constant, which can be absorbed into shifting $r$. This shift symmetry of equations (\ref{FullEinsteinM}) is broken by the boundary conditions (\ref{BCAug9}), so by integrating from the initial point $r=0$ to infinity, one finds a specific value of $c_0$ which depends only on $g$ and $q_1$. Functions (\ref{AdS2genDmain}) describe the asymptotic space $AdS_m\times H_{n+1}$, where $H_{n+1}$ is the maximally symmetric hyperbolic space with the metric
\bea\label{MetricHyper}
ds_H^2=g^2\frac{d-1}{m-1}dr^2+c_0\exp\left[\frac{2(d-1)}{\sqrt{n(m-1)}}g^2 r\right] d\Omega_m^2
\eea
and the Riemann tensor 
\bea
R_{abcd}=-\frac{1}{R_H^2}[g_{ac}g_{bd}-g_{ad}g_{bc}],\quad R_H=\frac{1}{g}\sqrt{\frac{n}{d-1}}\,.
\eea
The radius of $AdS_m$ is given by
\bea
R_{AdS}=\frac{1}{g}\sqrt{\frac{m-1}{d-1}}\,.
\eea
The numerical results for the critical value $C_\bullet$ are presented in figure \ref{FigQ1dots}, and in the next subsection we will derive the analytical expressions for $C_\bullet$ in the approximations of small and large $q_1$. 

\begin{figure}
 \includegraphics[width=1 \textwidth]{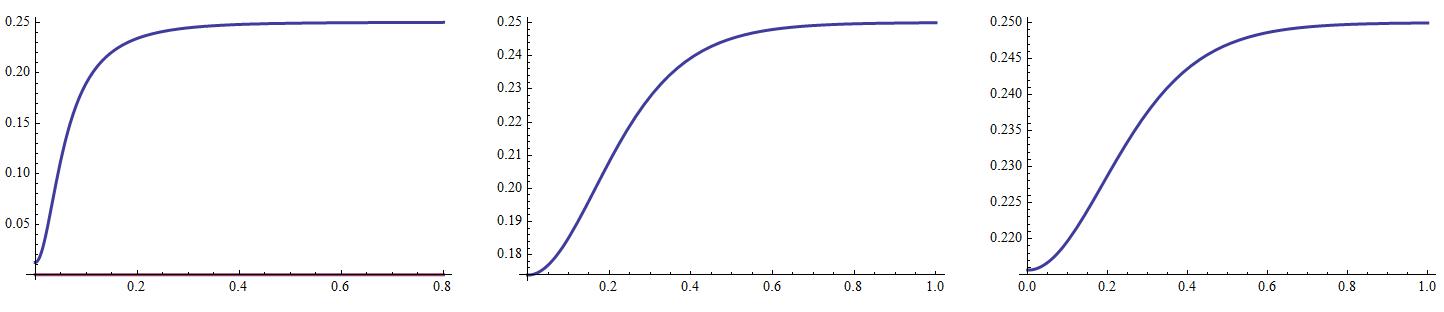}
 \begin{center}
(a)\qquad \qquad \qquad \qquad \qquad \qquad \qquad(b)\qquad \qquad \qquad \qquad \qquad
 \qquad  (c)
\end{center}
\caption{Plots of the $A(r)$ profiles describing solutions with $AdS_2\times S^3$ asymptotics for $q_1=0.1$ (a) , $q_1=1$ (b), $q_1=10$ (c). In all three cases, $g=1$.}
\label{FigQ1AdSH}
\end{figure}
\item $C_0>C_\bullet$: a flow to a naked curvature singularity at $r=r_1>0$.

If $C_0$ exceeds $C_\bullet$, then the warp factor $A(r)$ becomes a non--monotonic function. It grows at 
small $r$, but it turns around and goes to zero at some $r=r_1$. In the vicinity of this point, the leading contributions to functions $(A,C)$ are given by (\ref{Aug9L}),
\bea\label{Aug9Lv2}
C\simeq c (r_1-r)^{-\mu},\quad A\simeq a(r_1-r)^{\frac{2+n\mu}{m+1}},\quad
\mu=2\frac{(m+1)\sqrt{mn(d-2)}-2mn}{n[(m+1)^2+nm_-]}\,,
\eea
where $(a,c,r_1)$ are free parameters in the asymptotic expansions, which are fixed in terms of $C_0$ via integration of the system (\ref{FullEinsteinM}) between zero and $r_1$. Note that $\mu$ is bounded from above\footnote{Recall the shortcuts $m_\pm=m\pm 1$, as well as relation $d=m+n+1$.},
\bea
\frac{2+n\mu}{m+1}=2\frac{m_+\sqrt{mn(d-2)}+m_+^2-nm_+}{m_+[m_+^2+nm_-]}<2.
\eea
This implies that the distance between the origin and the singularity is finite,
\bea
\int_0^{r_1}\frac{dr}{\sqrt{A}}\,,\nonumber
\eea
so this branch should be discarded. Note that, according to (\ref{FirstOrdGenCnstr}) this singular branch persists for all values of $C_0$ above $C_\bullet$, and in the $C_0=\infty$ limit, the boundary value of $A$ is 
$A_0=mm_-/[(d-2)d_-g^2]$. 
\end{enumerate}

\noindent
To summarize, we have analyzed solutions of equations  (\ref{FullEinsteinM}) with initial conditions  (\ref{BCAug9}), (\ref{FirstOrdGenCnstr}) for all values of the parameter $C_0$. We found four critical points 
$(C_{min},C_\star,C_\#,C_\bullet)$ and identified eight scenarios (a)--(h). Short non--technical summaries of these situations are given in section \ref{SecSubSumMgn}. 

Our discussion in this subsection mostly relied on numerical solutions, and unfortunately this is the only option for generic values of $g$ and $q_1$ due to a complicated nature of the system (\ref{FullEinsteinM}). However, for small and large charges, one can obtain analytical approximations for the critical points $(C_{min},C_\star,C_\#,C_\bullet)$ and for the warp factors $(A,C)$. These results presented in the next subsection will also lead to insights into solutions with arbitrary $g$ and $q_1$. 

\subsection{Analytical approximations for large and small charges}
\label{SecApprox}

In this subsection we will analyze the system (\ref{FullEinsteinM}) for small and large values of $q_1$. Since charge is a dimensionful quantity, we need to clarify the meaning of these limits. The metric (\ref{TheAnsatz1}) implies that the warp factors $A$ and $C$, as well as coordinate $r$ have dimensions of length square, and $q_1$ has a dimension of length raised to the $(n-1)$--th power. The cosmological constant parameter $g$ was defined by
(\ref{LamDefG}) to have a dimension of inverse length. Therefore, the behavior of the system (\ref{FullEinsteinM})  is controlled by a dimensionless quantity
\bea
{\hat q}_1={q_1}{g^{n-1}}\,.
\eea
In this subsection we will analyze the limits of small and large ${\hat q}_1$.

\subsubsection{Large charge}
We begin with taking the limit of large ${\hat q}_1$ in the system (\ref{FullEinsteinM}). Since $q_1$ appears in  (\ref{FullEinsteinM}) only through the ratios $q_1/C^n$, the large charge limit must involve a rescaling of $C$. Specifically, writing
\bea
C=\left(\frac{q_1}{g}\right)^{\frac{2}{n}}E,
\eea
and sending ${\hat q}_1$ to infinity while keeping $(A,E,r,g)$ fixed, we find the limit of the system (\ref{FullEinsteinM}):
\bea\label{WinstLageQ1}
&&d_-g^2+\frac{n_- g^2}{2(d-2)E^n}-\frac{n\dot A\dot E}{4E}+\frac{nA{\dot E}^2}{4E^2}-\frac{m}{2}\ddot A-\frac{nA\ddot E}{2E}=0,
\nn
&&d_-g^2-\frac{m g^2}{2(d-2)E^n}-\frac{m_+{\dot A}{\dot E}}{4E}+\frac{(n-2)A{\dot E}^2}{4E^2}=0,
\\
&&d_- g^2A-m_-+\frac{(n-1)A g^2}{2(d-2)E^n}-\frac{m_-}{4}{\dot A}^2-\frac{nA{\dot A}{\dot E}}{4E}-\frac{1}{2}A{\ddot A}=0.\nonumber
\eea
Parameter $g$ can be absorbed into rescaling of $A$ and $r$ making them 
dimensionless.\footnote{Specifically, one should write $(A,r^2)= \frac{1}{g^{2}} ({\tilde A},{\tilde r}^2)$.} To avoid unnecessary clutter, we will not introduce a separate notation for the rescaled quantities, but rather set $g=1$ in (\ref{WinstLageQ1}). The limit of the boundary conditions (\ref{BCAug9}), (\ref{FirstOrdGenCnstr}) reads
\bea\label{BCAug9ForE}
E(0)=e,\quad A(0)=\frac{1}{g^2}\left[\frac{(d-1)(d-2)}{m(m-1)}-\frac{1}{2m(m-1)e^n}\right]^{-1}
{\hskip -0.3cm,}
\quad {\dot E}(0)={\dot A}(0)=0.
\eea
Integrating the system (\ref{WinstLageQ1})--(\ref{BCAug9ForE}), one can extract the values of the critical parameters $(C_\#,C_\bullet)$ discussed in the previous subsection. In the large ${\hat q}_1$ limit, the critical points depend on the charge and $g$ in a very simple way:
\bea\label{DefEcrit}
(C_{min},C_\star,C_\#,C_\bullet)=\left(\frac{q_1}{g}\right)^{\frac{2}{n}}
(e_{min},e_\star,e_\#,e_\bullet).
\eea
The values of $e_{min}$ and $e_\star$ can be extracted by taking the large charge limits in (\ref{CminDef}) and (\ref{AdS32exct2}):
\bea
e_{min}=\left[\frac{1}{2(d-1)(d-2)}\right]^{\frac{1}{n}},\quad
e_\star=\left[\frac{m}{2(d-1)(d-2)}\right]^{\frac{1}{n}}.
\eea
To determine $e_\#$ and $e_\bullet$, one has to solve the system (\ref{WinstLageQ1}) with the appropriate boundary conditions at infinity\footnote{Recall that $A/C$  approaches one at infinity for $C_0=C_\#$, and $A$ approaches the constant value (\ref{AdS2genDmain}) for 
$C_0=C_\bullet$.}. In particular the ratio $C/A$ goes to one at infinity for $C_0=C_\#$, but this means that $E/A$ goes to zero for $e=e_\#$, as it does for $e=e_*$. We conclude that in the large charge approximation,
\bea
e_\#=e_*\,.
\eea
Since equations (\ref{WinstLageQ1})  have a complicated dependence on $m$ and $n$, studying their analytical properties just to extract just one number $e_\bullet$ is not very instructive, so we opted for numerical integration. The resulting values of 
$(e_\star=e_\#,e_\bullet)$ for $d\le 10$ are summarized in Table \ref{Table1}, where every block has a format
\bea\label{FormatTbl1}
\begin{array}{|c|c|}
\hline
&d\\
\hline
&e_\star\\
m&e_\bullet\\
\hline
\end{array}
\eea
The corresponding values of $(C_{min},C_\star,C_\#,C_\bullet)$ are given by (\ref{DefEcrit}). 
\begin{table}
\begin{center}
\begin{tabular}{|c|c|c|c|c|c|c|}
\hline
&d=5&d=6&d=7&d=8&d=9&d=10\\
\hline
&0.2887&0.3684&0.4373&0.4735&0.5113&0.5428\\
m=2&0.4175&0.4290&0.4652&0.5001&0.5313&0.5586\\
\hline
&&0.2739&0.3684&0.4347&0.4848&0.5246\\
m=3&---&0.3807&0.4152&0.4612&0.5018&0.5362\\
\hline
&&&0.2582&0.3625&0.4347&0.4884\\
m=4&---&---&0.3515&0.4016&0.4554&0.5006\\
\hline
&&&&0.2440&0.3547&0.4317\\
m=5&---&---&---&0.3279&0.3890&0.4488\\
\hline
&&&&&0.2315&0.3467\\
m=6&---&---&---&---&0.3083&0.3775\\
\hline
&&&&&&0.2205\\
m=7&---&---&---&---&---&0.2918\\
\hline
\end{tabular}
\end{center}
\caption{Numerical values of $(e_\star,e_\bullet)$ in various dimensions specified in the format (\ref{FormatTbl1}).}
\label{Table1}
\end{table}

\subsubsection{Small charge}
\label{SecMgnSmChrg}

Let us now analyze the system (\ref{FullEinsteinM}) for ${\hat q}_1\ll 1$. In contrast to the large charge case, no simple scaling can eliminate $q_1$ while preserving the correct asymptotic behavior of functions $A$ and $C$. On the other hand, it is possible to divide the full range of $r$ into several overlapping regions and to construct analytic solutions in all of them. This leads to analytic expressions for $(C_{min},C_\star,C_\#,C_\bullet)$ in all dimensions. 

\bigskip

To perform the analysis in the small ${\hat q}_1$ regime, we focus on the rescaled system (\ref{RescaledSystemEqn}). The dependence on the charge $q_1$ and the cosmological parameter $g$ is encoded in one quantity ${\bar g}$ given by (\ref{gBarDef}):
\bea\label{gBarDefRept}
{\bar g}=gq^{\frac{1}{n-1}}.
\eea 
Functions $A$ and $C$ can be recovered from ${\bar A}$ and ${\bar C}$ using (\ref{ACrescaleBar}). To find the leading contribution to ${\bar C}_\star$, which encodes the boundary condition describing the $AdS_{m+1}\times S^n$ space, we can solve equation (\ref{AdS32exct2}) in the limit of small ${\hat q}_1$:
\bea\label{CstarLead}
C_\star\simeq \left[\frac{mq_1^2}{2(d-2)(n-1)}\right]^{\frac{1}{n-1}}\ \Rightarrow\ 
{\bar C}_\star\simeq \left[\frac{m}{2(d-2)(n-1)}\right]^{\frac{1}{n-1}}.
\eea
Numerical analysis shows that the flows to $AdS_{m+n+1}$ and $AdS_m\times H_{n+1}$ spaces happens when
\bea\label{CcritGbar}
{\bar C}_{\#}={\bar C}_\star+a_{\#} {\bar g}^m,\quad 
{\bar C}_{\bullet}={\bar C}_\star+a_{\bullet} {\bar g}^m,
\eea
where $(a_\#,a_\bullet)$ remain finite when $\bar g$ goes to zero. This illustrates the crucial difference between the large and small charge approximations: while the former case gave the uniform scaling (\ref{DefEcrit}) of all critical points, in the latter situation, both ${\bar C}_{\#}$ and ${\bar C}_{\bullet}$ contain important contributions with different powers of ${\bar g}$. Note that ${\bar C}_\star$ in the right hand sides of (\ref{CcritGbar}) denotes the exact solution of (\ref{AdS32exct2}) rather than the leading approximation (\ref{CstarLead})\footnote{In practice, it is important to keep all terms in ${\bar C}_\star$ at least up to 
${\bar g}^{m+1}$.}. For completeness we also present the leading contribution to the solution of equation (\ref{CminDef}):
\bea
C_{min}\simeq \left[\frac{q_1^2}{2n(n-1)}\right]^{\frac{1}{n-1}}\ \Rightarrow\ 
{\bar C}_{min}\simeq \left[\frac{1}{2n(n-1)}\right]^{\frac{1}{n-1}}\,.\nonumber
\eea
The goal of our analysis is to determine the values of  $(a_\#,a_\bullet)$. 

\bigskip
\noindent
We will proceed by dividing the $0\le {\bar r}<\infty$ part of the axis into three regions. All derivations are presented in the Appendix \ref{SecApp}. 
\begin{itemize}
\item {\bf Region 1}: ${\bar r}\ll \frac{1}{\bar g}$.

First we take the small ${\bar g}$ limit while keeping functions $({\bar A},{\bar C})$, as well as coordinate 
${\bar r}$ fixed. According to (\ref{CcritGbar}), in the leading order, the solution is $AdS_{m+1}\times S^n$, so recalling (\ref{AdS32exct1}), we write
\bea\label{Reg1AnstzMain}
{\bar A}={\bar A}_\star {{\bar r}^2}+\frac{1}{\Ab_\star}+\eps \Ab_1({\bar r}),\ 
{\bar C}=\Cb_\star+\eps \Cb_1({\bar r}),\ 
\Ab_\star=\frac{n-1}{2(d-2)m}\frac{1}{\Cb_\star^{n}}+\frac{d-1}{m}.
\eea
These expressions solve the system (\ref{RescaledSystemEqn}) for $\eps=0$, and the outcome of the numerical analysis (\ref{CcritGbar}) suggests that to recover the other critical points, $\eps$ should be of order 
${\bar g}^m$. We will assume that $\eps$ scales in this fashion, and this assumption will be justified by the final result. 

Expanding equations  (\ref{RescaledSystemEqn}) to the first order in $\eps$, we find linear equations for 
$(A_1,C_1)$. The boundary conditions, ${\dot{\bar A}}_1(0)={\dot{\bar C}}_1(0)=0$, eliminate some integration constants, leading to the unique expression for ${\bar C}_1$ in terms of the Legendre function\footnote{In the Appendix we also give the expression for $Q$ in terms of the hypergeometric function (\ref{Qhyper}), which, while messy, might resonate with some readers.}
$Q$:
\bea\label{C1solnMain}
\Cb_1({\bar r})=c_1\left[{\hat r}^2+1\right]^{\frac{1-m}{4}}Q^{(\beta)}_\alpha\left[i{\hat r}\right],\quad
\alpha=\frac{3m-1}{2},\quad \beta=\frac{m-1}{2}.
\eea
Here we defined a rescaled radial coordinate
\bea
{\hat r}\equiv \frac{(n-1)^2{\bar r}}{m^2\Cb_\star}.
\eea
Function ${\bar A}_1$ can be expressed in terms of Legendre functions as well, but the result is not very illuminating.

For even values of $m$, function (\ref{C1solnMain}) is a polynomial of degree $m$ in ${\hat r}^2$, and for odd $m$ it contains a mixture of polynomials and arctangents. The first few instances of functions 
${\bar C}_1$ are given by (\ref{C1evenM}) and (\ref{C1oddM}). The leading behavior at large ${\hat r}$ is 
always given by
\bea
{\bar C}_1({\bar r})=\gamma_m \Cb_1(0){\hat r}^m,
\eea
where $\gamma_m$ is a constant which can be easily extracted from (\ref{C1solnMain}) on the case--by--case basis. The first few instances are
\bea
\gamma_2=4,\quad \gamma_4=\frac{80}{3},\quad \gamma_6=\frac{896}{5},\quad
\gamma_3=\frac{21\pi}{4},\quad \gamma_5=\frac{165\pi}{16},\quad \gamma_7=\frac{4849845\pi}{32768}\,.
\eea
We conclude that the effect of the region 1 can be summarized as a map (\ref{Reg1Cfin}) between expansions at 
${\bar r}=0$ and ${\bar r}=\infty$:
\bea\label{Reg1CfinMain}
\Cb=\Cb_\star+a{\bar g}^m+O(\rb^2)\ \rightarrow \ \Cb=\Cb_\star+
a\gamma_m\left[\frac{(n-1)^2{\bar g}\rb}{m^2\Cb_\star}\right]^m+O(\rb^{m-1}).
\eea
The large--$\rb$ behavior of function $A$ is given by (\ref{Reg1Afin}):
\bea\label{Reg1AfinMain}
\Ab=\Ab_\star {\rb^2}+\frac{1}{\Ab_\star}-\frac{a\gamma_m}{{\bar g}^2} \frac{nm^2}{(m+1)(n-1)^2}\left[\frac{(n-1)^2{\bar g}\rb}{m^2\Cb_\star}\right]^{m+2}+\dots
\eea
Expansions (\ref{Reg1CfinMain}) and (\ref{Reg1AfinMain}) are applicable as long as ${\bar g}$ is smaller than any other scale in the problem. In particular, one runs into a potential trouble with the order of limits when ${\rb}$ becomes larger than any negative power of ${\bar g}$. A detailed study of subleading terms in the expansions (\ref{Reg1CfinMain})--(\ref{Reg1AfinMain}) indicates that these large $\rb$ approximation are applicable as long as ${\bar g}\rb\ll 1$. We need a different approximation to go beyond this limitation.

\item
{\bf Region 2}:  $\frac{1}{\bar g}\le {\bar r}\ll \frac{1}{{\bar g}^2}$.

The exit from region 1 at large $\bar r$, (\ref{Reg1CfinMain})---(\ref{Reg1AfinMain})\footnote{See also (\ref{Reg1Exit}) for a suggestive rewriting of these formulas.}, inspires a rescaling
\bea\label{Reg2Main}
{\bar A}=\frac{1}{{\bar g}^2}A_2({\bar g}{\bar r}),\quad
{\bar C}=C_2({\bar g}{\bar r}),\quad x={\bar g}{\bar r},
\eea
and taking the small ${\bar g}$ limit while keeping $x$ fixed. The resulting system (\ref{Reg2syst}) is solved in the Appendix \ref{SecApp}, and the effect of the region 2 can be summarized as a map between regions with small and large values of $x$ leading to 
\bea\label{ExitReg2Main}
{\bar A}= \frac{1}{{\bar g}^2 c_1}+...,\quad {\bar C}=c_1({\bar g}{\bar r})^2+...,
\quad
c_1=\frac{1}{{\bar A}_\star}\left[\frac{n_- a\gamma_m}{2{\bar C}_\star}\right]^{\frac{2}{m}}\left[\frac{(n-1)^2{\bar g}}{m^2{\bar C}_\star}\right]^2.
\eea
Here 
\bea\label{AppSum3aMain}
{\bar A}_\star=\frac{n-1}{2(d-2)m}{\bar C}_\star^{-n}+\frac{d-1}{m}.
\eea
Inversion of the last relation in (\ref{ExitReg2Main}) gives
\bea\label{AppSum4Main} 
a=\frac{2{\bar C}_\star}{n_- \gamma_m}\left[c_1{\bar A}_\star\right]^{\frac{m}{2}}
\left[\frac{m^2{\bar C}_\star}{(n-1)^2{\bar g}}\right]^m.
\eea
The expansions of $(A_2,C_2)$ in inverse powers of $x$ are fully determined by one parameter $c_1$ and dimensions $(m,n)$. Interestingly, function $A$ approaches a constant for large $x$. In figure \ref{FigQ1small} we confirm that such saturation indeed happens by integrating equations (\ref{FullEinsteinM}) with ${\bar g}=10^{-3}$ and $m=n=2$, but the general result holds in all dimensions. 

\begin{figure}
 \includegraphics[width=1 \textwidth]{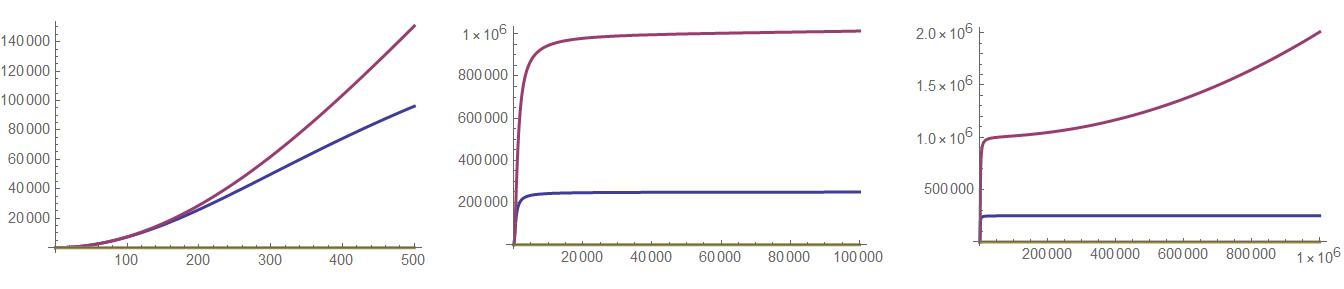}
 \begin{center}
(a)\qquad \qquad \qquad \qquad\qquad \qquad \qquad (b) \qquad \qquad  \qquad \qquad\qquad \qquad  (c)
\end{center}
\caption{Profiles of $A(r)$ for $(m,n)=(2,2)$, ${\bar g}=10^{-3}$ in regions 1 (a), 2 (b), 
and 3 (c) described in the text. Blue curve describes the flow to $AdS_5$, and the purple one describes the flow to $AdS_2\times H_3$.}
\label{FigQ1small}
\end{figure}

In our analysis of region 2, we assumed that $(A_2,C_2,x)$ remained finite when ${\bar g}$ went to zero, 
and this approximation breaks down when $x$ becomes larger than some negative power of ${\bar g}$. A detailed study of subleading terms in the expansions (\ref{ExitReg2Main}) indicates that these large $\rb$ approximation are applicable as long as ${\bar g}x\ll 1$. A new scaling is needed to overcome this limitation.

\item {\bf Region 3}

The exit from region 2 at large $\bar r$ suggests a rescaling
\bea
{\bar A}=\frac{1}{{\bar g}^2}A_3({\bar g}^2{\bar r}),\quad 
{\bar C}=\frac{1}{{\bar g}^2}
C_3({\bar g}^2 {\bar r}),\quad y={\bar g}^2{\bar r}.
\eea
Then expressions (\ref{ExitReg2}) imply that at small $y$ functions $(A_3,C_3)$ behave as
\bea\label{Reg3BCMain}
A_3=\frac{1}{c_1}+O(y),\quad C_3=c_1 y^2+O(y^3).
\eea
Equations for functions $(A_3,C_3)$ are analyzed in Appendix \ref{SecApp}. Depending on the value of $c_1$, function $A_3$ can grow at infinity, reach a finite value there, or hit zero at some finite point. We are particularly interested in the values of $c_1$ which give rise to the standard  $AdS_d$ and to 
$AdS_{m}\times H_{n+1}$:
\begin{enumerate}[a.]
\item Although all values of $c_1$ leading to growing function $A$ give rise to regular wormholes connecting two copies of asymptotically locally $AdS_d$ spaces\footnote{Recall option (d) on page \pageref{optDpg}.}, a special role is played by solutions with $C_0=C_\#$, which produce the standard $AdS_d$ asymptotics at both 
infinities\footnote{Recall option (e) on page \pageref{optEpg}.}. For this special initial condition, the ratio $C/A$ approaches one at infinity, and clearly this can happen only at one value of $c_1$ in (\ref{ExitReg2Main}). 
The corresponding exact solution for functions $(A_3,C_3)$ in region 3 is remarkably simple:
\bea
A_3=y^2+1,\quad C_3=y^2, \quad c_1=1.
\eea
In the original variables, this translates into the asymptotic expressions at large $r$:
\bea
{\bar A}\simeq{\bar g}^2 ({\bar r}-{\bar r}_0)^2+\frac{1}{{\bar g}^2},\quad 
{\bar C}\simeq{\bar g}^2 ({\bar r}-{\bar r}_0)^2.
\eea
The special value $c_1=1$ translates into 
\bea\label{ahashAns} 
a_\#=\frac{2{\bar C}_\star}{n_- \gamma_m}\left[{\bar A}_\star\right]^{\frac{m}{2}}
\left[\frac{m^2{\bar C}_\star}{(n-1)^2{\bar g}}\right]^m.
\eea
via (\ref{AppSum4Main}). For $C_0<C_\#$, function $A$ grows faster than $C$ at infinity, while for $C_0>C_\#$ the situation is reverse. 
\item The value of $C_\bullet$ was defined through the initial condition $C_0=C_\bullet$ that gives rise to the 
$AdS_m\times H_{n+1}$ asymptotic geometry with a constant $A$ at infinity. This warp factor already approaches a constant at the exit from region 2 (\ref{ExitReg2Main}), and for some value of $c_1$ it does not vary in region 3. The asymptotic value of $A$ from equation (\ref{AdS2genDmain}) determined the critical $c_1$, and the corresponding exact solution in region 3 is found in the Appendix \ref{SecApp}:
\bea
A_3=\frac{m-1}{d-1},\quad C_3=\frac{n}{4(d-1)}\left[w-\frac{1}{w}\right]^2,\quad 
w=\exp\Big[\frac{(d-1)y}{\sqrt{n(m-1)}}\Big].
\eea
The leading contribution to $C_3$ indeed reproduces (\ref{AdS2genDmain}). 
The value of $a_\bullet$ is determined by combining (\ref{AppSum4}) and (\ref{AppSum5}):
\bea\label{abulletAns} 
a_\bullet=\frac{2{\bar C}_\star}{n_- \gamma_m}\left[\frac{d_-}{m_-}{\bar A}_\star\right]^{\frac{m}{2}}
\left[\frac{m^2{\bar C}_\star}{(n-1)^2{\bar g}}\right]^m.
\eea
\end{enumerate}
This concludes the discussion of region 3.
\end{itemize}
\noindent
To summarize, we have solved equations (\ref{RescaledSystemEqn}) in three overlapping regions and determined the effects of the boundary conditions at zero on the behavior of the solution at infinity. Specifically, starting with 
\bea\label{CcritGbarV2}
{\bar C}(0)={\bar C}_\star+a {\bar g}^m,
\eea
we found the critical values $(a_\#,a_\bullet)$ describing the standard $AdS_d$ and $AdS_m\times H_{n+1}$. The answers are given by (\ref{ahashAns}) and (\ref{abulletAns}).
The numerical results for ${\bar g}=10^{-3}$, $m=n=2$ presented in figure \ref{FigQ1small} illustrate the general analytical properties of small--${\bar g}$ solutions derived for all values of $(m,n)$.

\bigskip
\noindent
This concludes our analysis of regular magnetic solutions covered by the ansatz (\ref{TheAnsatz1}). 
Before going to the dyonic case, which bears some qualitative similarities with the wormhole scenarios discussed here, we switch our attention to the purely electric case, where the structure of regular solutions is drastically different. The dyonic geometries will be discussed in section \ref{SecDyon}. 

\section{Electric solutions}
\label{SecElectr}
\renewcommand{\theequation}{4.\arabic{equation}}
\setcounter{equation}{0}

In this section we will analyze equations (\ref{FullEinstein}) with vanishing $q_1$. and nonzero $q_2$. In contrast to the magnetic case discussed in the previous section, the electric ansatz does not lead to wormhole geometries. The regular solutions have only one spacial infinity and they terminate at points where sphere $S^n$ collapses to zero size. In section \ref{SecExact} we have discussed several exact solutions of this type: the $AdS_{m}\times X$ geometries (\ref{AdS2xS3}),  (\ref{AdS2xR4}), (\ref{AdS2xH3}) and 
the neutral $AdS_{d}$ solution (\ref{AdSexact}). Here we will perform a complete analysis of all flows at arbitrary values of the electric charge $q_2$ and justify the summary presented in section \ref{SecFlowSumry}.

\subsection{The boundary problem for regular solutions}
\label{SecElectrBC}

We begin with quoting equations of motion (\ref{FullEinstein}) for $q_1=0$:
\bea\label{FullEinsteinE}
&&d_-g^2-\frac{m_- q_2^2}{2(d-2)A^m}-\frac{n\dot A\dot C}{4C}+\frac{nA{\dot C}^2}{4C^2}-\frac{m}{2}\ddot A-\frac{nA\ddot C}{2C}=0,
\nn
&&d_-g^2+\frac{n_-}{C}-\frac{m_-q_2^2}{2(d-2)A^m}-\frac{m_+{\dot A}{\dot C}}{4C}+\frac{(n-2)A{\dot C}^2}{4C^2}-
\frac{A\ddot C}{2C}=0,\\
&&d_-g^2-\frac{m_-}{A}+\frac{nq_2^2}{2(d-2)A^m}-\frac{m_-}{4}
\frac{{\dot A}^2}{A}-\frac{n{\dot A}{\dot C}}{4C}-\frac{1}{2}{\ddot A}=0.\nonumber
\eea
In contrast to the magnetic case, now $C$ can reach zero, and it does so in the 
$AdS_{m}\times S^{n+1}$ solution (\ref{AdS2xS3}). 
A wormhole would correspond to a solution where $C$ never vanishes. Repeating the analysis from section \ref{SecMagnBound}, we conclude that if $C$ reaches a non--zero minimum, then ${\dot A}$ must vanish at that point to avoid a singularity. This guarantees that both $C$ and $A$ are even functions under reflection around their common minimum. Looking at the expansions of the form
\bea\label{ElecExpand}
A&=&A_0+A_1 (r-r_0)^2+A_2 (r-r_0)^4+\dots,\\ 
C&=&C_0+C_1 (r-r_0)^2+C_2 (r-r_0)^4+\dots,\nonumber
\eea 
and substituting them into equations (\ref{FullEinsteinE}), we determine all coefficients in terms of $A_0$. The explicit expressions for $(A_1,C_0,C_1)$ are not very illuminating, but remarkably there is a simple relation between them:
\bea\label{ElecExpandConstr}
A_0C_1+\frac{m}{n-1}A_1C_0+1=0.
\eea
This implies that at least one of the four coefficients $(A_0,A_1,C_0,C_1)$ must be negative, and since both $A_0$ and $C_0$ must be positive to ensure the correct signature of the metric (\ref{TheAnsatz1}), we conclude that it is impossible to have a simultaneous minimum for $A$ and $C$. This implies that regular solutions must have a point where $C$ vanishes. An example of such a solution is given by $AdS_m\times S^{n+1}$(\ref{AdS2xS3}). In that case, function $C$ vanishes at $r=0$ and it reaches a {\it maximum} at $r=\frac{\pi}{2u}$. Near this critical point one finds
\bea\label{AdS2xS3NearPi}
A=A_*,\quad C=\frac{1}{A_* u^2}-\frac{1}{A_*}\left[r-\frac{\pi}{2u}\right]^2+
\frac{u^2}{3A_*}\left[r-\frac{\pi}{2u}\right]^4+\dots
\eea
Comparing this with (\ref{ElecExpand}), we can extract the values of various coefficients for 
$AdS_m\times S^{n+1}$,
\bea
(A_0,A_1,A_2)=(A_*,0,0),\quad (C_0,C_1,C_2)=
\left(\frac{1}{A_* u^2},\frac{1}{A_*},\frac{u^2}{3A_*}\right),\nonumber
\eea
and verify the constraint (\ref{ElecExpandConstr}). As guaranteed by the general analysis of that constraint, the $r=\frac{\pi}{2u}$ point for $AdS_m\times S^{n+1}$ is not a simultaneous minimum of $A$ and $C$.

To summarize, we have shown that there are no regular solutions of system (\ref{FullEinsteinE}) with growing $S^n$ at infinity unless function $C$ vanishes at some value of $r$. Since equations (\ref{FullEinsteinE})  are invariant under a shift of the radial coordinate, we can place the origin at that value, i.e., we can assume that $C(0)=0$ without loss of generality. An example of such solution is given by (\ref{AdSexact}), which describes the $AdS_d$ space in the absence of the electric charge. The goal of this section is to construct similar solutions for non--zero $q_2$.

\bigskip

To ensure regularity at $r=0$, we must require the radial direction and the sphere to combine into a patch of flat space. In the $A=B$ gauge used in (\ref{FullEinsteinE}) this implies that $C\simeq r^2/A_0$, then the system (\ref{FullEinsteinE}) gives the unique expansions for every value of $A_0$, and the first few terms are
\bea\label{ElctrSeries}
&&A=A_0+A_1 r^2+A_2 r^4+O(r^6),\quad C=\frac{r^2}{A_0}+C_1 r^4+O(r^6),\\
&&A_1=\frac{d_- g^2}{n_+}-\frac{m_-}{n_+A_0}+\frac{nq_2^2}{2n_+(d-2)A^m_0},\quad
C_1=\frac{m_-(1-A_1A_0)}{3nA_0^3}-\frac{q_2^2}{6nA_0^{m+2}}\,,\nonumber\\
&&A_2=\frac{m_-d_-A_1}{3A_0n_+(n+3)}\left[\frac{1}{A_0}-g^2-\frac{m(3n^2+4n-1)-n^2-2n+1}{4d_-m_-(d-2)A_0^m}q_2^2\right].\nonumber
\eea
The leading contributions to $(A,C)$ and symmetry of equations (\ref{FullEinsteinE}) under reflection of $r$ guarantee that only even powers of $r$ appear in the expansions. For $q_2=0$ and $A_0=1/g^2$, both series in (\ref{ElctrSeries}) terminate, reproducing the $AdS_d$ solution (\ref{AdSexact}). For $A_0=A_*$, the solution of (\ref{AdS2xS3star}), the series for $A$ terminates after $A_0$, while the expansion of $C$ can be summed to give (\ref{AdS2xS3}), (\ref{AdS2xR4}), or (\ref{AdS2xH3}), depending on the values of $q_2$ and $g$. 

\bigskip

We are interested in constructing regular solutions which start as (\ref{ElctrSeries}) near $r=0$. There are three logical options:
\begin{enumerate}[i)]
\item Function $C$ increases up to a certain value of $r$, then it decreases and goes to zero at $r=r_1$.
In the resulting solution, the $n$--sphere collapses at two points ($r=0,r_1$), but the geometry has a non--contractible $S^{n+1}$. Such solutions exist if 
\bea\label{q2Crit}
q_2<q_{cr}\,,
\eea
and the unique warp factors $(A,C)$ leading to regular metrics are given by (\ref{AdS2xS3}). 
\item Function $C$ increases indefinitely, but the AdS warp factor reaches a constant at $r=\infty$. 
Such solutions exist for all $q_2\le q_{max}$, and the asymptotic warp factors are given by (\ref{AdS2genD}). The asymptotic geometry is $AdS_m\times H_{n+1}$. For $q_2=q_{cr}$, there is an additional option (\ref{AdS2xR4as}) for the warp factors, which leads to a metric with the $AdS_m\times R^{n+1}$ asymptotics.
\item Functions $C$ and $A$ never vanish away from $r=0$, and they increase without bounds as $r$ goes to infinity. Such solutions exist for all values of $q_2$, and they have $AdS_d$ asymptotics. In the $q_2=0$ limit they recover the standard $AdS_d$ in global coordinates (\ref{AdSexact}). 
\end{enumerate}

Before analyzing these options in detail, it is instructive to compare them with our discussion on magnetic solutions presented in the previous sections. Since the sphere could not collapse in that case, option i) was not available. Instead, it was possible to have the $AdS_{m+1}\times S^n$ solution (\ref{AdS32exct1}), where $C$ remained constant. As discussed in section \ref{SecExact}, a decoupled spherical factor in the presence of a negative cosmological constant becomes possible due to a positive contribution to $R_S$ in (\ref{AdS32exct3a}) coming from the magnetic charge in the Freund--Rubin ansatz. An electric charge does not make such contribution, so the standard intuition about impossibility of an $X\times S^n$ space in the presence of a negative or vanishing cosmological constant holds. On the other hand, a flux dual to the electric field on $AdS_m$ provided a similar Freund--Rubin mechanism for $S^{n+1}$ leading to the option i). 

Options ii) and iii) had direct counterparts in the magnetic case. Specifically, option ii) was covered by item (g) in section \ref{SecSubMagnFlow} with a couple of differences. First, the item (g) existed for all values of the magnetic charge, while option ii) applies only for sufficiently small $q_2$. Second, the magnetic solutions (g) were constructed only numerically and through asymptotic expansions, while the electric $AdS_m\times H_{n+1}$ solution
(\ref{AdS2xH3}) is exact. Finally, in contrast to its magnetic version, configuration (\ref{AdS2xH3}) has only one 
spacial infinity and does not correspond to a wormhole. Option (iii) is analogous to items (d)--(f) in the magnetic case, but it has  only one spacial infinity and does not correspond to a wormhole. After this brief comparison, let us present the results for the electric case. Since some of our results will be numerical, we begin with reducing the number of parameters in the system (\ref{FullEinsteinE}) by performing a rescaling similar to (\ref{ACrescaleBar})--(\ref{gBarDef}).

\bigskip

To carry out a numerical study of the system (\ref{FullEinsteinE}), it is convenient to remove one of the two parameters  $(g,q_2)$ by rescaling the warp factors and the radial coordinate, following the same logic that led to (\ref{RescaledSystemEqn}) in the magnetic case. Specifically, introducing the counterparts of (\ref{ACrescaleBar})--(\ref{gBarDef}),
\bea\label{ACrescaleBarElctr}
A=q_1^{\frac{2}{m-1}}{\bar A}[r q_2^{-\frac{2}{m-1}}],\quad C=q_2^{\frac{2}{m-1}}
{\bar C}[r q_2^{-\frac{2}{m-1}}],\quad {\bar g}=gq_2^{\frac{1}{m-1}}\,,
\eea
we arrive at a rescaled version of the system (\ref{FullEinsteinE}):
\bea\label{RescaledSystemEqnElctr}
&&{\bar g}^2 d_--\frac{m_- }{2(d-2){\bar A}^m}-\frac{n\dot {\bar A}\dot{\Cb}}{4\Cb}+\frac{n\Ab{\dot \Cb}^2}{4\Cb^2}-\frac{m}{2}\ddot \Ab-\frac{n\Ab\ddot \Cb}{2\Cb}=0,
\nn
&&{\bar g}^2 d_-+\frac{n_-}{\Cb}-\frac{m_-}{2(d-2)\Ab^m}-\frac{m_+{\dot \Ab}{\dot \Cb}}{4\Cb}-\frac{(n-2)\Ab{\dot \Cb}^2}{4\Cb^2}-
\frac{\Ab\ddot \Cb}{2\Cb}=0,\\
&&{\bar g}^2 d_--\frac{m_-}{\Ab}+\frac{n}{2(d-2)\Ab^m}-
\frac{m_-}{4}\frac{{\dot \Ab}^2}{\Ab}-\frac{n{\dot \Ab}{\dot \Cb}}{4\Cb}-\frac{1}{2}{\ddot \Ab}=0.\nonumber
\eea
The dot above functions $(\Ab,\Cb)$ denotes derivatives with respect to the rescaled radial coordinate, 
${\bar r}=r q_2^{-\frac{2}{m-1}}$. The critical value (\ref{qCrit}) translates into
\bea\label{qCritBar}
{\bar g}_{cr}=(2m_-)^{\frac{2}{m-1}}\left[\frac{(m_-)^2}{(d-2)d_-}\right]^{\frac{1}{2}}\,.
\eea
Let us now describe the properties of the system (\ref{FullEinsteinE}), (\ref{RescaledSystemEqnElctr}).

\subsection{Classification of regular solutions}
\label{SecElectrAnlz}

Equations (\ref{FullEinsteinE}) can be integrated with boundary conditions (\ref{ElctrSeries}), and the outcome depends on three parameters, $(g,q_2,A_0)$. In contrast to the magnetic case, where all values of $q_1$ appeared on the same footing, now the available flows depend on the value of $q_2$. Specifically, there are two special points, $q_{cr}$ and $q_{max}$, so one has to analyze five separate cases, A-E:
\begin{enumerate}[A.]
\item $q_2<q_{cr}$
\begin{enumerate}[(a)]
\item $A_0<A_*$: singular solution with irregular collapse of the sphere.

\begin{figure}
 \includegraphics[width=1 \textwidth]{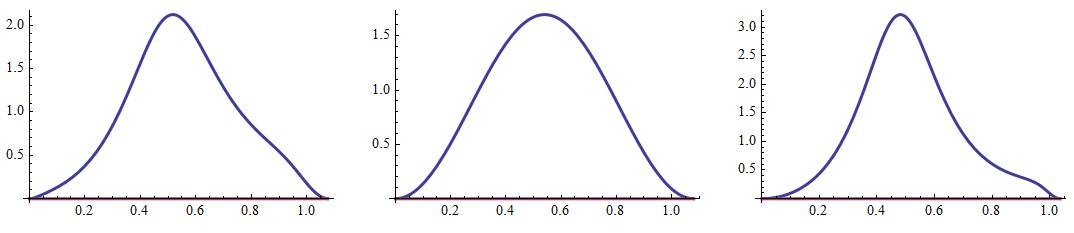}\\
 \includegraphics[width=1 \textwidth]{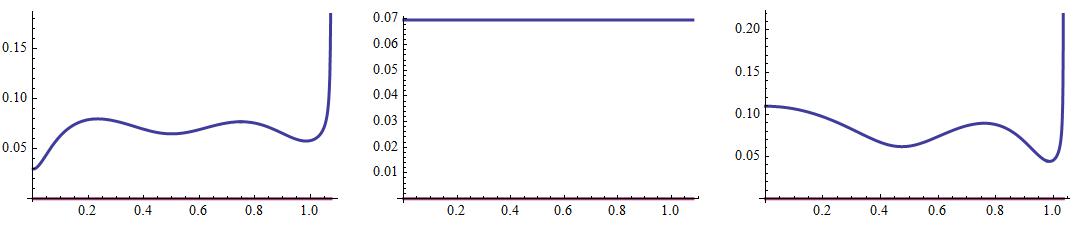}\\
 \begin{center}
(a)\qquad \qquad \qquad \qquad \qquad \qquad \qquad(b)\qquad \qquad \qquad \qquad \qquad
 \qquad  (c)
\end{center}
\caption{Warp factors $C(r)$ (top row) and $A(r)$ (bottom row) for $(m,n)=(2,2)$, $q_2<q_{cr}$ solutions with 
collapsing sphere. These numerical results are obtained for $g=1$ and $q_2=0.338$. Note that 
$q_{cr}=0.408$. \newline 
(a) $A_0<A_*$: unphysical solution with curvature singularity (\ref{ElecCaseAa});\newline 
(b) $A_0=A_*$: regular  $AdS_m\times S^{n+1}$ geometry;\newline 
(c) $A_*<A_0<A_\bullet$: unphysical solution with curvature singularity.\newline  }
\label{FigQ2Sing}
\end{figure}

The value of $A_1$ in (\ref{ElctrSeries}),
\bea\label{ElctrSeriesA1}
A_1=\frac{d_- g^2}{n_+}-\frac{m_-}{n_+A_0}+\frac{nq_2^2}{2n_+(d-2)A^m_0},
\eea
is positive. 
Derivative ${\dot A}$ is positive at $r=0$, then it becomes negative and turns positive again as $r$ increases. The warp factor $A$ diverges when $C$ collapses. Near the singularity one finds 
\bea\label{ElecCaseAa}
\hskip -1cm
&&A\simeq \frac{a}{(r-r_1)^{\mu}},\quad
C\simeq c(r-r_1)^{\nu},\\ 
&&\nu=\frac{m_+\mu+2}{n},\quad 
\mu=\frac{\sqrt{mn(d-2)}+n_--m}{(m_+)^2+nm_-}.\nonumber
\eea
Here $(a,c,r_1)$ are free parameters in the expansion near the singularity, and their values are determined in terms of $A_0$ by integrating the system from $r=0$ to $r=r_1$. An example of such singular solution is presented in figure \ref{FigQ2Sing}(a).

\item $A_0=A_*$: $AdS_m\times S^{n+1}$ solution.

The resulting geometry is (\ref{AdS2xS3}), 
\bea\label{AdS2xS3ppA}
A=A_*,\quad C=\frac{1}{A_* u^2}\sin^2[u r],\quad 
u=\left[\frac{m_-q_2^2}{2(d-2)nA^{m+1}_*}-\frac{d_-g^2}{nA_*}\right]^{\frac{1}{2}},
\eea
and the radii of the AdS space and the sphere are given by (\ref{RadAdS2S3}). Function $A$ is constant in the entire $0\le r<\frac{\pi}{u}$ region, in particular this implies that the coefficient $A_1$ in (\ref{ElctrSeries}) vanishes. The graphs for various functions are presented in figure \ref{FigQ2Sing}(b). 

\item $A_*<A_0<A_\bullet$: singular solution with irregular collapse of the sphere.

This situation is similar to option (a), but now the coefficient $A_1$ in (\ref{ElctrSeries}), (\ref{ElctrSeriesA1}) is negative, so the derivative of ${\dot A}$ changes from negative to positive as $r$ increases from zero to $r_1$. The warp factor $A$ still diverges at some $r=r_1$, and the leading behavior near the singular point is given by (\ref{ElecCaseAa}). Various functions for this case are presented in figure \ref{FigQ2Sing}(c).

\item $A_0=A_\bullet$: $AdS_m\times H_{n+1}$ solution (\ref{AdS2xH3}).

As in the case (b), the AdS warp factor $A$ is constant, but now the parameter $u$ in (\ref{AdS2xS3}) becomes imaginary, so one gets the maximally symmetric hyperbolic space $H_{n+1}$ instead of a sphere. The full solution is given by (\ref{AdS2xH3}):
 \bea\label{AdS2xH3ppA}
A=A_\bullet,\quad C=\frac{1}{A_\bullet v^2}\sinh^2[v r],\quad 
v^2=\frac{d_-g^2}{nA_\bullet}-\frac{m_-q_2^2}{2(d-2)nA^{m+1}_\bullet}.
\eea
\label{PageOptE}
\item $A_\bullet<A_0<A_\#$: asymptotically locally $AdS_d$ space.

This situation is analogous to option (d) from section \ref{SecSubMagnFlow}: both $A$ and $C$ grow quadratically at infinity, but function $C$ grows faster. For example, the large $r$--behavior of the warp factors in the (2,2) case is given by an electric counterpart of (\ref{ACasymp22}) with $b>1$,
\bea\label{ACasympElec}
A&=&(gr)^2+\frac{1+2b}{3bg^2}+\frac{1}{18(br)^2 g^6}\left[b^2-1+3g^2(b^2q_1^2+q_2^2)\right]\log\frac{a}{r}+o(r^{-3})\\
C&=&b(gr)^2+\frac{b-1}{6g^2}-\frac{b^2-1+3g^2(b^2q_1^2+q_2^2)}{36br^2 g^6}\log\frac{a}{r}+
\frac{2-2b^2-3(bgq_1)^2}{72br^2 g^6}+o(r^{-3})\nonumber
\eea
and expressions in other dimensions are similar. The leading contribution to the Riemann tensor doesn't depend on the charges, and in the (2,2) case it is still given by (\ref{RiemAsnpAdS}). A typical behavior of warp factors $(A,C)$ is shown in Figure \ref{FigQ2growAC}.

\begin{figure}
 \includegraphics[width=0.9 \textwidth]{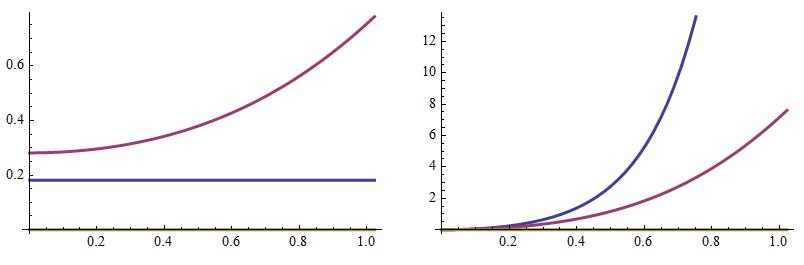}
 \begin{center}
(a)\qquad \qquad \qquad \qquad \qquad \qquad \qquad(b)
\end{center}
\caption{Warp factors $A(r)$ (a) and $C(r)$ (b) for $(m,n)=(2,2)$, $q_2<q_{cr}$ solutions with 
$A_0=A_\bullet$ (blue) and $A_\bullet<A_0<A_\#$ (red).}
\label{FigQ2growAC}
\end{figure}

\item $A_0=A_\#$: regular space with the standard $AdS_d$ asymptotics.

For this critical value of $A_0$, the asymptotic expansion is given by (\ref{ACasympElec}) with $b=1$. The solution can be interpreted as the unique geometry produced by an electric charge $q_2$ in the asymptotically $AdS_d$ space. For $q_2=0$ we get the standard $AdS_d$ (\ref{AdSexact}), and 
$A_\#=\frac{1}{g^2}$. For other values of $q_2$, parameter $A_\#$ can be determined numerically or from asymptotic expansions similar to the ones presented in section \ref{SecApprox}.   The resulting graph of $A_\#$ as a function of $q_2$ for the $m=n=2$ example is presented in Figure 
\ref{FigQ2growAC??}(b).

\item $A_0>A_\#$: asymptotically locally $AdS_d$ space.

This situation is similar to (e), but with $b<1$. 

\end{enumerate}
\item $q_2=q_{cr}$

For the critical value of the charge, 
\bea\label{qCritv2}
q_{cr}=\sqrt{2m_-}\left[\frac{(m_-)^2}{(d-2)d_-g^2}\right]^{\frac{m-1}{2}},
\eea
the $AdS_m\times S^{n+1}$ solution from item (b) above is replaced by 
$AdS_m\times R^{n+1}$, and the regions (a)-(c) are drastically different from their counterparts  from option A. To describe these regions, we begin with recalling that solutions of the form 
$AdS_m\times X$ have constant warp factors $A(r)=A_c$, where $A_c$ satisfies equation (\ref{AdS2xS3star}):
\bea\label{AdS2xS3starC2}
d_- g^2-\frac{m_-}{A_c}+\frac{nq_2^2}{2(d-2)A_c^m}=0.
\eea
As discussed earlier, for $q_2<q_{max}$, and in particular, for $q_2=q_{cr}$, this equation has two real positive solutions for $A_c$: $A_*$ and $A_\bullet$ with 
\bea\label{AdS2xR4Astar}
A_*=\frac{(m_-)^2}{(d-2)d_-g^2},\quad
A_\bullet >A_*\,.
\eea
\begin{figure}
\begin{tabular}{ccc}
 \includegraphics[width=0.45 \textwidth]{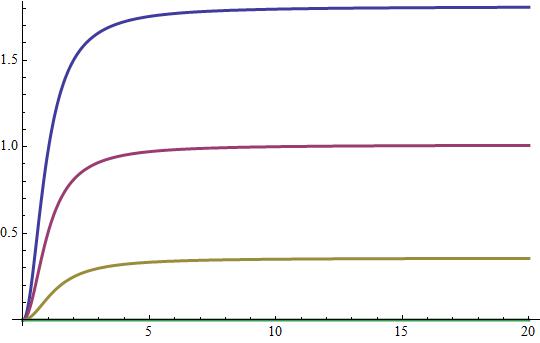}&\ \ &
  \includegraphics[width=0.45 \textwidth]{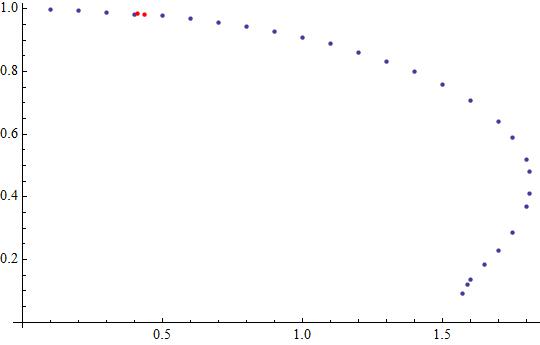}\\
(a)&&(b)
\end{tabular}
\caption{Electric solutions with $AdS_5$ and local $AdS_5$ asymptotics for $(m,n)=(2,2)$:\newline
(a) Ratios of $C/A$ as functions of $r$ for three values of $A_0$. The critical value $A_\#$ is determined by the condition 
$\lim_{r\rightarrow\infty}(C/A)=1$ (red curve).\newline
(b) Plot of $A_\#$ as a function of $q_2$. The critical values $q_2=(q_{cr},q_{max})$ are labeled by red points.
}
\label{FigQ2growAC??}
\end{figure}
As in option A, there are seven regions:
\begin{enumerate}[(a)]
\item $A_0<A_*$: solution with $AdS_m\times R^{n+1}$ asymptotics and oscillating $A$.

The value of $A_1$ in (\ref{ElctrSeries}),
\bea\label{ElctrSeriesA1b}
A_1=\frac{d_- g^2}{n_+}-\frac{m_-}{n_+A_0}+\frac{nq_2^2}{2n_+(d-2)A^m_0},
\eea
is positive, so the warp factors $A$ grows for small values of $r$ and saturates to $A_*$ at infinity. To understand the asymptotic behavior of the solution at large $r$, it is useful to introduce a perturbation of the $AdS_m\times R^{n+1}$ solution (\ref{AdS2xR4}),
\bea\label{CritPertAC}
A(r)=A_*+\eps a(r),\quad C(r)=\frac{r^2}{A_*}+\eps c(r),
\eea
and assume that the corrections become small at infinity. Then the leading terms in (\ref{FullEinsteinE}) 
produce a linear system for $(a(r),c(r))$. In particular, one of the equations is
\bea
r\ddot a+n\dot a+\frac{2g^2 d_-n_-}{A_*}ra=0.
\eea
This equation can be solved in terms of the Bessel functions, and at large $r$ one finds
\bea\label{OscilBessel}
a(r)=a_+\frac{e^{i\mu r}}{r^{n/2}}\left[1+\frac{n(n-2)}{8\mu r}+\dots\right]+
a_-\frac{e^{-i\mu r}}{r^{n/2}}\left[1-\frac{n(n-2)}{8\mu r}+\dots\right].
\eea
Here $(a_+,a_-)$ are integration constants and
\bea
\mu=\frac{d_-}{m_-}\sqrt{2n_-(d-2)}.
\eea 
The remaining equations can be solved for $c(r)$ and the result reads
\bea\label{OscilBesselcc}
c(r)=-\frac{mr^2}{A_*^2n_-}\left[a_+\frac{e^{i\mu r}}{r^{n/2}}+
a_-\frac{e^{-i\mu r}}{r^{n/2}}+\dots\right]+c_0 r.
\eea
The solution of the homogeneous equations for $c(r)$, which is parameterized by $c_0$, can be absorbed into a shift of the radial coordinate, so the solution (\ref{CritPertAC}) can be conveniently written as
\bea\label{CritPertACv2}
A(r)=A_*+\eps a(r-r_0),\quad C(r)=\frac{(r-r_0)^2}{A_*}+\eps c(r-r_0)|_{c_0=0}.
\eea
The approach to the asymptotic expressions $(A,C)=(A_*,\frac{(r-r_0)^2}{A_*})$ is parameterized by three constants $(a_+,a_-,r_-)$. Their values can be determined by numerical integration of the system (\ref{FullEinstein}) starting from zero, and an example of the full solution for $m=n=2$ is shown in figure \ref{FigQ2oscil}. As expected, functions $A(r)$ first grows from $A(0)$ to $A_*$, then it starts to oscillate around that value according to (\ref{OscilBessel}).

\item $A_0=A_*$: the exact $AdS_m\times R^{n+1}$ solution.

The metric is given by (\ref{AnstzMetr1}), (\ref{AdS2xR4}):
\bea\label{AdS2xR4ppA}
ds^2=A_*\, d\Sigma_m^2+\frac{dr^2}{A_*}+\frac{r^2}{A_*} d\Omega^2_{n},\quad
A_*=\frac{(m_-)^2}{(d-2)d_-g^2}\,.
\eea

\item  $A_*< A_0<A_\bullet$: solution with $AdS_m\times R^{n+1}$ asymptotics and oscillating $A$.

The situation is similar to branch (a), but now function $A$ decreases from $A(0)$ to $A_*$ before entering oscillations (\ref{OscilBessel}) around that value. An example of such solution for $m=n=2$ is shown in figure \ref{FigQ2oscil}.
\end{enumerate}
\begin{figure}
 \includegraphics[width=0.9 \textwidth]{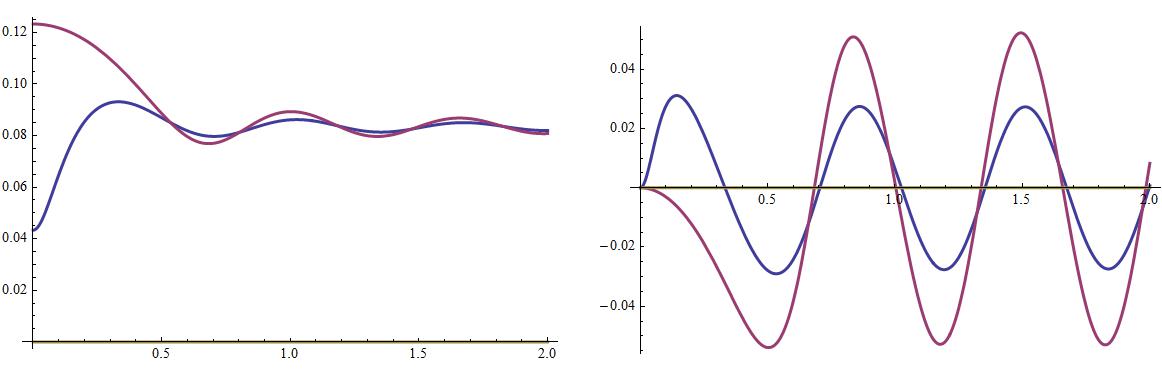}
 \begin{center}
(a)\qquad \qquad \qquad \qquad \qquad \qquad \qquad(b)
\end{center}
\caption{Warp factor $A(r)$ for $(m,n)=(2,2)$, $q_2=q_{cr}$:\newline 
(a) $A(r)$ for $A_0<A_*$ (blue) and $A_*< A_0<A_\bullet$ (red)\newline
(b) Functions $r{\dot A}(r)$ with the same color conventions: confirmation of 
oscillations (\ref{OscilBessel}).}
\label{FigQ2oscil}
\end{figure}
The regions (d)--(g) are identical to their counterparts from option A. 
\item $q_{cr}<q_2< q_{max}$

The situation is similar to option B, but now both solutions $A_c=(A_*,A_\bullet)$ of equation (\ref{AdS2xS3starC2}) describe $AdS_m\times H_{n+1}$. This leads to some minor modification of branches (a)--(c)

\begin{enumerate}[(a)]
\item $A_0<A_*$: solution with $AdS_m\times H_{n+1}$ asymptotics and oscillating $A$.

As in B.(a), the warp factor $A(r)$ first grows from $A_0$ to $A_*$ and then it starts to oscillate. 
Introducing a counterpart of (\ref{CritPertAC}),
\bea\label{CritPertAC1}
A(r)=A_*+\eps a(r),\quad C(r)=C_0 e^{2v r}+\eps c(r),
\eea
and treating $\eps$ as a small parameter, we find a linear equation for $a(r)$ which is valid for large values of $r$:
\bea\label{tempSep4}
\ddot a+nv{\dot a}+\frac{4}{A_*^2}[m^2+1-m(2+d_-g^2A_*)]a=0.
\eea
The solution of this equation is 
\bea\label{OscilBB1}
a(r)=a_0 e^{-nvr/2}\cos(\mu r+\phi_0),\quad
\mu=\left[\frac{9(m_-)^2}{4A_*^2}-\frac{d_-(9m+n-1) g^2}{4A_*}\right]^{\frac{1}{2}}.
\eea
Here we used the expression (\ref{AdS2xH3vv}) for $v$ in terms of $A_*$. Expansion of the remaining equations to the first order in $\eps$ leads to 
\bea\label{OscilBB2}
c(r)=a_0 c_0 e^{2vr-nvr/2}\cos(\mu r+\phi_0+\psi_0).
\eea
The constants $(c_0,\psi_0)$ are fully determined by the equations, but the explicit expressions for them are not illuminating. As in (\ref{CritPertAC}), we conclude that the solution at infinity is parameterized by three constants $(a_0,\phi_0,r_0)$:
\bea\label{CritPertAC1v2}
A(r)=A_*+\eps a(r-r_0),\quad C(r)=e^{2(r-r_0)v}+\eps c(r-r_0).
\eea
The values of these constants can be determined by numerical integration from zero.

For sufficiently large $q_2$, parameter $\mu$ in (\ref{OscilBB1}) becomes imaginary, so one gets exponential rather than oscillating approach to the asymptotic values of $A(r)$ and $C(r)$. The transition happens when $\mu=0$, and this gives
\bea\label{q2Trans}
A_*^{(trns)}=\frac{9(m_-)^2}{d_-(9m+n-1) g^2},\quad
q_2^{(trns)}=\left[\frac{2(d-2)m_-(n+8)}{n(9m+n-1)}
(A_*^{(trns)})^{m-1}\right]^{\frac{1}{2}}
\eea
The charge at the transition point, $q_2^{(trns)}$ satisfies inequalities
\bea
q_{cr} < q_2^{(trns)} < q_{max}\,.
\eea
The behavior of functions $(A(r),C(r))$ in the $q_2^{(trns)} < q_2 < q_{max}$ range is qualitatively similar to the $q_2 = q_{max}$ case, which is discussed below.

\item $A_0=A_*$: the $AdS_m\times H_{n+1}$ solution.

The metric is given by (\ref{AdS2xH3}):
 \bea
A=A_{*},\quad C=\frac{1}{A_{*} v^2}\sinh^2[v r],\quad 
v^2=\frac{d_-g^2}{nA_{*}}-\frac{m_-q_2^2}{2(d-2)nA^{m+1}_{*}}.
\eea
\item $A_*< A_0<A_\bullet$: solution with $AdS_m\times H_{n+1}$ asymptotics and 
oscillating $A$.

The situation is similar to branch (a), but in this case function $A(r)$ first {\it decreases} from $A_0$ to $A_*$ before entering the oscillatory regime (\ref{CritPertAC1}), (\ref{OscilBB1})--(\ref{OscilBB2}). 
\end{enumerate}
The regions (d)--(g) are identical to their counterparts from options A and B. 

\item $q_2= q_{max}$

When $q_2$ reaches the maximal value
 $q_{max}$ (\ref{AdS2xS3qMax}),
\bea\label{q2maxSep5}
q_{max}=\left[\frac{2(d-2)m_-}{nm}\right]^{\frac{1}{2}}\left[\frac{m_-^2}{d_-m g^2}\right]^{\frac{m-1}{2}},
\eea
the two transitions points, $A_*$ and $A_\bullet$, coincide, so the counterparts (b) and (c) from option C disappear as separate regions. The regions (d)--(g) are still identical to the previous cases, so we will discuss only the region (a).

\begin{enumerate}[(a)]
\item $A_0<A_*=A_\bullet$: solution with $AdS_m\times H_{n+1}$ asymptotics.

Function $A(r)$ grows from $A(0)=A_0$ to $A_*$ at infinity, but in contrast to options B and C, this growth happens monotonically. To see this, we introduce perturbations as in (\ref{CritPertAC1}). Equation (\ref{tempSep4}) still holds, but now its coefficients simplify due to relations\footnote{Recall (\ref{AdS2xS3qMax}) and (\ref{AdS2xS3}), as well 
as relation $v=iu$.}
\bea
A_*=\frac{m^2_-}{md_-g^2},\quad v=\frac{d_-g^2\sqrt{mn_-}}{nm_-}.
\eea
This leads to
\bea\label{tempSep4a}
\ddot a+nv{\dot a}=0\quad \Rightarrow\quad a=a_0 e^{-nvr}.
\eea
The second solution, $a=\mbox{const}$, does not become subleading at infinity, so it should be discarded. Solving equations for $c(r)$, we arrive at the final result for the perturbation
\bea\label{CritPertAC1D1}
A(r)=A_*+\eps e^{-nvr},\quad C(r)=C_0e^{2vr}\left[1+\eps c_1 e^{-nvr}\right].
\eea
The value of $c_1$ is determined from the equations, but the explicit cumbersome expression is not very illuminating. The large $r$ behavior (\ref{CritPertAC1D1}) has been confirmed by the full numerical solution in the $m=n=2$ case.

\end{enumerate}

\item $q_2>q_{max}$

Once the electric charge exceeds the maximal value (\ref{q2maxSep5}), any initial condition $A(0)=A_0$ generates a flow to a locally $AdS_d$ space. Specifically, any $A_0<A_\#$ results in option (e) on page \pageref{PageOptE}, while $A_0=A_\#$ and $A_0>A_\#$ lead to branches (f) and (g). 

Interestingly, the solution with $AdS_d$ asymptotics disappears at a sufficiently large value of charge, and the ratio $(C/A)$ at infinity (we denote it by $(C/A)_\infty$) always remains less than one. This phenomenon is illustrated in Figure \ref{FigQ2growAC??}(b) for the $(m,n)=(2,2)$ example, where it amounts to $b<1$ bound in the expansions (\ref{ACasympElec}). For sufficiently small charges, and in particular for $q_2<q_{max}$, the ratio $(C/A)_\infty$ is a monotonic function of $A(0)$, which goes from large positive values to zero as $A(0)$ increases. For large values of $q_2$, the 
$A(0)$--dependence of $(C/A)_\infty$ stops being monotonic: this function first grows, and then decreases again. In particular, the value $(C/A)_\infty=1$ can be crossed more than once, resulting in non--single-valuedness shown in figure \ref{FigQ2growAC??}(b). As $q_2$ increases, the values of $A(0)$ corresponding to two crossings are getting closer, and eventually the crossings disappear leaving only the $b<1$ option in (\ref{ACasympElec}). The same qualitative behavior happens in all dimensions. Therefore, while the asymptotically locally $AdS_d$ asymptotics are always possible, the standard AdS asymptotics cannot be imposed for sufficiently large electric charges. 
\end{enumerate}
This concludes our analysis of regular electric solutions covered by the ansatz (\ref{TheAnsatz1}). In contrast to the magnetic case discussed in the previous section, geometries with electric charge have only one asymptotic region, and they end at $r=0$, where sphere $S^n$ collapses. A brief summary of our findings for the electric case is presented in section \ref{SecSubSumElctr}. 

\section{Dyonic solutions}
\label{SecDyon}

In this section we will consider geometries that carry both the magnetic and electric charges. As discussed in section \ref{SecMagn}, the regular solutions of thus type must have a non--contractible sphere $S^n$, so qualitatively the geometries will be similar to the magnetic wormholes discussed in section \ref{SecMagn} rather than to the electric solutions considered in section \ref{SecElectr}. However, introduction of the electric charge leads to a rich phase structure of the geometries, which is much more complicated than the one discussed in section \ref{SecMagn}. In particular, as we will see, certain solutions appear and disappear as one changes relationship between three parameters $(g,q_1,q_2)$. 

We will begin in section \ref{SecDyonGen} by establishing a general analytical picture of various phases and relationships between them. In section \ref{SecDyonGen} we will perform a detailed study of these phases, using the $(m,n)=(2,2)$ case as an example, but the general features of this picture hold in all dimensions, although with different numerical coefficients. A brief summary of results obtained in this section can be found in section \ref{SecSubSumDyon}.

\subsection{The boundary problem and constraints on parameters}
\label{SecDyonGen}
\renewcommand{\theequation}{5.\arabic{equation}}
\setcounter{equation}{0}

In this section we will analyze the system (\ref{FullEinstein}) assuming that both $q_1$ and $q_2$ are non--zero. Looking at the list of the exact solutions presented in section \ref{SecExact}, one concludes that the only exact geometry relevant for the present situation is $AdS_m\times S^n\times R$ given by (\ref{SpecSolnNoRun}). It occurs only if charges are constrained by the relation 
\bea\label{AdSSRq}
\frac{m_-}{a^m}q_2^{-2/m_-}-\frac{n_-}{c^n}q_1^{-2/n_-}=2(d-2)d_- g^2,
\eea
where numerical parameters $(a,c)$ are given by (\ref{SpecSolnNRac}). For a fixed value of the cosmological constant, which is related to $g$ by (\ref{LamDefG}), expression (\ref{AdSSRq}) describes a curve in the 
$(q_1,q_2)$ plane. For example, in the $AdS_2\times S^2\times R$ case, this curve is given in (\ref{SpecSolnNoRun22}):
\bea\label{SpecCurve22}
 \frac{1}{q_2^2}-\frac{1}{q_1^2}=6g^2.
\eea
Even if the condition (\ref{AdSSRq}) is satisfied, the $AdS_m\times S^n\times R$ space is recovered only if $(A,C)$ are given by (\ref{SpecSolnNoRun}), and one can get nontrivial flows by picking different solutions of the system (\ref{FullEinstein}).

As discussed in section \ref{SecMagnBound}, to construct regular solutions with non-zero magnetic charge, one must impose the following condition\footnote{More generally, one needs ${\dot A}(r_0)={\dot C}(r_0)=0$, but this reduces to (\ref{ACbcDyon}) by performing a shift $r$, the symmetry of the system  (\ref{FullEinstein}).}
\bea\label{ACbcDyon}
{\dot A}(0)=0,\quad {\dot C}(0)=0.
\eea
In particular, this would lead to even functions $A(r)$ and $C(r)$, so any solution regular at $r\ge 0$ would directly extend to a regular wormhole with two identical asymptotics at $r=\pm\infty$. Therefore, we will focus on the solutions of the system (\ref{FullEinstein}) only at $r>0$ and treat (\ref{ACbcDyon}) as  boundary conditions.

By eliminating second derivatives from equations (\ref{FullEinsteinM}), one arrives at the dyonic version of the first order equation  (\ref{FirstOrdGen}):
\bea\label{FirstOrdGenDyon}
\hskip -0.4cm
\frac{nn_- A{\dot C}^2}{4C^2}+\frac{mn {\dot A}{\dot C}}{2C}+\frac{mm_-{\dot A}^2}{4A}+\frac{mm_-}{A}-\frac{nn_-}{C}+\frac{q_1^2}{2C^n}-\frac{q_2^2}{2A^m}-(d-2)d_-g^2=0.
\eea
In particular, combining this relation and the boundary conditions (\ref{ACbcDyon}), one gets the constraint on the boundary values of $(A,C)$:
\bea\label{BounCnstrDyon}
\frac{mm_-}{A_0}-\frac{nn_-}{C_0}+\frac{q_1^2}{2C_0^n}-\frac{q_2^2}{2A_0^m}-(d-2)d_-g^2=0.
\eea
In the purely magnetic case ($q_2=0$), similar logic led us to the conclusion that the boundary conditions for the system (\ref{FullEinsteinM}) were fully specified by one number $C_0$: the value of $A_0\equiv A(r)$ was fully determined in terms of $C_0$, and the derivatives at zero vanished due to (\ref{ACbcDyon}). Now there is a caveat: for nontrivial electric charge, the constraint (\ref{BounCnstrDyon}) might have more than one solution for $A_0$. Let us begin with an explicit discussion of the $m=2$ case before making the general conclusion based on more formal arguments. 

\bigskip

For $m=2$, equation (\ref{BounCnstrDyon}) has two complex solutions for $A_0$:
\bea\label{Apm22}
A_\pm=\frac{1}{D}\left[2\pm\sqrt{4-q_2^2 D}\right],\ \mbox{where}\  
D=2(d-2)d_- g^2+\frac{2nn_-}{C_0}-\frac{q_1^2}{C_0^n}\,.
\eea
We are interested in real and positive $A_0$, and this condition selects two windows:
\bea
&&0<D< \frac{4}{q_2^2}\ \Rightarrow\ A_\pm\,,\nn
&&D\le 0\  \Rightarrow\ A_-\,.
\eea
Recall that for $q_2=0$ we had only $A_+$ and the denominator $D$ had to be positive.

The $A_+$ and $A_-$ branches can collide and disappear when $D=\frac{4}{q_2^2}$. One of the places where this happens is the $AdS\times S\times R$ line in the $(q_1,q_2)$ plane. Indeed, equations (\ref{SpecSolnNoRun}) for $m=2$ give
\bea
D
=2(d-2)d_- g^2+\frac{n}{c^nq_1^{2/n_-}}-\frac{1}{c^n q_1^{2/n_-}}
=\frac{4}{q_2^2}.\nonumber
\eea
As we will see below, both this point and the full branch cut $D=\frac{4}{q_2^2}$ will play important roles in the structure of various solutions. In particular, the wormholes with $AdS_{m+1}\times S^n$ and $AdS_d$ asymptotics will terminate on the line (\ref{AdSSRq}), and the $A_+$ and $A_-$ solutions will be exchanged elsewhere on the branch cut. 

\bigskip

Using this discussion of the $m=2$ case as an inspiration, we observe that for general $m$ the 
constraint (\ref{BounCnstrDyon}) can be written as
\bea\label{Dec11}
\frac{2mm_-}{A_0}-\frac{q_2^2}{A_0^m}=D,
\eea
with $D$ still given by (\ref{Apm22}). For a fixed charge, the left hand side reaches the maximal value when
\bea\label{DefDmax}
A_m=\left[\frac{q_2^2}{2m_-}\right]^{\frac{1}{m-1}}\ \Rightarrow\ D_{max}=\frac{2(m_-)^2}{A_m}.
\eea
Therefore, there are no roots if $D>D_{max}$, and one root for $D=D_{max}$. To analyze the solutions at $D<D_{max}$, it is convenient to rewrite the left hand side of (\ref{Dec11}) in terms of $z=A_0^{-1}$ as 
$f(z)=\alpha z-\beta z^m$. Since $f''(z)$ is negative for all $z>0$, equation (\ref{Dec11}) may have at most two solutions with positive $A_0$. Furthermore, in the region where $f(z)$ is negative, $f'(z)$ is positive, so negative values of $D$ lead to one and only one root. These arguments extend the conclusion we reached for $m=2$ to all values of $m$: there are two roots if $0<D<D_{max}$, one root for $D=D_{max}$ and $D\le 0$, and no solutions for $D>D_{max}$. Therefore, the $m=2$ case is qualitatively similar to other dimensions. This will be important in the next subsection, where we will draw general qualitative conclusions  from an extensive numerical analysis of the $(m,n)=(2,2)$ configurations.

\bigskip

\begin{figure}
\begin{center}
 \includegraphics[width=0.6 \textwidth]{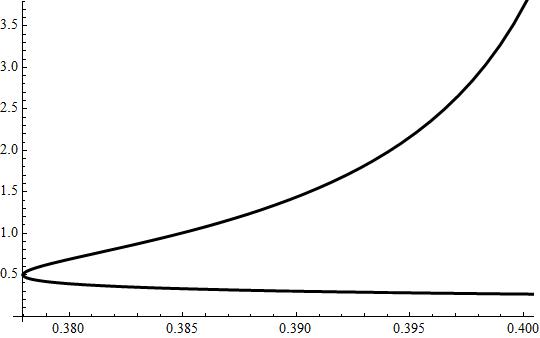}
 \end{center}
\caption{Branch cut in the $(q_2,C_0)$ plane for $m=n=2$. Physical region is to the left of this cut.}
\label{FigDyonBrCut}
\end{figure}

Let us now analyze the branch cut at which two solutions of equation (\ref{Dec11}) collide and disappear. This cut happens along a surface in the $(C_0,q_1,q_2)$ space:
\bea\label{BrnchCutSrf}
2(d-2)d_- g^2+\frac{2nn_-}{C_0}-\frac{q_1^2}{C_0^n}=D_{max},
\eea
with $D_{max}$ given by (\ref{DefDmax}). Straightforward substitution of (\ref{SpecSolnNoRun}) into the last relation demonstrate that the $AdS_m\times S^n\times R$ solution is located on the branch cut. This extends the $m=2$ result obtained earlier to all dimensions. Furthermore, the $AdS_m\times S^n\times R$ solution appears to be a rather special point on the cut (\ref{BrnchCutSrf}). It is instructive to keep $q_1$ fixed in (\ref{BrnchCutSrf}) and study $C_0$ as a function of $D_{max}$, which is in turn a function of $q_2$. The resulting curve for the $(m,n)=(2,2)$ case is presented in Figure \ref{FigDyonBrCut}, and it has two important properties: $C_0$ is always greater than some critical value, and that value corresponds to the $AdS_2\times S^2\times R$ solution. As we will now demonstrate, these features extend to all dimensions. 

For fixed values of $q_1$ and $g$, equation (\ref{BrnchCutSrf}) can be written in a schematic form
\bea\label{tmpAlBetX}
\alpha+\beta x-\gamma x^n=q_2^{-\frac{2}{m-1}}\,,
\eea
where $x=C_0^{-1}$. For positive values of $x$, the left hand side of the last expression reaches a maximum at 
\bea
x_*=\left[\frac{\beta}{n\alpha}\right]^{\frac{1}{n-1}}\quad\Rightarrow\quad
\alpha+\beta x_*-\gamma x_*^n=\alpha+\frac{n-1}{n}\beta x_*\,.
\eea
This implies that the branch cut exists only for $q_2\ge q_{2,min}$, where $q_{2,min}$ corresponds to $x=x_*$. Furthermore, for $q_2$ slightly above $q_{2,min}$, there are two solutions for $x$, and at $q_2=q_{2,min}$ they merge into one. As $q_2$ increases, the separation between two roots of equation (\ref{tmpAlBetX}) increases, and for $q_2=\alpha^{-\frac{m-1}{2}}$ one of these roots crosses zero. Then  the corresponding $C_0$ goes to infinity, and as $q_2$ increases further, there is only only physical solution for $C_0$.  Substitution of the explicit expression for the parameters 
$(\alpha,\beta,\gamma)$ from (\ref{BrnchCutSrf}) leads to the conclusion that the special point 
$(q_1,q_{2,min},C_*)$ is precisely the $AdS_m\times S ^n\times R$ solution (\ref{SpecSolnNoRun}). This 
analysis implies that the branch cuts in all dimensions have the same qualitative behavior as the example presented in figure \ref{FigDyonBrCut}.

We conclude this subsection by summarizing the boundary conditions leading to regular geometries. 

\begin{itemize}
\item The warp factors $A(r)$ and $C(r)$ satisfy the system of coupled second order 
differential equations (\ref{FullEinstein}).
\item Regular wormhole geometries must have ${\dot A}(0)={\dot C}(0)=0$. In particular, this ensures that 
$(A,C)$ are even functions of $r$, so the solutions have the same asymptotics at $r=\pm\infty$
\item The boundary values, $A(0)$ and $C(0)$, satisfy the algebraic constraint (\ref{BounCnstrDyon}). For a given $C_0$, this equation has at most two real positive solutions for $A_0$. If there are two such solutions, we will label them as $A_+$ and $A_-$. In the $m=2$ case, the explicit expressions are given by (\ref{Apm22}). 
\item Parameters $A_+$ and $A_-$ become equal on the surface (\ref{BrnchCutSrf}) in the $(C_0,q_1,q_2)$ space with $D_{max}$ given by (\ref{DefDmax}). On this surface equation (\ref{BounCnstrDyon}) has only one solution, and beyond this surface (i.e., when the left hand side of (\ref{BrnchCutSrf}) is larger than $D_{max}$), it is impossible to satisfy the constraint (\ref{BounCnstrDyon}) with real $(A_0,C_0)$.
\item For fixed values of $(q_1,g)$, the surface (\ref{BrnchCutSrf}) becomes a line in the $(C_0,q_2)$ plane. It passes only through the points where $q_2\ge q_{2,min}$, where $q_{2,min}$ is the solution of the equation (\ref{AdSSRq}) describing the $AdS_m\times S^n\times R$ space. Figure \ref{FigDyonBrCut} shows the behavior of the branch cut $(m,n)=(2,2)$ case, but qualitative features of this picture hold in all dimensions. 
\end{itemize}
In the next subsection we will analyze all physically interesting solutions with these boundary conditions. 

\subsection{Phase diagram for wormhole geometries}

To give a pictorial representation of various phases associated with the system (\ref{FullEinstein}), it is convenient to fix the values of $(g,q_1)$ and study the behavior of the initial conditions in the 
$(q_2,C_0)$ plane. 

First we observe that the $q_2=0$ line reproduces the system discussed in section \ref{SecMagn}: it has three special points $(C_*,C_\#,C_\bullet)$ corresponding to the flows to $AdS_{m+1}\times S^n$, $AdS_d$, and $AdS_{m}\times S^{n+1}$. The solutions are singular if $C_0<C_*$ or $C_0>C_\bullet$. When $q_2$ is slightly above zero, the same three flows persist, but the corresponding initial values $(C_1,C_2,C_3)$ acquire $q_2$ dependence. Furthermore, the $AdS_{m+1}\times S^n$ is no longer an exact solution, but rather one arrives at this asymptotic by starting with $C_0=C_1>C_*$. Therefore, all flows appear on the same footing. 

As $q_2$ increases, there are two logical possibilities: either the three flows persist for all values of $q_2$ or they cease to exist beyond some critical value. We will now demonstrate that the flows exist only in the following ranges:
\bea\label{q2Ranges}
AdS_{m+1}\times S^n:&& q_2\le q_{S},\nn
AdS_{d}:&& q_2\le q_{AdS},\quad q_H<q_{AdS}<q_S,\\
AdS_{m}\times H_{n+1}:&& q_2\le q_{H}.\nonumber
\eea
For example, the line describing the boundary condition for the $AdS_{m+1}\times S^n$ flow starts at the $(0,C_*)$ point in the $(q_2,C_0)$ plane, then it continues as a curve $(q_2,C_1(q_2))$ until $q_2$ reaches $q_S$, and eventually it disappears by hitting the branch cut (\ref{BrnchCutSrf})\footnote{Naively, one may expect that $q_S$ is located on the branch cut itself, but as we will see below, the situation is more interesting, and some values of $q_2<q_S$ give rise to more than one $AdS_{m+1}\times S^n$ solution with different values $C_1(q_2)$, and the branch cut is reached at $q_2=q_S^{bc}<q_S$.}. To proceed, we recall that for nonzero electric charge, the constraint (\ref{BounCnstrDyon}) has at most two real positive solutions for $A_0$, which we labeled as $(A_+,A_-)$ in the previous subsection\footnote{The explicit expression for $A_\pm$ in the $m=2$ case are given by (\ref{Apm22}).}, and we begin with analyzing the initial conditions corresponding to $A_+$. 

\begin{figure}
 \begin{center}
 \includegraphics[width=0.7 \textwidth]{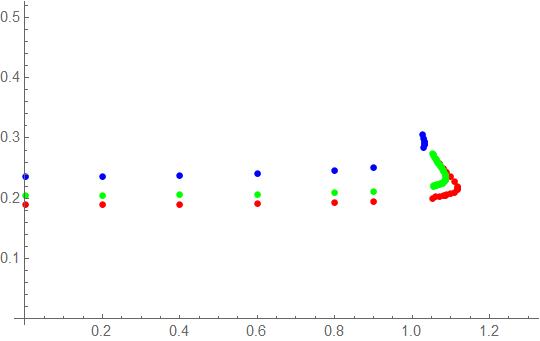}
\end{center}
\caption{Curves in the $(q_2,C_0)$ plane describing flows to various asymptotic regions for $m=n=2$, $q_1=g=1$ in the $A_+$ branch. Notation and color conventions are explained in the the text.}
\label{FigDyonApBig}
\end{figure}

\begin{figure}
 \includegraphics[width=1 \textwidth]{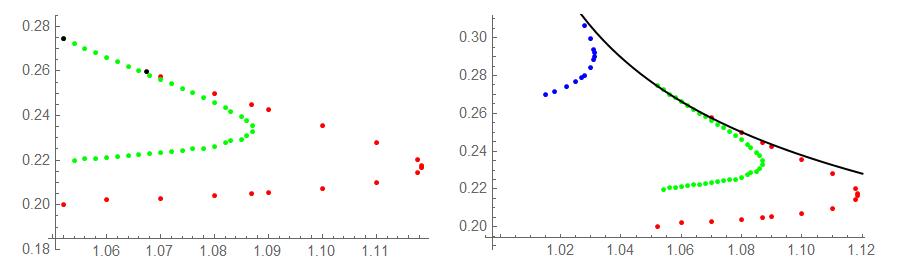}
\caption{Curves in the $(q_2,C_0)$ plane describing flows to various asymptotic regions for $m=n=2$, $q_1=g=1$ in the $A_+$ branch. Notation and color conventions are explained in the the text. The locations of the branch cut are denoted by black dots of a black line.}
\label{FigDyonApDetail}
\end{figure}

As shown in section \ref{SecMagn}, in the absence of the electric charge $q_2$, the flows with initial conditions $C_0<C_*$ lead to naked curvature singularities. These arguments extend to non--zero $q_2$ as well, so in this case the flow to $AdS_{m+1}\times S^n$ must start with $C_0=C_1(q_2)>C_*$, and all $C_0<C_1(q_2)$ lead to curvature singularities. Recall that $C_*$ is the solution of equation (\ref{AdS32exct2}), and it doesn't depend on $q_2$. On the other hand, the equation (\ref{BrnchCutSrf}) for the branch cut implies that any initial condition $C_0$ must satisfy the inequality
\bea\label{ineqDec17}
2(d-2)d_- g^2+\frac{2nn_-}{C_0}-\frac{q_1^2}{C_0^n}\le D_{max}=
2(m_-)^2\left[\frac{2m_-}{q_2^2}\right]^{\frac{1}{m-1}}\,.
\eea
Therefore, the $AdS_{m+1}\times S^n$ flow may exist only if $C_*$ satisfies the last inequality\footnote{This is a necessary, but not a sufficient condition.}. To demonstrate that such a flow cannot exist for arbitrarily large $q_2$, it is sufficient to show that the $q_2=\infty$ version of the inequality (\ref{ineqDec17}),
\bea\label{ineqDec17a}
2(d-2)d_- g^2+\frac{2nn_-}{C_*}-\frac{q_1^2}{C_*^n}\le 0,
\eea
is inconsistent with the defining equation (\ref{AdS32exct2}). The last statement follows from combining (\ref{ineqDec17a}) with (\ref{AdS32exct2}) to arrive at a contradiction:
\bea
2(d-2)d_-m_- g^2+\frac{n^2_-m_-}{C_*}\le 0.
\eea
To summarize, we have demonstrated that the curve describing the initial conditions 
$(q_2,C_1(q_2))$ in the $(q_2,C_0)$ plane starts at $(0,C_*)$ and terminates on the branch cut (\ref{BrnchCutSrf}) without going to arbitrarily large values of $q_2$. In other words, we have justified the first line in (\ref{q2Ranges}). To prove the statements in the second and third lines, we note that the lines describing solutions with different asymptotics can't cross. The discussion presented in section \ref{SecMagn} implies that $q_2=0$, there is an inequality
\bea
C_1(0)<C_2(0)<C_3(0),
\eea
and we just showed that $C_1(q_2)$ exists only for $q_2\le q_S$, so it has to disappear under the branch cut at $q_2= q^{(bc)}_S\le q_S$. Combination of these facts implies (\ref{q2Ranges}), as well as an inequality
\bea\label{q2RangesBc}
q^{(bc)}_H<q^{(bc)}_{AdS}<q^{(bc)}_S\,.
\eea
Numerical examples showing the shapes of $(C_1,C_2,C_3)$ which confirm these relations for 
$(m,n)=(2,2)$ are shown in figures \ref{FigDyonApBig}, \ref{FigDyonApDetail}, but our analytical proof of (\ref{q2Ranges}) and (\ref{q2RangesBc}) holds in all dimensions. 

Let us now comment on solutions with $A_-$. These branches appear at $q_2=q^{(bc)}$, and they exist only in some ranges, for example, $q^{(+)}_S\le q_2\le q^{(bc)}_S$. This can be proven using the logic similar to one in the last paragraph. In contrast to the $A_+$ lines, the $A_-$ curves must begin and end on the branch cut since they cannot extend to arbitrarily small values of $q_2$. Numerical analysis presented below demonstrates that all lines with $A_-$ terminate on the 
$AdS_{m}\times S^{n}\times R$ points (\ref{SpecSolnNoRun}). This is not very surprising since perturbations of $AdS_{m}\times S^{n}\times R$ necessarily lead to radial dependence in various 
warp factors, so perturbing in various directions, one can get running in $A$, $C$, or both of them. 

\bigskip

After this general qualitative discussion, let us summarize the full structure of phases in the dyonic system and illustrate it by the numerical results for the $(m,n)=(2,2)$ case. We begin with solutions described by the boundary conditions 
\bea\label{Dec17BC}
A(0)=A_+,\quad C(0)=C_0,\quad {\dot A}(0)={\dot C}(0)=0,
\eea
where $A_+$ is the largest positive solution of the constraint (\ref{BounCnstrDyon}) on $A_0$.

\bigskip
\noindent
{\bf Solutions with $A(0)=A_+$.}

\begin{enumerate}[A.]
\item $q_2<q_{H}^{(bc)}$\\
In this range of charges, the behavior is qualitatively similar to the one discussed in section 
\ref{SecMagn}. The main difference is that now the boundary values $(C_1,C_2,C_3)$ acquire $q_2$--dependence, but in the $q_2=0$ limit they reduce to $(C_*,C_\#,C_\bullet)$:
\bea
C_1(q_2=0)=C_*,\quad C_1(q_2=0)=C_\#,\quad C_3(q_2=0)=C_\bullet. 
\eea
Recall that $q_{H}^{(bc)}$ was defined as the value at which the initial condition for $AdS_m\times H_{n+1}$ crosses the branch cut.  Let us briefly summarize various branches:

\begin{enumerate}[(a)]
\item $C_0\le C_{min}$: unphysical branch with a negative warp factor $A$.

The critical point $C_{min}$ is determined by solving the algebraic equation of $n$--th degree:
\bea\label{CminDefIntro}
\frac{nn_-}{C_{min}}-\frac{q_1^2}{2(C_{min})^n}+(d-2)d_-g^2=0.
\eea

\item $C_{min}<C_0<C_1$: a flow to a naked curvature singularity at $r=r_1>0$. 

\item $C_0=C_1$: a flow to the $AdS_{m+1}\times S^n$ solution (\ref{AdS32exct1}), (\ref{AdS32exct3}), (\ref{AdS32exct3a}). In contrast to the $q_2=0$ case, function $C$ has a nontrivial $r$--dependence.

\item $C_1<C_0<C_2$: a flow to an asymptotically locally $AdS_d$. 

\item  $C_0=C_2$: a flow to the standard $AdS_d$ at infinity. 

\item $C_2<C_0<C_3$: a flow to an  asymptotically locally  $AdS_d$. 

\item $C_0=C_3$: a flow to $AdS_m\times H_{n+1}$ at infinity.
\item $C_3<C_0\le C_{bc}$: a flow to a naked curvature singularity at $r=r_1>0$. Here 
$C_{bc}$ is the location of the branch cut for the given values of $(g,q_1,q_2)$. i.e., the solution of equation (\ref{BrnchCutSrf}) for $C_0$.
\item $C_0> C_{bc}$: system  (\ref{FullEinstein}) has no real solutions.
\end{enumerate}

\item $q_{H}^{(bc)}<q_2<q_{AdS}^{(bc)}$\\
The items (a)--(g) are the same as above, but now one crosses a second copy of $AdS_m\times H_{n+1}$ before reaching the branch cut. This leads to the following modifications:

\begin{enumerate}[(h)]
\item $C_3<C_0< C_{4}$: a flow to a naked curvature singularity at $r=r_1>0$. 
\end{enumerate}
(i) $C_0= C_4$: a flow to $AdS_m\times H_{n+1}$ at infinity.
\begin{enumerate}[(j)]
\item $C_4<C_0< C_{bc}$:  a flow to an  asymptotically locally  $AdS_d$. 
\end{enumerate}
\begin{enumerate}[(k)]
\item $C_0> C_{bc}$: system  (\ref{FullEinstein}) has no real solutions.
\end{enumerate}
At $q_2=q_{H}^{(bc)}$, the second $AdS_m\times H_{n+1}$ emerges from the branch cut (at that point $C_4=C_{bc})$, and as $q_2$ increases, the values of $C_3$ and $C_4$ are getting closer.
\item $q_{AdS}^{(bc)}<q_2<q_{S}$\\
As $q_2$ increases between $q_{AdS}^{(bc)}$ and $q_{S}$, there are two competing effects: additional branches emerge from the branch cut and two phases of the same type (e.g, $C_3$ and $C_4$) can collide and disappear. Therefore the $q_2$--dependence of the phase diagram becomes rather complicated, and instead of describing it in words, we present typical examples in figure \ref{FigDyonApDetail}. Although these numerical results were obtained for $(m,n)=(2,2)$, similar pictures hold in all dimensions. We use the following notation for the initial values of $C_0$:
\bea\label{defCk}
AdS_{m+1}\times S^n:&& C_1,\ C_6,\nn
AdS_{d}:&& C_2,\ C_5,\\
AdS_{m}\times H_{n+1}:&& C_3,\ C_4.\nonumber
\eea
If $C_k$ and $C_{l}$ exist for a particular set of parameters $(g,q_1,q_2)$, then they obey 
inequalities $C_k\le C_l$ if $k<l$. The pairs $(C_3,C_4)$, $(C_2,C_5)$, $(C_1,C_6)$ can collide and disappear as shown in figure \ref{FigDyonApDetail}. 
\item $q_2>q_{S}$\\
The last pair of flows disappears at $q_2=q_{S}$, when $C_1$ collides with $C_6$. Immediately to the left of that point, there are two flows to $AdS_{m+1}\times S^{n}$. Once the point $q_{S}$ is crossed, there are no physically interesting solutions, and one encounters only three regions:
\begin{enumerate}[(a)]
\item $C_0\le C_{min}$: unphysical branch with a negative warp factor $A$.
\item $C_{min}<C_0\le C_{bc}$: a flow to a naked curvature singularity at $r=r_1>0$. 
\item $C_0> C_{bc}$: system  (\ref{FullEinstein}) has no real solutions.
\end{enumerate}
\end{enumerate}
This concludes our discussion of the $A_+$ branch. The numerical results for 
$(C_1,\dots,C_6)$ as functions of $q_2$ are presented in figures \ref{FigDyonApBig}, \ref{FigDyonApDetail}.

\bigskip
\noindent
{\bf Solutions with $A(0)=A_-$.}

\begin{figure}
\begin{tabular}{ccc}
 \includegraphics[width=0.4 \textwidth]{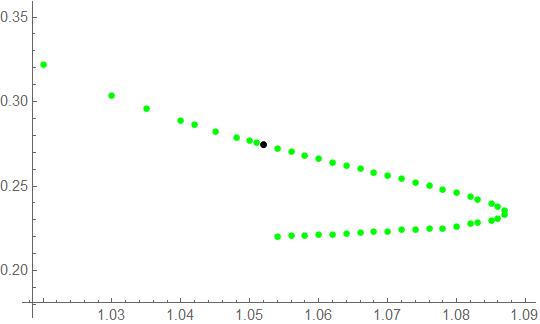}&\ 
 \includegraphics[width=0.4 \textwidth]{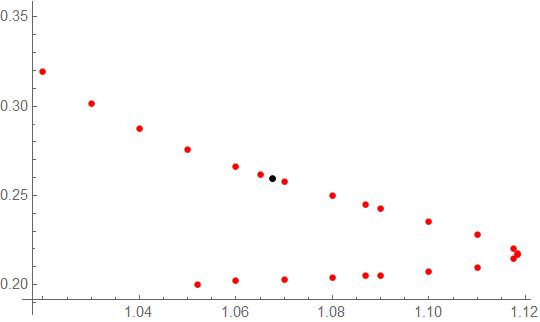}\\
 (a)&(b)
 \end{tabular}
\caption{Curves in the $(q_2,C_0)$ plane describing flows to various asymptotic regions for $m=n=2$, $q_1=g=1$ in the transition between $A_+$ and $A_-$ branches. Notation and color conventions are explained in the the text. Black dots denote the locations of the branch cut.}
\label{FigDyonAmMrg}
\end{figure}

\begin{figure}
 \begin{center}
 \includegraphics[width=0.7 \textwidth]{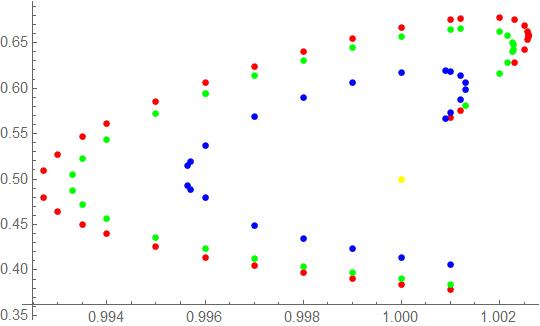}
\end{center}
\caption{Curves in the $(q_2,C_0)$ plane describing flows to various asymptotic regions for $m=n=2$, $q_1=g=1$, in the $A_-$ branch. Notation and color conventions are explained in the the text.}
\label{FigDyonAmElps}
\end{figure}

We will now consider the solutions of the system (\ref{FullEinstein}) with the boundary conditions (\ref{Dec17BC}), where $A_+$ is replaced by $A_-$, the second real solution of the constraint the constraint (\ref{BounCnstrDyon})\footnote{Recall that in the previous subsection we have demonstrated that this constraint can have at most two real positive solutions.}. Since in this case the structure of the phases is rather involved, we just give a qualitative description of various flows and refer to the Figures \ref{FigDyonAmMrg}, \ref{FigDyonAmElps} for the quantitative results. 

\begin{enumerate}[I.]
\item Flows to $AdS_{m+1}\times S^n$.\\
The line of flows to $AdS_{m+1}\times S^n$ begins at $q_2=q_S$ and terminates at $q_2=q_{cr}$, the $AdS_{m}\times S^n\times R$ solution. As can be seen from figure \ref{FigDyonAmElps}, this line first goes to the left, where it bounces from some $q_2<q_{cr}$. The line then goes above the top edge of the branch cut, and after bouncing from some $q_2>q_{cr}$, it goes to the critical point. The number of solutions for $C_0$ that give such flows is a complicated function of $q_2$. 

\item Flows to $AdS_{d}$.\\
The line of flows to $AdS_{d}$ begins at $q_2=q_{AdS}$ and terminates at $q_2=q_{cr}$. Just as the  $AdS_{m+1}\times S^n$ curve, this line bounces back and forth, but it always stays inside the area bounded by the $AdS_{m+1}\times S^n$ curve and the branch cut. Figure \ref{FigDyonAmElps} shows this for the $(m,n)=(2,2)$ example. 

\item Flows to $AdS_{m}\times H_{n+1}$.\\
The line of flows to $AdS_{m}\times H_{n+1}$ begins at $q_2=q_{H}$ and terminates at $q_2=q_{cr}$. This curve always stays inside the area bounded by the $AdS_{d}$ line and the branch cut. Figure \ref{FigDyonAmElps} shows this for the $(m,n)=(2,2)$ example. 
\end{enumerate}
This concludes our discussion of the $A_-$ branch. 

\section{Discussion}

In this article we have constructed regular geometries connecting two $AdS_p\times S^q$  or 
$AdS_d$ asymptotic regions via wormhole--like bridges. Our solutions are supported by a negative cosmological constant and magnetic or dyonic fluxes. The latter case exhibits a rich phase structure of flows between geometries with a wide variety of asymptotic behaviors. In the purely electric case, regular solutions have only one asymptotic region, and we have constructed such geometries as well. All our findings are summarized in section \ref{SecSetup}.

It would be interesting to extend our results in several directions. First, the products of maximally symmetric spaces, which arise both at infinities and in the near--bridge limit of our geometries, can often be embedded as supersymmetric solutions into supergravity theories. In cases when it is possible\footnote{While construction works in all dimensions, there are no supegravities with $d>11$.}, it would be interesting to see whether our flows preserve some supersymmetries. Second, it would be nice to extend our results to rotating geometries by relaxing the ansatz (\ref{TheAnsatz1}) and allowing dependence on two or more coordinates. Finally, to connect to the recent work on transversable wormholes \cite{EREPR}, it would be interesting to look for solutions with multiple bridges connecting the same pair of asymptotic regions and to analyze stability of such configurations. While we expect the solutions constructed in this article to be stable due to conservation of the magnetic flux (the sphere supporting the bridge cannot collapse without creating an infinite flux density), it would be interesting to demonstrate this stability directly by studying propagation of particles and waves on our wormhole--like geometries.

\section*{Acknowledgements}

This work was supported in part by the DOE grant DE-SC0015535. RD also received partial support from the Graduate Research Fellowship at the University at Albany.

%\end{document}
\appendix

\section{Wormholes with small magnetic charge}
\label{SecApp}
\renewcommand{\theequation}{A.\arabic{equation}}
\setcounter{equation}{0}

In this appendix we provide some technical details supporting the discussion of solutions with small magnetic presented in section \ref{SecMgnSmChrg}. Our goal is to find the values of parameters $(a_{\#},a_{\bullet})$ in the expressions (\ref{CcritGbar}) for the initial conditions describing flows to $AdS_{m+n+1}$ and $AdS_m\times H_{n+1}$. As outlined in section \ref{SecMgnSmChrg} we will do this by solving equations (\ref{FullEinstein}) in three separate regions, and we will show that it is possible to match the resulting functions in the overlaps if the magnetic charge is sufficiently small. Specifically, we will assume that ${\bar g}$ defined by (\ref{gBarDef}) is much smaller than one.

Numerical analysis of the flows to $AdS_{d}$ and $AdS_m\times H_{n+1}$ asymptotics suggests that for small ${\bar g}$ the boundary conditions have the form (\ref{CcritGbar}),
\bea\label{CcritGbarApp}
{\bar C}_{\#}={\bar C}_\star+a_{\#} {\bar g}^m,\quad 
{\bar C}_{\bullet}={\bar C}_\star+a_{\bullet} {\bar g}^m,
\eea
where parameters $(a_{\#},a_{\bullet})$ remain finite in the small ${\bar g}$ limit. Motivated by  this observation, we impose the boundary condition
\bea\label{CbarApp}
{\bar C}(0)={\bar C}_\star+a {\bar g}^m
\eea
with finite $a$ and demonstrate that the desired flows are indeed recovered for some values of this parameter. Note that we used the numerical data only as inspiration, and eventually the assumption (\ref{CbarApp}) is justified by the result of the analytical discussion presented below. 
For completeness we also give the expression for $C(0)$ corresponding to (\ref{CbarApp}):
\bea
{C}(0)={C}_\star+a q^{(m+2)/(n-1)}{g}^m.\nonumber
\eea
We will now proceed with solving equations (\ref{RescaledSystemEqn}) with the boundary
conditions (\ref{CbarApp}). To avoid unnecessary clutter, in this appendix we will not put bars over variables, i.e., we will make the following replacements:
\bea\label{ReplaceApp}
{\bar r}\rightarrow r,\quad {\bar A}\rightarrow A,\quad {\bar C}\rightarrow C.
\eea
To phrase this differently, we will look for solutions of system (\ref{FullEinsteinM}) with $q_1=1$ and small values of ${\bar g}$. Let us consider three separate regions.

\bigskip
\noindent
{\bf Region 1}

First we take the small ${\bar g}$ limit while keeping all functions fixed. In the leading order the solution is $AdS_{m+1}\times S^n$, so we write
\bea\label{Reg1Anstz}
A=A_\star {r^2}+\frac{1}{A_\star}+\eps A_1(r),\quad C=C_\star+\eps C_1(r),\quad A_\star=\frac{n-1}{2(d-2)m}C_\star^{-n}+\frac{d-1}{m}.
\eea
Parameter $C_\star$ satisfies a rescaled version of equation (\ref{AdS32exct2}),
\bea\label{Reg1Anstz1}
2(n-1)-\frac{m}{d-2}C_\star^{1-n}+2(d-1){\bar g}^2 C_\star=0,
\eea
and the leading contribution to the solution is
\bea\label{Reg1AnstzLeadC}
C_\star\simeq\left[\frac{m}{2(d-2)(n-1)}\right]^{\frac{1}{n-1}}\,.
\eea
For $\eps=0$, expressions (\ref{Reg1Anstz})--(\ref{Reg1Anstz1}) give the exact solution of the system (\ref{FullEinsteinM}), and our ansatz (\ref{CbarApp}) suggests that  
$\eps\propto {\bar g}^m$. Expanding equations (\ref{FullEinsteinM}) to the first order in 
$\eps$ and substituting the leading approximation (\ref{Reg1AnstzLeadC}) in the result\footnote{Subleading terms in (\ref{Reg1Anstz1}) become important if one looks at higher powers in $\eps$, and the required accuracy in expanding (\ref{Reg1Anstz1}) depends on $m$ and $n$.}, we find linear equations for 
$A_1$ and $C_1$:
\bea\label{Aug4}
&&\left[r^2+\frac{m^4C_\star^2}{(n-1)^4}\right]{\ddot C}_1+(m+1)r {\dot C}_1-2m^2 C_1=0,\\
&&C_\star^2\frac{m^2m_-}{n_-^2} \frac{d}{dr}\frac{A_1}{r}+n\left[r^2+\frac{m^4C_\star^2}{(n-1)^4}\right] \frac{d}{dr}\frac{C_1}{r}=0.
\nonumber
\eea
The most general solution reads
\bea\label{Aug4a}
C_1(r)=\left[{\hat r}^2+1\right]^{\frac{1-m}{4}}\left\{c_1 Q^{(\beta)}_\alpha\left[i{\hat r}\right]+c_2 P^{(\beta)}_\alpha\left[i{\hat r}\right]\right\},\quad {\hat r}\equiv \frac{(n-1)^2r}{m^2C_\star}\,.
\eea
Here $(P^{(\beta)}_\alpha,Q^{(\beta)}_\alpha)$ are the Legendre functions with
\bea
\alpha=\frac{3m-1}{2},\quad \beta=\frac{m-1}{2}\,.
\eea
The expression for $A_1$ in terms of Legendre functions is not very illuminating. Note that the homogeneous solution of the second equation in (\ref{Aug4}), $A_1=r$, should be discarded since it doesn't satisfy the initial condition ${\dot A}(0)=0$. The even solution in (\ref{Aug4a}) is
\bea\label{C1soln}
C_1(r)=c_1\left[{\hat r}^2+1\right]^{\frac{1-m}{4}}Q^{(\beta)}_\alpha\left[i{\hat r}\right].
\eea
It is useful to recall the expression for $Q^{(\beta)}_\alpha$ in terms of the hypergeometric function:
\bea\label{Qhyper}
Q^{(\beta)}_\alpha(z)=\frac{\sqrt{\pi}\Gamma[\alpha+\beta+1]}{2^{\alpha+1}\Gamma[\alpha+\frac{3}{2}]}\frac{e^{i\beta\pi}(z^2-1)^{\alpha/2}}{z^{\alpha+\beta+1}}F\left[\frac{\alpha+\beta+1}{2},\frac{\alpha+\beta+2}{2};\alpha+\frac{3}{2};\frac{1}{z^2}\right].
\eea
For even values of $m$, function (\ref{C1soln}) is a polynomial of degree $m$ in ${\hat r}^2$. The first few instances relevant for our discussion are
\bea\label{C1evenM}
m=2:&& C_1({r})=c_1\sqrt{\frac{\pi}{2}}\left[1+4{\hat r}^2\right],\nn
m=4:&& C_1({r})=c_1\sqrt{\frac{\pi}{2}}\left[6+96{\hat r}^2+160{\hat r}^4\right],\\
m=6:&& C_1({r})=c_1\sqrt{\frac{\pi}{2}}\left[80+2880{\hat r}^2+13440{\hat r}^4+14336{\hat r}^6\right].\nonumber
\eea
For the first two odd values of $m$ we find\footnote{To have a better physical interpretation we changed the sign of $c_1$ in comparison to (\ref{C1soln}).}:
\bea\label{C1oddM}
m=3:&& C_1({r})=c_1\left[\frac{5}{3}+\frac{35{\hat r}^2}{2}+\frac{1}{{\hat r}^2+1}+\frac{5{\hat r}}{2}(3+7{\hat r}^2)\arctan{\hat r}\right],\nn
m=5:&& C_1({r})=c_1\left[\frac{273}{5}+\frac{231}{8}{\hat r}^2(17+39{\hat r}^2)-\frac{27{\hat r}^2+29}{({\hat r}^2+1)^2}\right.\\
&&\qquad\qquad\qquad \left.+\frac{63{\hat r}}{8}(15+110{\hat r}^2+143{\hat r}^4)\arctan{\hat r}
\right].\nonumber
\eea
The leading behavior of $C_1$ at large values of ${\hat r}$ is given by
\bea
C_1(r)\simeq\gamma_m C(0){\hat r}^m\,,
\eea
where
\bea\label{GamaNumb}
\gamma_2=4,\quad \gamma_4=\frac{80}{3},\quad \gamma_6=\frac{896}{5},\quad
\gamma_3=\frac{21\pi}{4},\quad \gamma_5=\frac{165\pi}{16},\quad \gamma_7=\frac{4849845\pi}{32768}\,.
\eea
To summarize, we conclude that if $C$ starts as 
\bea
C=C_\star+\eps+O(r^2)
\eea
near $r=0$, then for large values of $r$ in the region 1 we find 
\bea
C=C_\star+\eps\gamma_m\left[\frac{(n-1)^2r}{m^2C_\star}\right]^m+\dots
\eea
Recalling that $\eps\propto {\bar g}^m$, the effect of region 1 can be summarized as
\bea\label{Reg1Cfin}
C=C_\star+a{\bar g}^m+O(r^2)\ \rightarrow \ C=C_\star+a\gamma_m\left[\frac{(n-1)^2{\bar g}r}{m^2C_\star}\right]^m+\dots
\eea
The large--$r$ behavior of function $A$ is determined by (\ref{Reg1Anstz}) and  (\ref{Aug4}):
\bea\label{Reg1Afin}
A=A_\star {r^2}+\frac{1}{A_\star}-\frac{a\gamma_m}{{\bar g}^2} \frac{nm^2}{(m+1)(n-1)^2}\left[\frac{(n-1)^2{\bar g}r}{m^2C_\star}\right]^{m+2}+\dots
\eea
Note that in the region 1 we used an approximation of small ${\bar g}$ (or small $\eps$ in (\ref{Reg1Anstz})) with finite $r$, and this regime breaks down when $r$ becomes comparable with ${\bar g}^{-1}$ . At that point corrections to (\ref{Reg1Cfin}) and (\ref{Reg1Afin}) which scale as higher powers of ${\bar g}$ become important. To extend the solution beyond region 1, we rewrite (\ref{Reg1Cfin})--(\ref{Reg1Afin}) in a suggestive form
\bea\label{Reg1Exit}
C&=&C_\star+a\gamma_m\left[\frac{(n-1)^2{\bar g}r}{m^2C_\star}\right]^m+\dots,\nn
A&=&\frac{1}{{\bar g}^2}\left[({\bar g}{r})^2 A_\star+\frac{{\bar g}^2}{A_\star}-\frac{a\gamma_m}{{\bar g}^2} \frac{nm^2}{(m+1)(n-1)^2}\left[\frac{(n-1)^2{\bar g}r}{m^2C_\star}\right]^{m+2}+\dots\right].
\eea
This indicates that beyond region 1, the natural variable is $x={\bar g}r$, and the constant term $\frac{{\bar g}^2}{A_\star}$ disappears. 

\bigskip
\noindent
{\bf Region 2}

Equation  (\ref{Reg1Exit}), describing the exit from region 1 at large $r$, suggests making a rescaling of $r$ coordinate and warp factors,
\bea
A=\frac{1}{{\bar g}^2}A_2({\bar g}r),\quad C=C_2({\bar g}r),\quad x={\bar g}r,
\eea
and taking the small ${\bar g}$ limit in (\ref{FullEinsteinM}) while keeping $x$ fixed:
\bea\label{Reg2syst}
&&\frac{n_- }{2(d-2)C_2^n}-\frac{n\dot A_2\dot C_2}{4C_2}+\frac{nA_2{\dot C}_2^2}{4C_2^2}-\frac{m}{2}\ddot A_2-\frac{nA_2\ddot C_2}{2C_2}=0,
\nn
&&\frac{n_-}{C_2}-\frac{m}{2(d-2)C_2^n}-\frac{m_+{\dot A}_2{\dot C}_2}{4C_2}-\frac{(n-2)A_2{\dot C}_2^2}{4C_2^2}-
\frac{A_2\ddot C_2}{2C_2}=0,\\
&&\frac{n_-A_2}{2(d-2)C_2^n}-\frac{m_-}{4}{\dot A}_2^2-\frac{nA_2{\dot A}_2{\dot C}_2}{4C_2}-\frac{1}{2}A_2{\ddot A}_2=0.\nonumber
\eea
Since this system does not have an explicit $x$--dependence, we can lower the order of equations by using $A_2$ as an independent variable and defining
\bea
Y(A_2)=\frac{dx}{dA_2}, \quad C_2=C_2(A_2).
\eea
Then the last equation in (\ref{Reg2syst}) becomes
\bea
\frac{n_-A_2}{2(d-2)C_2^n}-\frac{m_-}{4Y^2}-\frac{nA_2C_2'}{4C_2Y^2}+\frac{A_2Y'}{2Y^2}=0.
\eea
Defining a new function $Z$ by 
\bea
Y=C_2^{n/2} Z,
\eea
we arrive at a decoupled first order equation for $Z(A_2)$:
\bea
2A_2 Z'-{m_- Z}+\frac{2n_-}{d-2}A_2Z^3=0.
\eea
The general solution reads
\bea
Z=\sqrt{\frac{m(d-2)}{2(n-1)}}\frac{A_2^{m/2}}{[A_2^{m+1}+c]^{1/2}}\,,
\eea
where $c$ is an integration constant. To recover the expressions (\ref{Reg1Exit}) at small values of $x$, function $Z$ should diverge when $A_2$ goes to zero. This fixes the integration constant to $c=0$, leading to the final expression
\bea
Y=\left[\frac{m(d-2)}{2(n-1)}\frac{C_2^n}{A_2}\right]^{\frac{1}{2}}\,.
\eea
Rewriting the system (\ref{Reg2syst}) in terms of $\{Y(A_2),C_2(A_2)\}$, and substituting the last expression, we verify that the third equation is satisfied, while the first two become
\bea\label{Aug2}
&&(n+1)A_2 C_2'+(m-2)C_2-\frac{2A_2C_2C_2''}{C_2'}=0,\nn
&&-\frac{m^2C_2}{n_-A_2}+\frac{2(d-2)mC_2^n}{A_2}+\frac{2A_2[C_2']^2}{C_2}-\left[(m+2)C_2'+2A_2 C_2''\right]=0.
\eea
The most general solution of the first equation in (\ref{Aug2}) is
\bea
C_2=\left[c_2\Big(1-(c_1A_2)^{m/2}\Big)^2\right]^{-\frac{1}{n-1}}\,,\nonumber
\eea
where $(c_1,c_2)$ are integration constants. 
The second equation in  (\ref{Aug2}) determines the value of $c_2$ leading to the final answer:
\bea\label{SolnReg2}
Y=\left[\frac{m(d-2)}{2(n-1)}\frac{C_2^n}{A_2}\right]^{\frac{1}{2}},\quad 
C_2=\left[\frac{2(d-2)(n-1)}{m}\Big(1-(c_1A_2)^{m/2}\Big)^2\right]^{-\frac{1}{n-1}}\,.
\eea
Recalling that $Y=dx/dA_2$, we can integrate these expressions to find
\bea\label{Aug2a}
x&=&b\int \frac{dA_2}{\sqrt{A_2}}\Big[1-(c_1A_2)^{m/2}\Big]^{-\frac{n}{n-1}}=2a\sqrt{A_2}F\left[\frac{1}{m},\frac{n}{n-1},1+\frac{1}{m},
(c_1 A_2)^{m/2}\right],\nn
\\
b&=&\left[\frac{m(d-2)}{2(n-1)}\right]^{\frac{1}{2}}\left[\frac{2(d-2)(n-1)}{m}\right]^{-\frac{n}{2(n-1)}}
=\left[\frac{m}{2(n-1)}\right]^{\frac{2n-1}{2(n-1)}}(d-2)^{-\frac{1}{2(n-1)}}\,.
\nonumber
\eea
Here $F$ is the hypergeometric function. 

The final integration constant $c_1$ can be expressed in terms of the parameter $a$ in (\ref{Reg1Cfin}) by matching the small $A_2$ expansion of (\ref{Aug2a}) with (\ref{Reg1Exit}). 
To do that, we combine equations (\ref{Reg1Exit}) into a relation between $C$ and $A$:
\bea
C&=&C_\star+a\gamma_m\left[\frac{(n-1)^2{\bar g}}{m^2C_\star}\right]^m
\left[\frac{A}{A_*}\right]^{m/2}+\dots
\eea
Comparing this with expansion of (\ref{SolnReg2})
\bea\label{SolnReg2exp}
C_2=\left[\frac{m}{2(d-2)n_-}\right]^{\frac{1}{n-1}}\left[1+\frac{2}{n-1}(c_1A_2)^{m/2}+\dots\right]
=C_\star+\frac{2C_\star}{n-1}(c_1A_2)^{m/2}+\dots,
\eea
we find
\bea\label{c1Reg2}
c_1=\frac{1}{A_\star}\left[\frac{n_- a\gamma_m}{2C_\star}\right]^{\frac{2}{m}}\left[\frac{(n-1)^2{\bar g}}{m^2C_\star}\right]^2\,.
\eea
In this approximation, the values of $A_\star$ and $C_\star$ are given by (\ref{Reg1Anstz}) and (\ref{Reg1AnstzLeadC}). 

Region 2 continues to large values of $x$, where $A_2$ approaches $1/c_1$. This implies that 
\bea\label{ExitReg2}
A= \frac{1}{{\bar g}^2 c_1}+\dots,\quad C=c_1({\bar g}r)^2+\dots
\eea
The $(m,n)$--dependent expansions in inverse powers of $r$ are fully determined by one parameter $c_1$. Note that $A$ saturates to a constant when $x={\bar g}r$ becomes large but still remains ${\bar g}$--independent. Equations (\ref{Reg2syst}) and their consequences (\ref{SolnReg2})--(\ref{ExitReg2}) don't apply when $r$ becomes as large as ${\bar g}^{-2}$, so we need a separate analysis in a new region. 

\bigskip
\noindent
{\bf Region 3}

The exit from region 2 at large $r$ suggest a rescaling
\bea
A=\frac{1}{{\bar g}^2}A_3({\bar g}^2r),\quad C=\frac{1}{{\bar g}^2}C_3({\bar g}^2 r),\quad y={\bar g}^2r.
\eea
Then expressions (\ref{ExitReg2}) imply that at small $y$ functions $(A_3,C_3)$ behave as
\bea\label{Reg3BC}
A_3=\frac{1}{c_1}+O(y),\quad C_3=c_1 y^2+O(y^3).
\eea
The equations become
\bea\label{Reg3syst}
&&d_--\frac{n\dot A\dot C}{4C}+\frac{nA{\dot C}^2}{4C^2}-\frac{m}{2}\ddot A-\frac{nA\ddot C}{2C}=0,
\nn
&&d_-+\frac{n_-}{C}-\frac{m_+{\dot A}{\dot C}}{4C}-\frac{(n-2)A{\dot C}^2}{4C^2}-
\frac{A\ddot C}{2C}=0,\\
&&d_-A-m_--\frac{m_-}{4}{\dot A}^2-\frac{nA{\dot A}{\dot C}}{4C}-\frac{1}{2}A{\ddot A}=0.\nonumber
\eea
Every value of $c_1$ in (\ref{Reg3BC}) determines the unique solution of this system. We are particularly interested in two flows (a)-(b) from the list below:
\begin{enumerate}[(a)]
\item The geometries with the standard $AdS_d$ asymptotics have $C/A=1$ at infinity. This can happen only at one value of $c_1$ in (\ref{Reg3BC}), and the corresponding exact solution of (\ref{Reg3syst}) is remarkably simple:
\bea
A_3=y^2+1,\quad C_3=y^2, \quad c_1=1.
\eea
In the original variables this translates into
\bea
A={\bar g}^2 (r-r_0)^2+\frac{1}{{\bar g}^2},\quad C={\bar g}^2 (r-r_0)^2.
\eea
The expression for the shift $r_0$ can be determined by taking subleading terms in the analysis presented above, but it is not illuminating. We just mention its scaling with ${\bar g}$: 
$r_0\propto {{\bar g}}^{-1}$.
\item The geometries with $AdS_p\times X$ asymptotics have constant $A$ at infinity, and given the boundary conditions (\ref{Reg3BC}), it is reasonable to assume that $A$ remains constant in the entire region 3. The solution with constant $A_3$ indeed exists, and it is unique:
\bea
A_3=\frac{m-1}{d-1},\quad C_3=\frac{n}{4(d-1)}\left[w-\frac{1}{w}\right]^2,\quad 
w=\exp\Big[\frac{(d-1)y}{\sqrt{n(m-1)}}\Big].
\eea
This determines the value of $c_1$ in (\ref{Reg3BC}). In terms of the original variables we find the large $r$ behavior
\bea
A=\frac{1}{{\bar g}^2}\frac{m-1}{d-1},\quad C=\frac{1}{{\bar g}^2}\frac{n}{4(d-1)}\exp\left[\frac{2(d-1){\bar g}^2(r-r_0)}{\sqrt{n(m-1)}}\right],
\eea
where $r_0\propto {{\bar g}}^{-1}$ is a constant shift that can be extracted from subleading orders in powers of ${\bar g}$.
\item Any other value of $c_1$ leads to nontrivial expansions
\bea
A_3&=&\frac{1}{c_1}+\frac{d-1-(m-1)c_1}{n+1}\left[y^2+\frac{c_1(c_1-1)(m-1)(d-1)}{3(n+1)(n+3)}y^4+\dots\right],\nn
C_3&=&c_1 y^2+\frac{c_1^2(c_1-1)(m-1)(d-1)}{3n(n+1)}y^4+\dots
\eea
\item There is an unphysical exact solution that does not satisfy the boundary conditions
\bea
A_3=\frac{m-1}{d-2}+y^2,\quad C_3=-\frac{n-1}{d-2}-\frac{n-1}{m-1}y^2.
\eea
\end{enumerate}
To construct physically interesting solutions, we will be interested only in options (a) and (b). 

\bigskip
\noindent
{\bf Summary}\\
Let us summarize the results derived in this appendix. Since this summary will be used in the main text, we will write it in proper variables without making the replacement (\ref{ReplaceApp}) used throughout this appendix.
\begin{itemize}
\item
Solution in region 1 connects the initial condition at $r=0$,
\bea\label{AppSum1}
{\bar C}={\bar C}_\star+a{\bar g}^m+O(r^2),\quad
{\bar C}_\star=\left[\frac{m}{2(d-2)(n-1)}\right]^{\frac{1}{n-1}},
\eea
with the behavior of the warp factor at $1\ll {\bar r}\ll {\bar g}^{-1}$:
\bea\label{AppSum2}
{\bar C}={\bar C}_\star+a\gamma_m\left[
\frac{(n-1)^2{\bar g}{\bar r}}{m^2{\bar C}_\star}\right]^m+\dots
\eea
Numerical factors $\gamma_m$ are determined by taking limits of the hypergeometric function, and some examples are given in equation (\ref{GamaNumb}).
\item Solution in region 2 connects the initial condition (\ref{AppSum2}) with the behavior of the warp factors at ${\bar g}^{-1}\ll {\bar r}\ll {\bar g}^{-2}$:
\bea\label{AppSum3}
{\bar A}= \frac{1}{{\bar g}^2 c_1}+...,\quad {\bar C}=c_1({\bar g}{\bar r})^2+...,
\quad
c_1=\frac{1}{{\bar A}_\star}\left[\frac{n_- a\gamma_m}{2{\bar C}_\star}\right]^{\frac{2}{m}}\left[\frac{(n-1)^2{\bar g}}{m^2{\bar C}_\star}\right]^2.
\eea
Here 
\bea\label{AppSum3a}
{\bar A}_\star=\frac{n-1}{2(d-2)m}{\bar C}_\star^{-n}+\frac{d-1}{m}.
\eea
Inversion of the last relation in (\ref{AppSum3}) gives
\bea\label{AppSum4} 
a=\frac{2{\bar C}_\star}{n_- \gamma_m}\left[c_1{\bar A}_\star\right]^{\frac{m}{2}}
\left[\frac{m^2{\bar C}_\star}{(n-1)^2{\bar g}}\right]^m\,.
\eea
\item Solution in region 3 connects the initial condition (\ref{AppSum3}) with the asymptotic behavior at infinity. There are only two physically interesting cases:
\bea\label{AppSum5}
AdS_d\ \mbox{asymptotics}:&&c_1=1,\nn
AdS_m\times H_{n+1}\ \mbox{asymptotics}:&&c_1=\frac{d-1}{m-1}.
\eea
Substitution of these values into (\ref{AppSum4}) gives the expressions for $a_{\#}$ and $a_{\bullet}$ in  (\ref{CcritGbarApp}).
\end{itemize}
This construction, along with explicit formulas for $A$ and $C$ scattered throughout this appendix, provide the analytic solutions for wormholes with $AdS_d$ and $AdS_m\times H_{n+1}$ asymptotics in the small charge approximation.

\end{document}